\documentstyle[12pt]{article}
\newcommand{\newsection}[1]{
\addtocounter{section}{1}
\setcounter{equation}{0}
\setcounter{subsection}{0}
\addcontentsline{toc}{section}{\protect
\numberline{\arabic{section}}{{\rm #1}}}
\vglue .6cm
\pagebreak[3]
\noindent{\bf  \thesection. #1}\nopagebreak[4]\par\vskip .3cm}
\newcommand{\newsubsection}[1]{
\addtocounter{subsection}{1}
\addcontentsline{toc}{subsection}{\protect
\numberline{\arabic{section}.\arabic{subsection}}{#1}}
\vglue .4cm
\pagebreak[3]
\noindent{\it \thesubsection. #1}\nopagebreak[4]\par\vskip .3cm}

\renewcommand{\theequation}{\thesection.\arabic{equation}}
\renewcommand{\thefootnote}{\fnsymbol{footnote}}
\newcommand{\ben}{\begin{enumerate}}
\newcommand{\een}{\end{enumerate}}
\newlength{\extraspace}
\newlength{\extraspaces}
\newcounter{dummy}
\newcommand{\bc}{\begin{center}}
\newcommand{\ec}{\end{center}}
\newcommand{\be}{\begin{equation}
\addtolength{\abovedisplayskip}{\extraspaces}
\addtolength{\belowdisplayskip}{\extraspaces}
\addtolength{\abovedisplayshortskip}{\extraspace}
\addtolength{\belowdisplayshortskip}{\extraspace}}
\newcommand{\ee}{\end{equation}}

\newcommand{\ba}{\begin{eqnarray}
\addtolength{\abovedisplayskip}{\extraspaces}
\addtolength{\belowdisplayskip}{\extraspaces}
\addtolength{\abovedisplayshortskip}{\extraspace}
\addtolength{\belowdisplayshortskip}{\extraspace}}
\newcommand{\ea}{\end{eqnarray}}
\newcommand{\nonu}{\nonumber \\[.5mm]}
\newcommand{\is}{& \!\! = \!\! &}
\newcommand{\ban}{\begin{eqnarray*}
\addtolength{\abovedisplayskip}{\extraspaces}
\addtolength{\belowdisplayskip}{\extraspaces}
\addtolength{\abovedisplayshortskip}{\extraspace}
\addtolength{\belowdisplayshortskip}{\extraspace}}
\newcommand{\ean}{\end{eqnarray*}}
\newcommand{\baa}{                         
\addtocounter{equation}{1}
\setcounter{dummy}{\value{equation}}
\setcounter{equation}{0}
\renewcommand{\theequation}{\thesection.\arabic{dummy}\alph{equation}}
\begin{eqnarray}
\addtolength{\abovedisplayskip}{\extraspaces}
\addtolength{\belowdisplayskip}{\extraspaces}
\addtolength{\abovedisplayshortskip}{\extraspace}
\addtolength{\belowdisplayshortskip}{\extraspace}}
\newcommand{\eaa}{                                       
\end{eqnarray}
\setcounter{equation}{\value{dummy}}
\renewcommand{\theequation}{\thesection.\arabic{equation}}}

\input epsf
\newcommand{\figuurnum}{\arabic{fignum}}
\newcommand{\tabelnum}{\arabic{tabel}}
\newcounter{fignum}
\newcounter{tabel}

\newcommand{\figuurplus}[3]{
\addtocounter{fignum}{1}
\hspace{-3mm}{\it fig.}\ \figuurnum.
\begin{figure}[t]\begin{center}
\leavevmode\hbox{\epsfxsize=#2 \epsffile{#1.eps}}\\[3mm]
\parbox{10cm}{\small \bf Fig.\ \figuurnum: \it #3}
\end{center} \end{figure}\hspace{-1.5mm}}
\newcommand{\fig}{{\it fig.}\ }

\newcommand{\tabelplus}[2]{
\addtocounter{tabel}{1}
\hspace{-3mm}{\it table}\ \tabelnum.
\begin{figure}[t]\begin{center}
#1\\[3mm]
\parbox{10cm}{\small \bf Table \tabelnum: \it #2}
\end{center} \end{figure}\hspace{-1.5mm}}

\newcommand{\insertfig}[2]{\leavevmode \vcenter{\hbox{\epsfxsize=#2 cm
\epsffile{#1.eps}}}}
\newcounter{tabnum}
\newcounter{xxx}
\newcommand{\bl}{\begin{list}{({\it\roman{xxx}})}{\usecounter{xxx}}}
\newcommand{\el}{\end{list}}
\renewcommand{\d}{{{\partial}}}
\newcommand{\pp}[1]{{\partial \over \partial #1}}             
             
\newcommand{\ppt}[1]{{\partial \over \partial t}}            
\newcommand{\ppx}[1]{{\partial \over \partial x}}            
\newcommand{\pqt}[1]{{\partial^2 \over \partial t^2}}            
\newcommand{\pqx}[1]{{\partial^2  \over \partial x^2}}            
\newcommand{\Pp}[2]{{\partial #1 \over \partial #2}}

\newcommand{\twomatrix}[4]{{\left(\begin{array}{cc}#1 & #2\\
#3 & #4 \end{array}\right)}}
\newcommand{\twomatrixd}[4]{{\left(\begin{array}{cc}
\displaystyle #1 & \displaystyle #2\\[2mm]
\displaystyle  #3  & \displaystyle #4 \end{array}\right)}}
\newcommand{\vectord}[2]{{#1 \choose #2}}
\newcommand{\ie}{{\it i.e.\ }}
\newcommand{\etc}{{\it etc.\ }}

\newcommand{\eg}{{\it e.g.\ }}

\renewcommand{\l}{\langle}
\newcommand{\r}{\rangle}
\newcommand{\Bl}{\Bigl\langle}
\newcommand{\Br}{\Bigr\rangle}

\renewcommand{\.}{\cdot}

\newcommand{\Pf}{{\rm Pf}\,}

\newcommand{\half}{{\textstyle{1\over 2}}}

\newcommand{\Z}{{\bf Z}}
\newcommand{\R}{{\bf R}}

\newcommand{\C}{{\bf C}}

\newcommand{\M}{{\bf M}}
\renewcommand{\H}{{\bf H}}

\newcommand{\cO}{{\cal O }}            
\newcommand{\cH}{{\cal H }}            
\newcommand{\cA}{{\cal A }}            
\newcommand{\cG}{{\cal G }}

\newcommand{\ra}{\rightarrow}

\newcommand{\lra}{{\mathop{\leftrightarrow}}}

\newcommand{\opra}{\mathop{\rightarrow}}

\newcommand{\gives}{{\ \Rightarrow\ }}
\newcommand{\inv}{^{-1}}

\renewcommand{\P}{{\bf P}}

\newcommand{\Tr}{{\rm Tr}\,}

\newcommand{\subg}{_{\strut g}}

\newcommand{\subzero}{_{\strut 0}}

\newcommand{\th}{^{\mit th}}

\newcommand{\dbar}{{\overline{\partial}}}

\newcommand{\cbar}{{\overline{c}}}
\newcommand{\ebar}{{\overline{e}}}

\newcommand{\ibar}{{\overline{\imath}}}
\newcommand{\jbar}{{\overline{\jmath}}}
\newcommand{\kbar}{{\overline{k}}}

\newcommand{\qbar}{{\overline{q}}}

\newcommand{\xbar}{{\overline{x}}}
\newcommand{\ybar}{{\overline{y}}}
\newcommand{\zbar}{{\overline{z}}}

\newcommand{\Gbar}{{\overline{G}}}

\newcommand{\Jbar}{{\overline{J}}}

\newcommand{\Lbar}{{\overline{L}}}

\newcommand{\Qbar}{{\overline{Q}}}

\newcommand{\Tbar}{{\overline{T}}}

\newcommand{\Zbar}{{\overline{Z}}}

\newcommand{\mubar}{{\overline{\mu}}}

\newcommand{\psibar}{{\overline{\psi}}}

\newcommand{\phibar}{{\overline{\phi}}}
\newcommand{\thetabar}{{\overline{\theta}}}
\newcommand{\taubar}{{\overline{\tau}}}
\newcommand{\etabar}{{\overline{\eta}}}
\newcommand{\lambdabar}{{\overline{\l}}}

\newcommand{\cM}{{\cal M}}
\newcommand{\cF}{{\cal F}}
\newcommand{\cN}{{\cal N}}
\newcommand{\cD}{{\cal D}}
\newcommand{\Diff}{{\rm Diff}}

\newcommand{\Int}{\mathop{\int}}
\newcommand{\Lie}[1]{{{\cal L}_{#1}}}

\renewcommand{\Im}{{\rm Im\,}}

\newcommand{\End}{{\rm End}}
\newcommand{\Vir}{{\mbox{\it Vir}}}

\newcommand{\option}[4]{ \left\{ \begin{array}{ll}
#1, & \hbox{#2}, \\[2mm]
#3, & \hbox{#4}. \end{array} \right. }
\newcommand{\threeoptions}[6]{ \left\{ \begin{array}{ll}
#1, & \hbox{#2}, \\[2mm]
#3, & \hbox{#4}, \\[2mm]
#5, & \hbox{#6}. \end{array} \right. }
\newcommand{\Aut}{{\rm Aut\,}}
\newcommand{\ext}{{\raisebox{.2ex}{$\textstyle\small\bigwedge$}}}

\def\a{\alpha} 
\def\b{\beta} 
\def\g{\gamma} 
\def\G{\Gamma}
\def\e{\epsilon}
\def\h{\eta}
\def\th{\theta} 
  
\def\k{\kappa}
\def\l{\lambda} 
\def\L{\Lambda} 
\def\m{\mu}
\def\n{\nu}
\def\r{\rho} 
\def\s{\sigma} 
\def\t{\tau}
\def\f{\phi} 
\def\F{\Phi} 
\def\w{\omega}
\def\W{\Omega} 
\def\v{\varphi} 

\def\<{\langle}
\def\>{\rangle}
\newcommand{\hep}[1]{{\tt hep-th/#1}}

\newcommand{\id}{{\bf 1}}

\newfont{\gothic}{eufm10 scaled\magstep1}

\newcommand{\Xhat}{{\widehat X}}
\newcommand{\cL}{{\cal L}}
\newcommand{\cP}{{\cal P}}
\newcommand{\cB}{{\cal B}}
\newcommand{\cT}{{\cal T}}

\newcommand{\adot}{{\dot a}}

\newcommand{\Mbar}{{\overline{\cM}}}
\newcommand{\bn}{{\bf n}}
\newcommand{\mod}[1]{{\ \hbox{(mod $#1$)}}}
\renewcommand{\R}{{\bf R}}
\renewcommand{\C}{{\bf C}}
\renewcommand{\P}{{\bf P}}
\renewcommand{\Z}{{\bf Z}}

\newcommand{\bps}{{\mbox{\small\sc BPS}}}
\newcommand{\third}{{\textstyle {1\over 3}}}
\newcommand{\pslash}{{p \hspace{-5pt} \slash}}
\newcommand{\dslash}{{\d \hspace{-7pt} \slash}}
\newcommand{\Dslash}{{D \hspace{-8pt} \slash}}
\renewcommand{\hat}{\widehat}
\renewcommand{\bar}{\overline}
\newcommand{\sdim}{{\rm sdim\,}}
\renewcommand{\Re}{{\rm Re}}
\newcommand{\Sym}{{\rm Sym}}
\newcommand{\cS}{{\cal S}}
\newcommand{\vol}{{\rm vol}}
\newcommand{\lltimes}{\times \hspace{-6pt} 
\raisebox{.2ex}{\mbox{$\scriptscriptstyle |$}} \;}

\begin{document}


\begin{flushright}
March 1997\\
\end{flushright}
\vspace{3cm}
\thispagestyle{empty}

\begin{center} 
{\Large\sc Les Houches Lectures on\\[5mm] Fields, Strings and
Duality\footnote{Lectures given at the Les Houches Summer School on
Theoretical Physics, Session LXIV: {\sl Quantum Symmetries,} Les
Houches, France, 1 Aug - 8 Sep 1995.}}\\[13mm]

{\sc Robbert Dijkgraaf}\\[2.5mm]
{\it Department of Mathematics}\\
{\it University of Amsterdam\\
Plantage Muidergracht 24,\\
1018 TV Amsterdam}\\
{\tt rhd@wins.uva.nl}\\
[18mm]

{\sc Abstract}

\end{center}

\noindent 

Notes of my 14 `lectures on everything' given at the 1995 Les Houches
school. An introductory course in topological and conformal field
theory, strings, gauge fields, supersymmetry and more. The presentation
is more mathematical then usual and takes a modern point of view
stressing moduli spaces, duality and the interconnectedness of the
subject. An apocryphal lecture on BPS states and D-branes is added. 

\vfill

\newpage
\renewcommand{\Large}{\normalsize} 
\tableofcontents

\newpage

\renewcommand{\thefootnote}{\arabic{footnote}}
\newsection{Introduction}

This is an almost literal write-up of the 14 lectures I gave at the 1995
Les Houches school {\it Quantum Symmetry}, plus an extra lecture on BPS
states and D-branes that summarizes some important developments that
have taken place after the school. It should be regarded as a
broad-brushed sketch of modern views on quantum field theory and string
theory, stressing moduli spaces, duality and the interconnectedness of the
subject. I have taken perhaps a more mathematical point of view than is
usual in these kind of introductory texts. I also tried to emphasize the
common themes of various fields, perhaps at the expense of completeness
and many details.

In the last decades, in a long series of abstractions and
generalizations, string theory has emerged as the leading candidate for
a fundamental theory of nature. Part of the motivation is
(unfortunately, from a physical point of view) the intrinsic
mathematical beauty of the theory. In fact, both directly and indirectly
string theory has influenced various mathematical fields, and {\it vice
versa} string theory has been influenced by recent mathematical
development. In the last two years we have witnessed a marked change in
pace in all this. In a confluence of a wide variety of ideas, many of
them dating back to the 70s and 80s, the structure, internal consistency
and beauty of string theory has greatly improved. We have now a much
better and clearer picture of what string theory is about. Somehow the
original concept of one-dimensional strings generalizing
zero-dimensional point particles has become less central. The extended
nature of the string and the infinite tower of oscillations that comes
with it, serves more to naturally regularize the important
field-theoretic massless excitations that include the gravitons and
gauge bosons that mediate the unified forces. Indeed, it is not at all
clear that the massive modes will form stable asymptotic states that can
be (in principal) observed.

More than just a generalization of field theory, the crucial fact seems
to be that string theory leads by itself naturally to the most important
physical principles, such a quantum mechanics, general relativity, gauge
theories and supersymmetry, without assuming these from the beginning.
Indeed, it seems that any quantum theory that combines these four
ingredients has to be a string theory. One therefore gets a the feeling
that we are essentially dealing with a unique, very constrained object.
The precise mathematical definition of this object still eludes us, but
the remarkable internal consistency has been a constant source of new
insights in the nature of quantum systems, the dynamics of gauge
theories, the geometry of space-time, and last but not least a wonderful
inspiration for beautiful mathematics.

Crucial in all the recent developments has been the concept of {\it
duality.} In fact, duality has been a powerful idea in physics for a
long time, both in statistical mechanics and field theory. In many
respects duality in field theory and string theory can be used as an
organizing principle, as I will try to do in these lectures.

The purpose of these notes is to give an introduction to many of the
important elementary concepts from a point of view that is considerable
more mathematical than usual, as was dictated by the particular audience
of this Les Houches school. We will start at a (pedantically) low level
and try to work our way up at a steady pace. In my lectures I used the
analogy of a mountain walk (in a leisurely, Swiss style) and I gave the
road map of \figuurplus{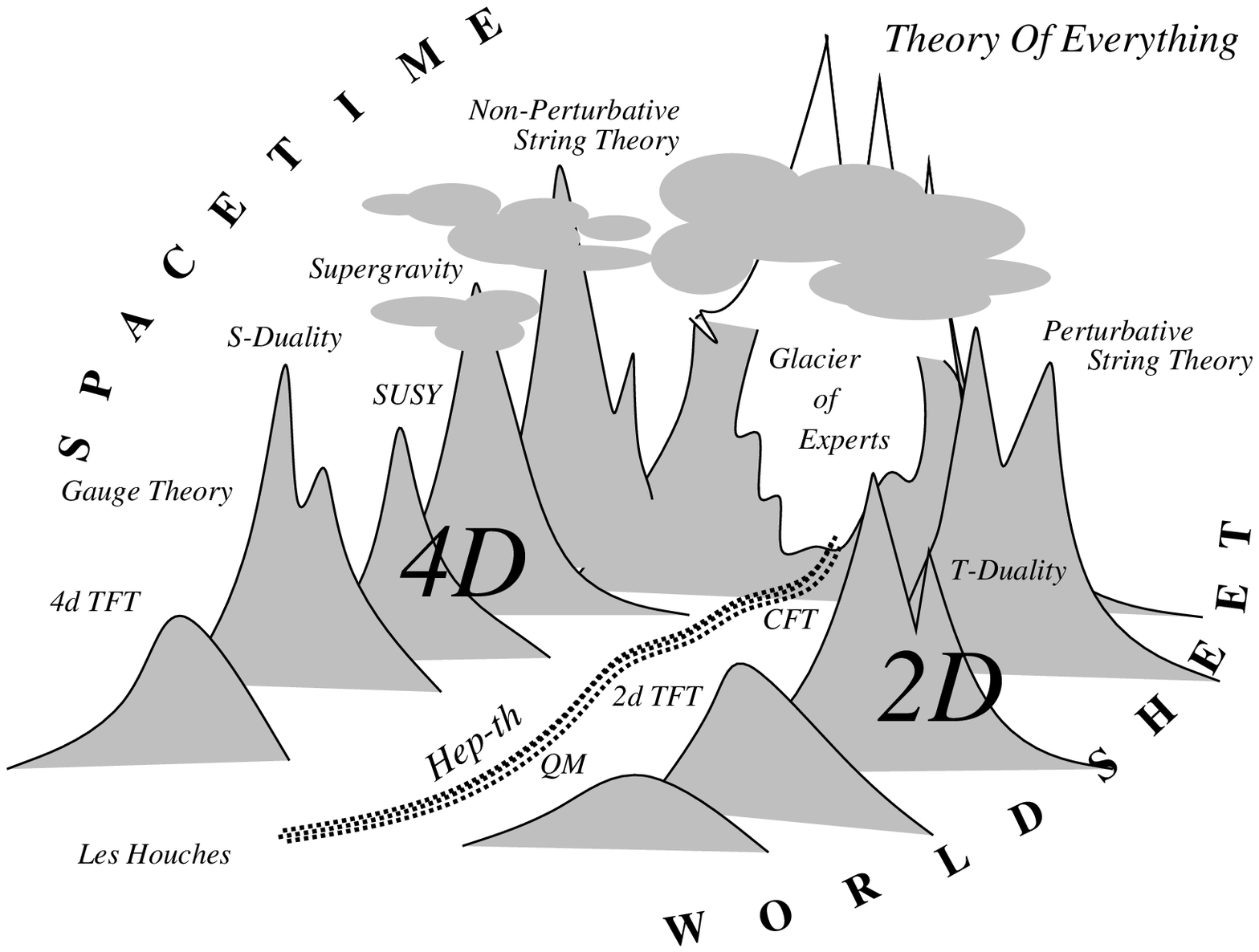}{13cm}{A road map for the lectures.} as a
summary of the course.

Let me describe it in some detail. The scenery consists of two mountain
ranges that we can label as {\it 2D} or {\it World-Sheet} and as {\it
4D} or {\it Space-Time}. They both lead to the great summit of the {\it
Theory of Everything.} Unfortunately, it is permanently hidden in a
thick layer of clouds, so we have no guarantee that our trail will
actually reach this summit. From the slopes we see the thick {\it
Glacier of Experts.} Now and then we witness enormous avalanches on the
glacier when big pieces of knowledge drop down and into the ferocious
preprint flow called {\it Hep-th} that comes running down the glacier.
This treacherous stream is very difficult to navigate, at least if you
are a typical Les Houches student. We will walk two trails that explore
the scenery, avoiding the glacier and the flow of preprints, that will
each take half of the course of lectures.

The first one traverses the two-dimensional range. If the first few
miles of this trail look familiar and uninteresting, I beg the reader
patience, since the surroundings will change quickly. We will start with
the small hills of {\it Quantum Mechanics,} after which we will move to
the barren scenery of {\it 2D Topological Field Theory,} a bare bones
version of quantum field theory. In the next stage we will put some
flesh to these bones, and explore {\it Conformal Field Theory.} Then we
are ready to meet the first examples of {\it T-Duality} in the context
of toroidal sigma models. After all this we make our first attempt to
scale {\it String Theory}. However, we exclusively take a perturbative
point of view using the language of Riemann surfaces. This will slow
down our pace and, more important, our view. Slightly disappointed with
the vista, we return to base camp and plan a second effort, avoiding the
Riemann surfaces.

In our second walk we explore the four-dimensional world, keeping
consistently a space-time perspective. Now we can use non-perturbative
techniques to our advantage. We start by looking at {\it S-Duality} in
abelian gauge theory. After a little detour to study four-dimensional
{\it Topological Field Theory,} we then move on to {\it Supersymmetry},
in particular supersymmetric gauge theories, spending a lot of time on
exploring the vacuum structure. Here the most prominent feature is the
{\it Seiberg-Witten Mountain Pass} that allows us to leap forward into the
non-perturbative world. The next mountain, {\it Supergravity}, we will
mostly skip because the guide feels very uneasy at its slippery slopes.
Instead we start immediately our second ascent of {\it String
Theory}, now using the powerful tools of duality. After this strenuous
trip, we spend some time at high altitude to look for stringy {\it BPS
States}, that we can take home as a souvenir of our adventure and show
our mathematical friends. Slightly dizzy of spending too much time at
these great heights, we return back home, leaving the guide exhausted.
(He had to recover for more than a full year, before he could even start
recounting this experience!)

Needless to say, it is unthinkable to make this trip and cover all
territory in fine detail, particularly if one walks in the high
mountains. Sometimes we have to jump over a few crevasses, and we might
not have always kept a careful balance. Whenever the reader gets lost, I
hope the (always incomplete) references to the extensive literature
can help to find the way back to the trail. The reader might even try
a refreshing dip in the {\it Hep-th} river in the end. Enjoy our trip! 

\newsection{What is a quantum field theory?}

Before we start the actual lectures, let us make a few very general
philosophical remarks about quantum field theory --- a hard to grasp
subject, despite decades of physical and mathematical efforts. 

\newsubsection{Axioms vs. path-integrals}

There are basically two ways in which one can approach quantum field 
theories: either in terms of axioms or in terms of quantization.

{\it 1. Axioms.} Here we do not refer to any classical theory. One
simply looks for a consistent set of amplitudes or correlation
functions, satisfying certain well-defined sets of axioms. Here one can
think of such diverse examples as the Wightman's formulation of
axiomatic field theory \cite{wightman} or Segal's geometric axioms of
conformal field theory \cite{segal}. Although this approach is much
preferred from a mathematical point of view, it is, unfortunately,
highly non-trivial to define realistic four-dimensional QFTs along these
lines. In fact, only for simpler systems such as topological field
theories or two-dimensional theories can this powerful approach be used
to its full extent and satisfaction. 

{\it 2. Quantization.} The textbook approach \cite{qft}. Here one starts with some
classical action $S(\f)$ for a classical field $\f(x)$ and subsequently
tries to make sense of the path-integral (over some appropriately
defined big enough, infinite-dimensional space of field configurations)
\be
 A=\int \cD\f \, e^{iS(\f)/\hbar} \; P(\f),
\label{path}
\ee
with $P(\f)$ some local expression representing the correlation function. One
usually first tries to interpret this definition in perturbation theory
in Planck's constant $\hbar$. In a certain regime, the amplitudes can
have a meaningful, although most likely asymptotic, expansion of the
form
\be
A \sim \sum_n A_g \hbar^g,
\ee
where the perturbative
coefficients $A_g$ are computed as sums over all Feynman diagrams
$\G$ with a fixed number of $g$ loops \cite{qft}
\be
A_g = \sum_{\G} {w_\G \over \# \Aut(\G)}.
\ee
Here the individual weight $w_\G$ of the diagram $\G$ is computed using
the Feynman rules and we divide by the order of the symmetry group of
the diagram (as one should do in all generating functions as a matter of
principle). This relation with graphs explains why quantum field
theories can describe point-particles and their interactions. The
expansion of the path-integral in this combinatorical, diagrammatical
fashion is of course a very general property of the asymptotics of
integral of weights $e^{iS(\f)}$, where $S(\f)$ is a polynomial in $\f$. A
famous finite-dimensional example of such an integral is the Airy function.

From this perturbative point of view there is no essential difference
between a field theory and a string theory. String theories just give
rise to amplitudes that allow expansions where the coefficients $A_g$
are given as sums of {\it surfaces} with $g$ handles. In general the
weight of a particular surface is computed as an integral over the
moduli space of inequivalent conformal metrics $\cM_g$ on
the surface \cite{gsw}
\be
A_g = \int_{\cM_g} w_g.
\ee
The beauty of perturbative string theory, as contrasted with
perturbative point-particle field theory, is that at fixed order $g$ in
the above expansion there is only one surface with $g$ handles to
consider, whereas the number of connected graphs growths as $g!$. So
string theory simplifies the combinatorics. In the
usual low-energy limit reduction of string theory to the supergravity
quantum field theory, these various diagrams can be recovered as
particular limits of degenerated surfaces in the ``corners'' of the
moduli space $\cM_g$. 

It is clear that this expansion in terms of surfaces will not be
generated in any obvious fashion by a path-integral of the form
(\ref{path}). Indeed, string field theory, which tries to do exactly
that, is a particularly complicated (and some would say, unnatural)
effort to capture the surface expansion is terms of graphs. In order to
do this successfully, one has to cut up the moduli space of Riemann
surfaces $\cM_g$ in cells, labeled by graphs, such that the Feynman sum
over graphs reduces to an integral over $\cM_g$. It is quite remarkable
that this can be done in a completely well-defined and mathematically
very beautiful way \cite{zwiebach}, and this formalism makes use of the
most advanced concepts in the quantization of gauge systems, such as the
BV-formalism \cite{henneaux-teitelboim}. But all in all, at this moment
the conceptual advances in this approach are less striking.

However strong the intuition behind the semi-classical picture is, both
in field theory and in string theory, the perturbative point of view
does not adequately capture everything about a quantum field theory, as
has become clear from our growing understanding of the phenomenon of
duality.

\newsubsection{Duality}

Indeed, recently a particular powerful framework to think about quantum
field theories has emerged, that is somehow complementary to the
axiomatic or path-integral approach and that stresses the concept of a
{\it moduli space} of quantum field theories. A central role is played
by duality symmetries. (For a nice physical introduction see
\eg \cite{jeff}.) Through these dualities the semi-classical
quantization point of view looses some of its uniqueness.

Just as is the case for many other fields of mathematics (in particular
algebraic geometry) it is useful to consider {\it families} of the
objects that one is studying. The space that parametrizes the objects is
usually called a moduli space, and can be studied on its own. For
example, in the case of Riemann surfaces we obtain the moduli space
$\cM_g$, first introduced by Riemann and one of the most intricated and
beautiful spaces in algebraic geometry, see \eg \cite{texel}. Other
famous moduli spaces are the spaces of vector bundles, self-dual
connections, holomorphic maps \etc

So, following this line of thought, it is natural to consider a family
of quantum field theories. This gives us a parametrized set of
amplitudes $A(\l)$, where the moduli $\l_1,\l_2,\ldots$ take value in a
moduli space $\cM$. Depending on the particular kind of quantum field
theory that we are interested in, the space $\cM$ will carry certain
geometric structures. Indeed, one of the important issues in these
lectures will be how properties of the QFT translate into properties of
the moduli space.

The classical action underlying the QFT can be recovered in perturbation
theory if there exists a point (or, more generally, a codimension one
subspace) in $\cM$ where one of the moduli vanishes, say $\l=0$, and
where the coefficients $A_g$ in the perturbative expansion of $A$ in $\l$ 
\be
A(\l) \sim \sum A_g \l^g
\label{expan}
\ee
can be computed by Feynman diagrams. Out of the Feynman rules we can
then ``in weak coupling'' reconstruct the action $S(\f)$. After the fact
we can then identify $\l$ as Planck's constant. The classical action
thus captures the description of the QFT in the neighbourhood of $\l=0$.
The zero-locus $\l=0$ is then parametrized by the remaining moduli,
that get now reinterpreted as a set of classical coupling
constants that appear in the action $S(\f)$.

And as we stressed above, this perturbative information is in general
incomplete (\eg has a zero radius of convergence) and has to be be
supplemented by non-perturbative corrections. We can think of the usual
semiclassical approach as giving some preferred set of local coordinates
on $\cM$ in the neighbourhood of $\l=0$ (technically the normal bundle
of the locus $\l=0$).

However, it is very well possible that there is another point $\l_0$
in the moduli space with analogous properties, see
\figuurplus{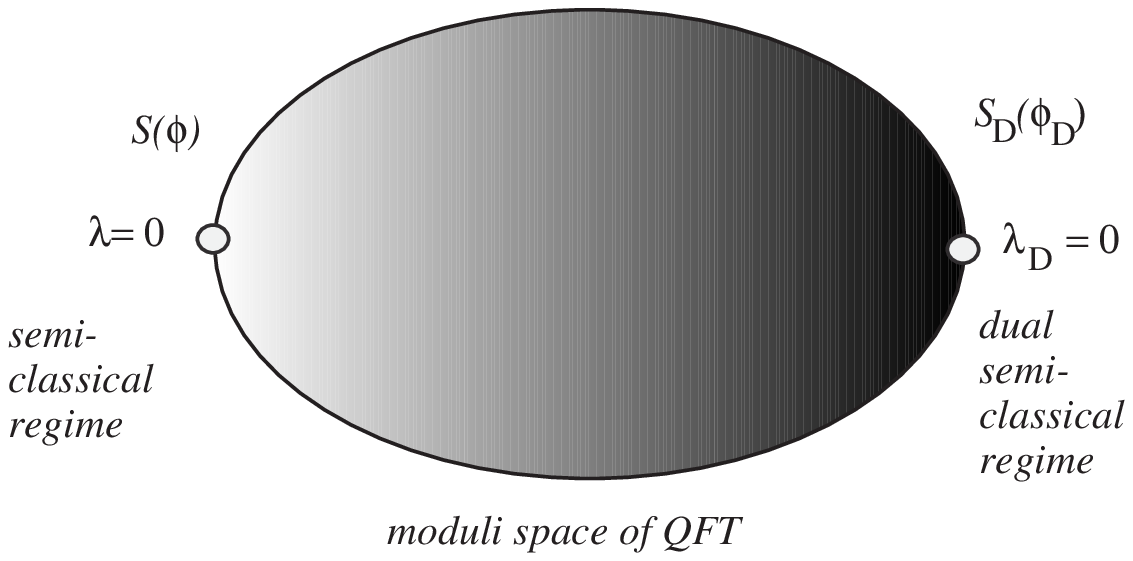}{10cm}{In the moduli space of a QFT we could have two
possible semi-classical, perturbative regimes, described by fields
$\f$ with a classical action $S(\f)$ and dual fields $\f_D$ with a
classical action $S_D(\f_D)$. The transformation relating the
variables in the two different patches is a duality transformation.}
That is, there can exist a second expansion parameter $\l_D=\l-\l_0$
and a second expansion
\be
A(\l_0+\l_D)\sim \sum B_g\l_D^g
\ee
with exactly the same properties as the original, semi-classical
expansion (\ref{expan}). Out of the coefficients $ B_g$ we can now
recover an {\it a priori} different classical action $S_D(\f_D)$,
possibly using a completely different set of fields $\f_D$. In that case
we have two alternative ``quantizations'' that give rise to the same
family of quantum field theories, and we speak of a dual description,
with dual fields and a dual action. Note that we also have two
candidates for Planck's constant, $\l$ and the dual expansion 
parameter $\l_D$.
This makes the concept of quantization ambiguous, to say the least.

There are various examples of such dual formulations, some known
for a long time, some only discovered recently. A rather famous case
of a duality transformation is the
identification between free bosons and free fermions in two dimensions,
 or more generally the identification between the
Sine-Gordon model and the massive Thirring model \cite{boson},
see \eg the textbook \cite{rajaraman}. The two models describe
respectively a scalar field $\v(x)$ with an interaction of the form
\be
\l\inv\cos(\l^{1\over2}\v)-1)
\ee
and a (massive) fermion field $\psi(x)$ with interaction
\be
\l_D (\psibar\psi)^2.
\ee
The claim is that these models are physically equivalent, where the 
coupling constants $\l,\l_D$ are related by
\be
\l = {1\over 1+ \l_D}
\ee
In particular weak coupling in the Sine-Gordon model corresponds to
strong coupling in the Thirring model.
In the solution of Coleman the fermion field $\psi$ is written in 
terms of the boson $\v$ as
the non-polynomial vertex operator
\be
\psi(x) = :e^{i\v(x)}:
\ee
This operator can be seen to create a soliton or kink in the Sine-Gordon
model. {\it Vice versa,} the bosons are recovered in the fermionic model
by ``bosonizing'' the charge current as
\be
\d\v(x) = \psibar\psi(x).
\ee

This two-dimensional example captures some important characteristics of
the more general picture. Quite often the dual variables describe
solitonic objects in the original weakly coupled system. These solitons
become infinitely massive at weak coupling. At strong coupling however,
the solitonic degrees of freedom can be come very light and take over
the role as preferred classical fields. 

For instance, the Montonen-Olive duality \cite{montonen-olive,jeff} of
four-dimensional $N=4$ non-abelian super Yang-Mills theory interchanges
the gauge bosons, that are typically thought of as fundamental fields,
with the 't Hooft-Polyakov monopoles \cite{monopole} that appear as
solitonic solutions to the classical field equations, thereby also
interchanging electric and magnetic fields. As such it is a quantum
realization of the classical electric-magnetic duality of (abelian)
Maxwell theory, that we will discuss at great length in \S9. These kind
of dualities have been generalized successfully to string theory
recently, see \eg
\cite{font,sen-duality,hull-townsend,witten-duality,schwarz-sl2}
and the reviews \cite{string-duality-reviews,polch-review}.
Indeed, string theory is at present our most fruitful source of these
bizarre quantum equivalences. We will see examples of this later in the
course.

An additional phenomenon that is nicely illustrated by the example of
Montonen-Olive duality, is the concept of {\it self-duality}.
Namely, it might be the case that the ``new'' dual theory we discovered 
at the second expansion point is actually identical to the old one,
\ie
\be
S(\f;\l')=S_D(\f_D;\l_D').
\ee
Here $\l'$ denote the remaining moduli that parametrize the classical
action and the subspace $\l=0$.
In the case that such a quantum identification exists,
the moduli space $\cM$ is clearly not parametrizing
inequivalent quantum field theories, since the points $\l=0$ and
$\l_D=0$ represent the same theory. We therefore have to divide by a
further group $G$ that identifies the dual descriptions. This group $G$
is called the {\it duality group}. It can be regarded as the natural
automorphism group of the QFT. The proper moduli space is then the
quotient of $\cM$ by $G$. 

If the dual coupling constants $\l$ and $\l_D$ are related as
\be
\l_D = 1/ \l,
\ee
or more generally, if the limit $\l_D\ra 0$ corresponds to
$\l\ra\infty$, we speak of an {\it S-duality.} This will then relate
weak coupling $\l=0$ to strong coupling $\l=\infty$. A typical example
in statistical physics of such a duality is the Kramers-Wanier duality
of the Ising model, another one is the Montonen-Olive duality mentioned
above. In string theory these S-dualities are particularly powerful.

The main lesson that duality teaches us, is that ``quantized fields'' might
not be the ultimate way to think about quantum field theories! It might
be necessary to cover the moduli space $\cM$ of quantum field theories
(or strings for that matter) by local patches, in a way analogous to the
coordinate patches by which we cover a manifold. In certain patches we
have a formulation in terms of fields, actions, and path-integrals. The
precise choice of fields and action might change from patch to patch.
The transition functions are the duality transformations.
There might also be parts of the moduli space where no such
semi-classical description exist at all, and we are left with the
``abstract'' QFT. 

This general description is of course terribly academic at this point.
Only when we will meet various examples of duality groups during the
lectures, it will become clear how powerful this notion of duality is.
With these warnings out of the way, let us now start the proper
lectures. 

\newsection{Quantum mechanics}

We start on familiar grounds --- quantum mechanics.
In it simplest form quantum mechanics consists of a Hilbert space $\cH$
and a unitary map 
\be
\F(t):\cH \ra \cH
\ee
describing the time evolution of the system during a time interval $t$. 
The composition law of time evolution tells us these transition
amplitudes satisfy the relation 
\be
\F(t_1)\circ \F(t_2)=\F(t_1+t_2),
\label{comp}
\ee
which implies the existence of an hermitian Hamiltonian $H$ with
$\F(t)=e^{itH}$. We will often consider the Euclidean case, obtained
by the analytic continuation $t \ra it$, where $\F(t) = e^{-tH}.$ 

The Euclidean partition function is defined as 
\be
Z(t)=\Tr e^{-tH}
\ee
(if this makes
mathematical sense)
and will then be a good way to encode the spectrum and degeneracies
of the system. Indeed, if the spectrum of $H$ is discrete with eigenvalues
$\e_n$ and finite degeneracies $d(n) = \dim \cH_n$ of the corresponding 
eigenspaces $\cH_n$ we have
\be
Z(t) =\sum_n d(n) e^{-\e_nt}.
\ee

A particular useful quantum mechanical model is the point particle
moving on a space $X$ and its supersymmetric cousin that we will
consider in a moment. In this case we have coordinates and momenta
$(q^\m,p_\m)$ that take value in the standard phase space $T^*X$, the
total space of the tangent bundle to the manifold. In canonical
quantization the Hilbert space is given as $\cH=L^2(X)$ and we realize
the operator $p_\m$ as $-i\pp {q^\m}$.
 
In the path-integral approach to quantization, we pick a Lagrangian 
of the form
\be
L = p_\m\dot q^\m - H(p,q)
\ee
with $H(p,q)$ the Hamiltonian function. With this Lagrangian we
compute the matrix elements
\be
K_t(y,x) = \<y|\F(t)|x\> 
\ee
of the transition amplitude $\F(t)$ as the Feynman-Kac
path-integral over all paths with $q(0)=x$, $q(t)=y$
\be
K_t(y,x) = \int\cD p\cD q\, e^{-\int_0^t L}.
\ee
The additivity of time translations (\ref{comp}) gets reflected in the
composition law of these kernels
\be
K_{t_1+t_2}(z,x) =  \int_{X}\!\! dy\; K_{t_1}(z,y) K_{t_2}(y,x).
\label{add}
\ee

The simplest case is actually the one with vanishing Hamiltonian, $H=0$,
which one could call topological quantum mechanics. This is an almost
empty system. As we see by doing the integral over $p$, the
path-integral localizes to $\dot q=0$, \ie to constant configurations
$q\in X$. So the evolution is indeed trivial with 
\be
K_t(x,y) = \delta(x,y)
\ee
and $\F(t)=\id$. The only
ingredient is this model is therefore the Hilbert space $\cH=L^2(X)$.

For the usual point particle we pick a metric $g_{\m\n}$ on $X$ and
define $H=p^2$. The Hamiltonian is therefore represented as the
Laplacian on $X$,
\be
H=-\Delta.
\ee
In that case $K_t(x,y)$ reduces to the usual heat-kernel, satisfying
\be
\Delta_x K_t(x,y)=\d_tK_t(x,y),\qquad K_0(x,y)=\delta(x,y).
\ee
An important example is the case of a torus, $X=T^n$, represented
as the quotient space $\R/2\pi\L$, with $\L$ a lattice. Since the
spectrum of $p$ is now given by the dual lattice $\L^*$, the partition
function of this model becomes a theta-function (see \S7.3)
\be
Z(t) = \Tr_{\cH} e^{-tH} = \sum_{p\in \L^*} e^{-2\pi t p^2}
\ee

\newsubsection{Supersymmetric quantum mechanics}

$N=1$ supersymmetric quantum mechanics leads to spinors on the manifold
$X$ and has proven to be of fundamental importance to understand for
example index theorems. In topological applications, the $N=2$
supersymmetric point particle is more useful (see \eg
\cite{tft-review}, or for a recent analysis \cite{froehlich}). 
Here one introduces besides the bosonic variable
$q,p$ two more fermionic variables $\theta,\thetabar$ satisfying 
\be
\theta^2=\thetabar^2=0,\qquad \theta\thetabar= -\thetabar\theta,
\ee
and a Lagrangian
\be
L =  p\dot q + i\thetabar \dot\theta - H.
\ee
In canonical quantization we have $\{\th,\bar\th\}=1$ with
the reality condition $\th^\dagger = \bar\th$, so that we can
consider $\th$ as a coordinate and $\bar\th$ as the conjugated
variable
\be
\thetabar = \pp \theta,
\ee
which acts as an operator on wave functions
\be
\psi(q,\th) \in \cH = L^2(\R^{1|1}) \cong \W^*(\R).
\ee
Supersymmetric wave functions can either be viewed as functions on the 
vector space $\R^{1|1}$, which is by definition the vector
space with one even and one odd coordinate,
or they can be viewed as differential forms on $\R$. In this latter
point of view one considers $\th$ as the one-form $dq$ and expand
the wave function as
\be
\psi(q,\th) = \psi^{(0)}(q) + \psi^{(1)}(q) \th .
\ee

More generally, for a superparticle on a general $n$-manifold $X$, 
the Hilbert space is in this way interpreted as the total space of 
normalizable differential forms on $X$,
\be
\cH = L^2(\Xhat) \cong  \W^*(X).
\ee
Here the $n|n$ dimensional
supermanifold $\Xhat$ is defined as $\Pi TX$, where the parity
transformation $\Pi$ declares the fibers of the tangent bundle $TX$ to
be anti-commuting. The second isomorphism is implied by the more general
expansion of a super wave function as
\be
\psi(q,\th) = \sum_{k=0}^n \psi^{(k)}_{\m_1,\ldots,\m_k}(q) \th^{\m_1} 
\cdots \th^{\m_k}.
\ee
The coefficients $\psi^{(k)}_{\m_1,\ldots,\m_k}$ are completely
antisymmetric and are naturally interpreted as differential $k$-forms on
$X$. 

Note that the Hilbert space has natural $\Z$ gradation. We can define fermion
number $F$ by given $\th$ charge one and consequently $\bar\th$ charge
minus one. Fermion number will then correspond to the notion of degree for
differential forms. 

The supersymmetry transformations are given by 
\be
\delta q^\m\ = \theta^\m,\qquad \delta \thetabar_\m = ip_\m,
\ee
so that $Q=i\theta^\m p_\m$. On wave functions $Q$ equals the exterior
differential $d$,
\be
Q = \theta^\m \pp {q^\m} = d.
\ee
We thus find that $Q^2=0$ and its cohomology gives the de Rahm 
cohomology of $X$
\be
H^*_Q(\cH) = H_{dR}^*(X).
\ee
There is also the hermitian conjugated operator $Q^*$, defined by
(here the metric
$g_{\m\n}$ and its inverse $g^{\m\n}$ enter)
\be
Q^* = g^{\m\n} \bar\th_\m \pp {q^\n} = d^*
\ee
These operators give the one-dimensional $N=2$ supersymmetry algebra
\be
\{Q,Q^*\}= H \ \iff \  \{d,d^*\}=\Delta.
\ee
The supersymmetric ground states satisfy
\be
Q|\psi\>=Q^*|\psi\>=0,
\ee
and thus their wave functions
 correspond to harmonic differential forms,
\be
d\psi=d^*\psi=0.
\ee
The space $\cH_0$ of supersymmetric
ground states is thus canonically identified as
\be 
\cH_0 = \mbox{Harm}^*(X).
\ee
We can compute the superdimension of the Hilbert space
(defined as the dimension of the even part minus the dimension of the
odd part) in terms of the Witten index \cite{witten-index}
\be
\sdim \cH = \Tr (-1)^F = \sdim \cH_0.
\ee
It equals the Euler number of the space $X$
\be
\sdim \cH = \chi(X) = \sum_k (-1)^k \dim H^k(X).
\ee
These relations between supersymmetry and algebraic topology
were first stressed by Witten \cite{witten-topology} and permeate 
the whole subject.

\newsubsection{Quantum mechanics and perturbative field theory}

Perturbative quantum field theories that describe particles and their
interactions can be obtained in ``first
quantized'' form from quantum mechanics by making the metric on the
world-line of the particle dynamical. We integrate the amplitudes over
the world-line time $t\geq 0$.

For example, the usual free field massless
propagator, associated to a line segment, takes the form
\be
\hbox{---------} = \int_0^\infty \!\!dt\, e^{-tH} 
= - {1\over \Delta} = {1\over p^2}.
\ee

To include interactions one considers ``one-dimensional quantum
gravity'' where the space-time is an arbitrary graph $\G$. The space of
metrics modulo diffeomorphisms on such a graph is parametrized by an
assignment of lengths $t_i\geq0$ to all the edges of the graph. So, we
sum over all graphs and integrate over the lengths with some particular
weights. Of course, we can think of these lengths as the Schwinger
parameters of the Feynman diagrams of quantum field theory. As such they
are completely equivalent to the moduli of Riemann surfaces that appear
in string theory and that we discuss at great length later on. 

For the one-loop amplitude one finds in this way
\be
\insertfig{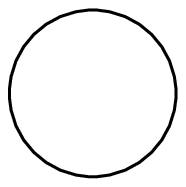}{1.2}
\ = \half \int_0^\infty {dt\over t}\Tr e^{-tH} = -\half \Tr \log
\Delta= - \half \log \det \Delta.
\ee
(The extra factor $1/t$ in the integrand has to do with the action of
the circle group on the graph. We have to factor by the volume of this
isometry group, which is given by $t$. The factor $\half$ is a
reflection of the $\Z_2$ symmetry of the circle graph, that flips the
orientation.) 

Simlarly, a more general one-loop scattering amplitude, with external
momenta $p_1,\ldots,p_n$ satisfying $p_i^2=0$, can be
represented as
\be
\insertfig{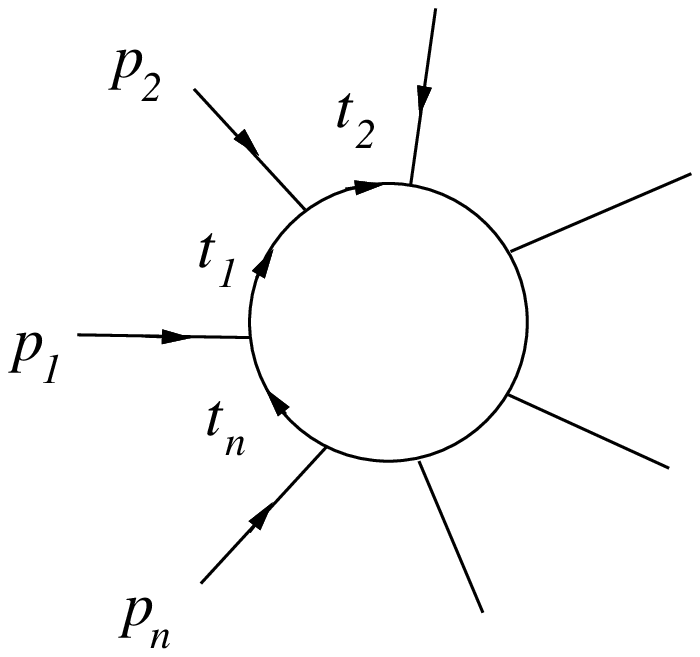}{3.5}\ \ \ = 
\int dt_1\cdots\int dt_n\. {\rm Tr}\left(e^{ip_1x}e^{-t_1H}e^{ip_2x}
e^{-t_2 H}\cdots e^{ip_nx} e^{-t_nH}\right)
\label{scat}
\ee

A more modern presentation of this first-quantized picture of QFT
would use the notions of ghosts and BRST
operators. We add to the QM system two anticommuting fields $b$ and $c$ with
action
\be
S_{gh} = \int dt\, b\dot c,
\ee
which gives canonical anticommutation relations $\{b,c\}=1$ and
a two-dimensional ghost Hilbert space $\cH_{gh}$. It consists of
the ghost vacuum of degree zero defined by $b|0\>=0$ and the state
$|1\> = c|0\>$ in degree one. 

The full Hilbert space is now defined as the tensor product
of the ``matter'' Hilbert space $\cH_m =L^2(\R^n)$, for which we choose
a momentum basis $|p\>$, $p\in\R$, and the ghost Hilbert space $\cH_{gh}$
\be
\cH = \cH_m \otimes \cH_{gh} = \cH^{(0)} \oplus \cH^{(1)}.
\ee
As indicated it has graded pieces in degree zero and one. The BRST
operator is given by the degree one operator
\be
Q=cH.
\ee
On a general state $Q$ acts as
\ba
Q|p\>\otimes |0\>  \is p^2 |p\>\otimes |1\>,\nonu
Q|p\>\otimes |1\>  \is 0.
\ea
The space $V$ of ``physical'' states is now defined as the 
cohomology of $Q$,
\be
V = H^*_Q(\cH).
\ee
One easily verifies that these physical states are of the form
\be
|phys\> = |p\> \otimes |1\>,\qquad p^2=0.
\ee
They are created out of the
vacuum, $|vac\>=|0\> \otimes |0\>$, by the action of the vertex
operators $\f_p=c \. e^{ipx}$ with $p^2=0$,
\be
|phys\> = \f_p |vac\>.
\ee
Note that any expectation value of the form
\be
\<phys'|\{Q,\cO\}|phys\>
\ee
vanishes, since $Q$ annihilates both the bra and the ket.
In particular we have the algebra
\be
\{Q,b\} = H,
\ee
which tells us that within physical correlation functions the
Hamiltonian vanishes. This vanishing of the Hamiltonian is the
signature of a theory of quantum gravity.
We can also introduce operators
\be
\f_p^{(1)} = \{b,\f_p\} =e^{ipx}
\ee
In terms of these operators the general one-loop amplitude 
(\ref{scat}) can then be written as
\be
\int dt_1\cdots\int dt_n\. {\rm Tr}\left(\f_{p_1} e^{-t_1H}\f_{p_2}^{(1)}
e^{-t_2 H}\cdots \f_{p_n}^{(1)} e^{-t_nH}\right)
\ee
We will see later that this particular way of writing field theory
amplitudes will be generalized in string theory.
 
As an aside we note that, for more complicated graphs, the Schwinger
parameter spaces of the individual graphs can be glued together to
form a connected moduli space that is quite analogous to the moduli
space $\cM_g$ of Riemann surfaces that will be a central point in the
coming lectures. To every diagram $\G$ we can associate a cell whose
points are parametrized by the length of the edges. These cells can be
glued together in the following fashion. If one of the lengths becomes
zero, the graph will change topology. We now glue the cell of this new
graph to the side of the cell of the old graph, as is illustrated in
\figuurplus{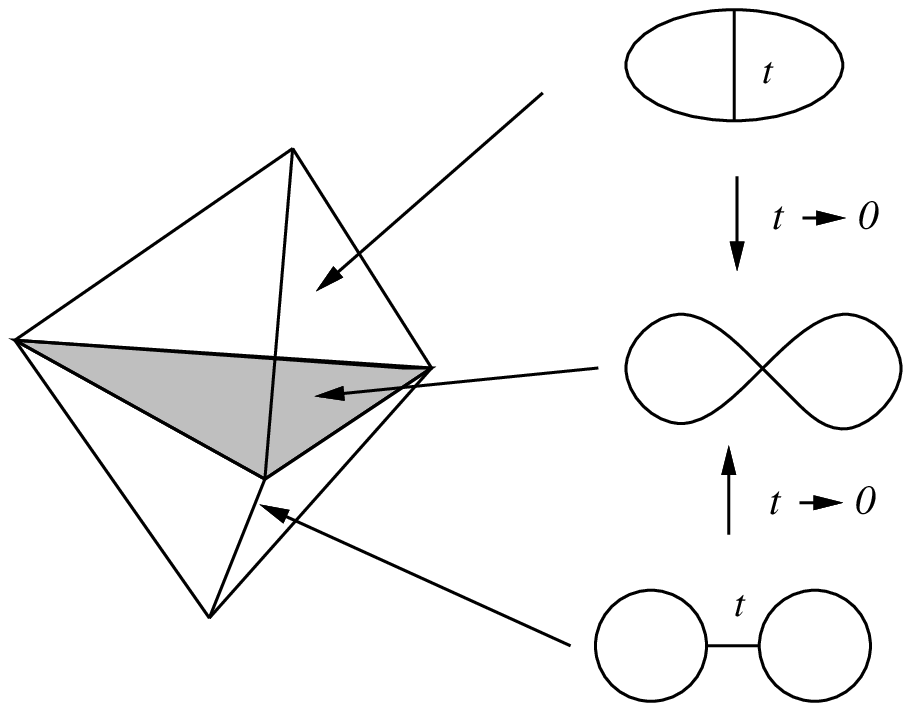}{6.5cm}{The 
Schwinger parameters of various Feyman graphs contributing to a
particular amplitude can be glued together to form one connected
moduli space.}. In this way we construct a cell complex that is a
Feynman diagram analogue of $\cM_g$
\cite{kontsevich-combi,pert}.

\newsection{Two-dimensional topological field theory}

After our discussion of quantum mechanics and its relation to field
theory, we now turn to to two-dimensional quantum field theories and
their relation to string theory. The simplest type of quantum field
theory is a topological field theory (TFT) which has the defining
property that all amplitudes are independent of the local Riemannian
structure. Two-dimensional TFTs will be the subject of this lecture.
Here we will concentrate on the properties of one particular quantum
field theory. Later in \S8 we will generalize our analysis to include
families of such theories.

\newsubsection{Axioms of topological field theory}

The axiomatic formulation of topological field theories has been given
by Atiyah \cite{atiyah} following Segal \cite{segal} 
and uses the language of categories and functors.
(A particularly thorough exposition of this approach is given in
\cite{quinn}.)

Let us recall that a {\it category} contains a set of objects $x$
and a set of arrows or morphisms $f: x\ra y$, which should be regarded
as abstract quantities satisfying an associative composition law. That
is, given two arrows $f:x\ra y$ and $g:y\ra z$, we can form the
composite arrow $g\circ f : x \ra z$, such that $(h\circ g)\circ f =
h\circ (g\circ f)$. A category further presumes that for each object $x$
an identity arrow $\id_x: x\ra x$ exists, satisfying $f\circ \id_x =
\id_y \circ f = f$. A {\it functor} between two categories is a map that maps
objects to objects, morphisms to morphisms, that respects all relations.

A good example is the category $\bf Vect$ of (say, complex, possibly
graded) vector spaces, where the objects are obviously vector spaces and
the morphisms correspond to linear maps between the vector spaces. This
is actually an example of a so-called abelian tensor category, where the
objects and morphisms can also be multiplied using the (associative and
commutative) tensor product $\otimes$, and have an inverse, the linear
dual $V^*$. If we have maps $\F_1: V_1\ra W_1$ and $\F_2: V_2 \ra W_2$,
we can form the tensor product map $\F_1\otimes \F_2: V_1 \otimes V_2
\ra W_1 \otimes W_2$. The unit is $\C$. 

To define a topological quantum field theory, we start with the category
${\bf Man}(d)$ of $d$-dimensional manifolds, where the the objects are
smooth, compact, oriented manifolds $X$ ({\it not} defined up to
isomorphism, and thus equipped with a given parametrization in local
coordinates) and where the morphisms 
\be
M:X\ra Y
\ee
are bordisms. That is, the morphism $M$ is a smooth, oriented manifold
of dimension $d+1$ with the property that it has two boundary
components, isomorphic to $X$ and $Y$, where the natural orientation
coming from the orientation of $M$ agrees with the orientation
of $Y$ but disagrees with the orientation of $X$,
\be
\d M =(- X) \cup Y.
\ee
(So $-X$ indicates the same manifold $X$, now with 
the opposite orientation.) We will call the components
$X$ and $Y$ ``incoming'' and ``outgoing'' respectively. In a figure we have
$$
\insertfig{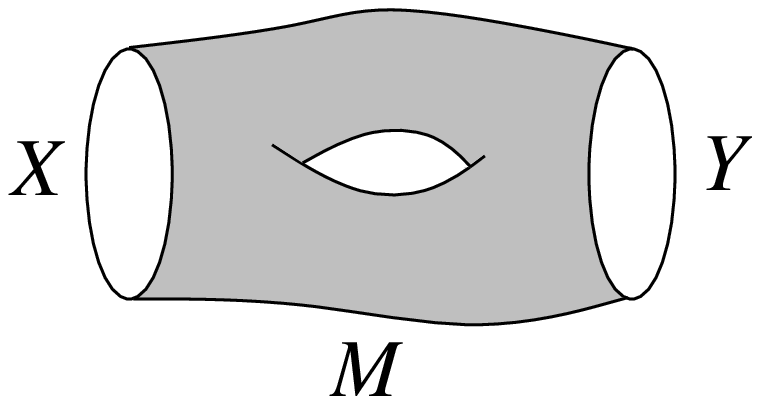}{3.2}
$$
The properties of a
category require that there exists a composition law of two bordisms
$M:X\ra Y$ and $N:Y\ra Z$, 
$$
\insertfig{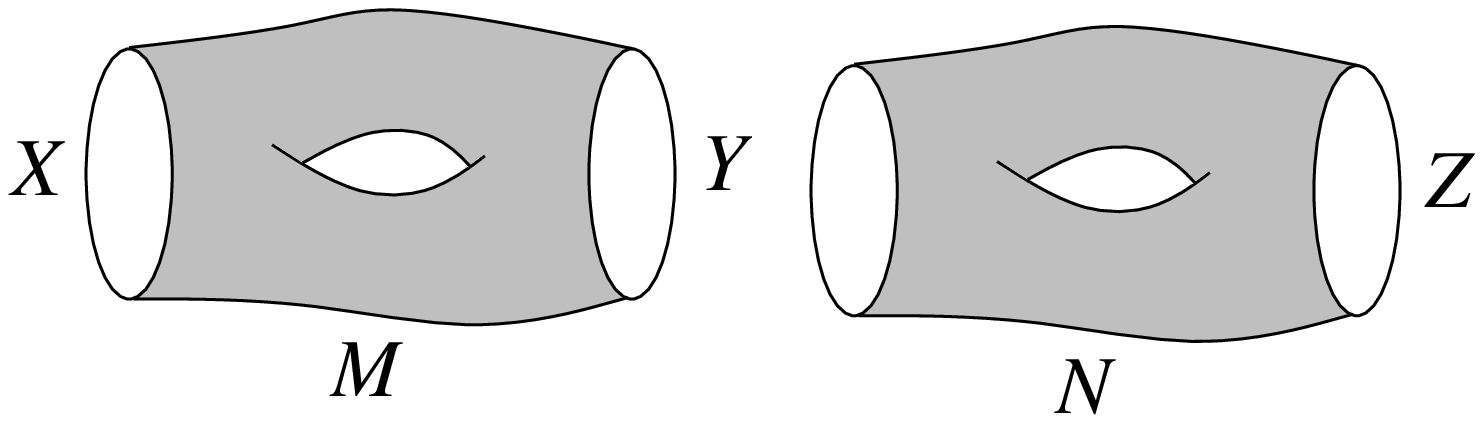}{6}
$$
producing a new bordism
\be
N\circ M: X \ra Z.
\ee
This composition law is obviously given by ``gluing'' the two
boundaries, 
$$
\insertfig{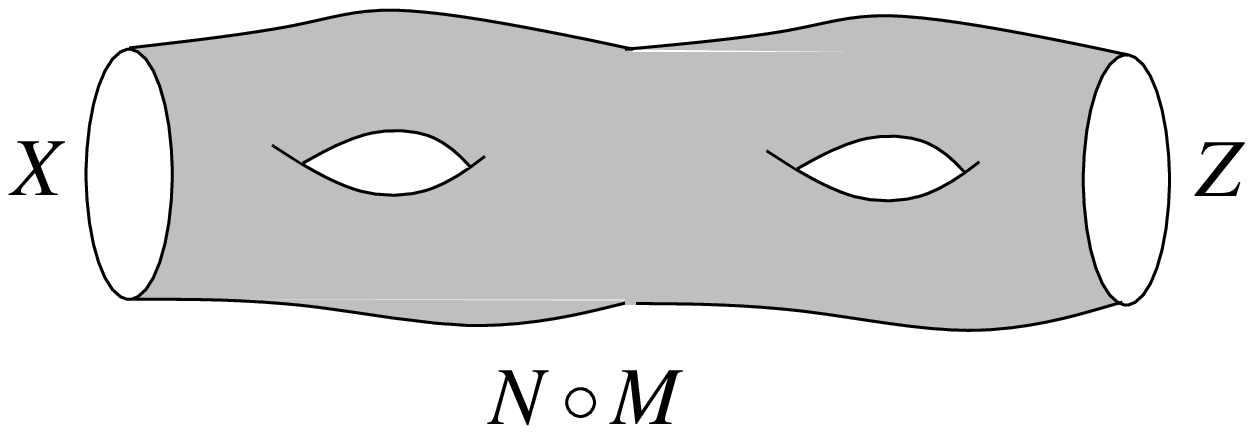}{4.6}
$$
that is, by identifying the two boundary components $Y$
and $-Y$. We can do this in a unique way, since both copies of $Y$
come with a parametrization.  The identity morphism is given by the
cylinder,
\be
\id_X: X\times [0,1].
\ee
These definitions make ${\bf Man}(d)$ into a category.

We have one extra operation that is not standard in categories.
We also want to be able to glue two boundary components of
a single irreducible manifold: if the two boundary components
of $M$ contain a common factor $X$, we want to define the
partial trace 
\be
\Tr_X :M \ra \Tr_X( M).
\ee
This is best explained in a picture
\be
\Tr:\ \ \insertfig{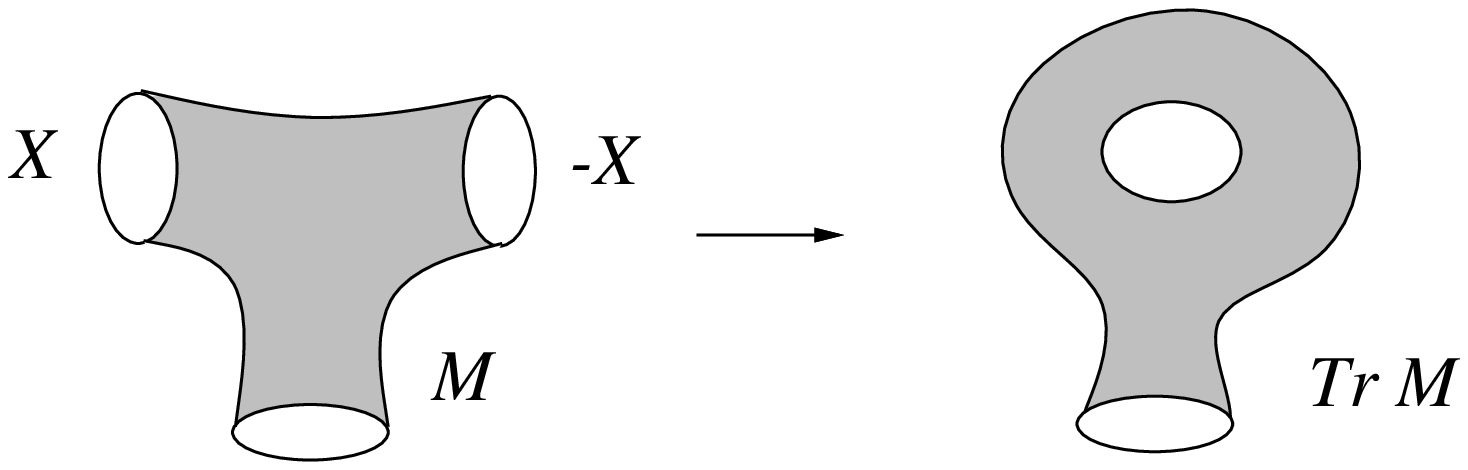}{7}
\ee
 Note further that ${\bf Man}(d)$ 
is also a tensor category, with the product given by
the disjoint union $\cup$ and unit the empty set $\emptyset$.

A $d+1$ dimensional {\it topological field theory} (TFT)
can now be defined as a functor
\be
\F : {\bf Man}(d) \ra {\bf Vect}
\ee
from the category of $d$-manifolds to the category of vector spaces,
satisfying certain extra properties.  Concretely this means that to any space $X$
we associate a vector space $V_X$,
\be
\F: X \ra V_X,
\ee
and to any bordism $M$ a linear map $\F_M$
\be
\left(M:X\ra Y\right)
\ \ \gives\ \  
\left(\F_M : V_X \ra V_Y\right).
\ee
The fact that $\F$ is a functor tells us that the amplitudes $\F_M$
should satisfy a factorization law
\be
\F_{N\circ M}=\F_N \circ \F_M.
\ee 
In case that the relevant operator is trace-class (or when all
vector spaces $V_X$ are finite-dimensional) we have an extra condition: 
\be
\F_{\Tr_X M}  = \Tr_{V_X} \F_M.
\ee
Since both category are abelian tensor
categories, we further demand that $\F$ respects the products $\otimes$
and $\cup$,
\be
V_{X\cup Y} = V_X \otimes V_Y,\qquad
V_{-X} = V_X^*,\qquad
V_\emptyset = \C.
\ee
and
\be
\F_{M\cup N}= \F_M \otimes \F_N,
\qquad \F_{-M}= \F_M^*,\qquad
\F_\emptyset = \C.
\ee

The existence of these amplitudes
$\F_M$ can be physically motivated by the following path-integral
argument. If $\f(x)$
is a local set of fields in the theory, a state in the ``Hilbert space''
$V_X$ will be a wave function $\Psi(\f_X)$ on the space of field
configurations $\f_X(x)$ on the spacelike manifold $X$. The path-integral
on $M$ with fixed values $\f_X,\f_Y$ at the boundaries $X$ and $Y$ then
gives the kernel $K_M(\f_Y,\f_X)$ of the evolution operator $\F_M$
\be
K_M(\f_Y,\f_X) = \!\!\!\! \mathop{\int}_{\f|_X=\f_X,\ \f|_Y=\f_Y} \!\!\!
\!\!\! \!\!\!\cD\f \,  e^{-S(\f)}
\ee
The transition amplitude then relates an ``incoming'' wave function
$\Psi_{in}(\f_X)$ to an ``outgoing'' wave function $\Psi_{out}(\f_Y)$ as
\be
\Psi_{out}(\f_Y) = \int\cD\f_X\, K_M(\f_Y,\f_X) \Psi_{in}(\f_X)
\ee
in a huge generalization of the evolution operator $e^{-tH}$ and its
kernel $K_t(y,x)$ of quantum mechanics. The gluing law then corresponds
to the composition law (\ref{add}) in quantum mechanics
\be
K_{N\circ M}(\f_Z,\f_X) = \int \cD\f_Y\, K_N(\f_Z,\f_Y) K_M(\f_Y,\f_X),
\ee
that tells us that evolving along $M$ and then evolving along $N$ is
equivalent to evolving along $N\circ M$.

Some further scattered remarks:

(1) We assume a natural action of the permutation group, possibly graded
if we are dealing with a fermionic theory (the vector spaces $V_X$
can be odd), if several boundary components are isomorphic.

(2) By definition the vector space associated with the empty set
(a legitimate boundary!) is $\C$. Therefore, in case the $d+1$
dimensional manifold $M$ is closed, $\d M = \emptyset$, the corresponding
morphism $\F_M:\C \ra\C$ will be given by multiplication by a constant 
$Z_M$, the {\it partition function}. In path-integral language we have
the identification
\be
Z_M = \int\cD\f \, e^{-S(\f)}
\ee
where one integrates over all field configurations on $M$.

(3) If $M$ has a single outgoing boundary $Y$, so that $M:\emptyset\ra Y$, the
transition amplitude $\F_M$ is a map $\C\ra V_Y$ and thus defines a 
special state 
\be
\insertfig{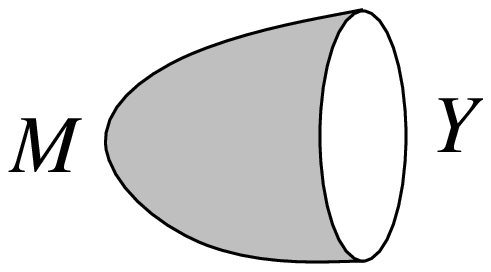}{2.2}\ \ :\ \ \Psi_M \in V_Y.
\ee
This can also be understood in path-integral language. The wave function
is given by
\be
\Psi_M(\f_Y) = \Int_{\f|_Y=\f_Y} \cD\f\, e^{-S(\f)}.
\ee
Here we integrate over all possible extensions of the field $\f_Y$
on $Y$ to a field $\f$ over all of $M$. In concrete situations, the
field configurations on $Y$ might be classified by certain topological
indices that indicate whether or not this extension exists. (Here one
can think about a spin structure for a fermion, or a vector bundle
for a gauge field.) The vector space $V_Y$ then splits into
superselection sectors, and depending how the data can be extended over
$M$, the vector $\Psi_M$ will sit in one of these superselection
sectors.

3. If a manifold $M$ can be split into two parts $M_1$ and $M_2$ with
common boundary $\d M_1=X = -\d M_2$, we can compute the partition
function as the inner product of the two states representing the
two halves,
\be
\insertfig{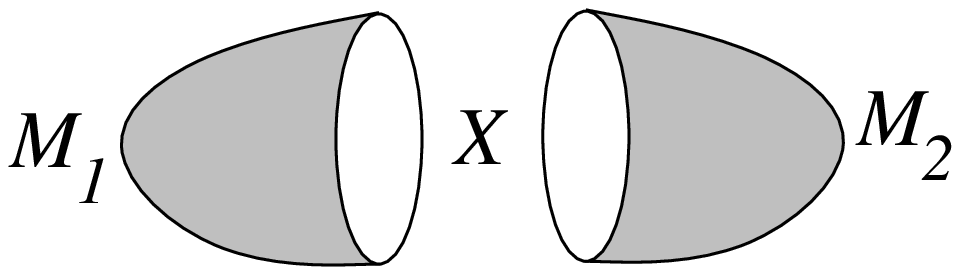}{4}\ \ : \ \  Z_M = \<\Psi_{M_1},\Psi_{M_2}\>.
\label{bloep}
\ee
This way of writing the partition function can be helpful if one wants
to construct a new manifold $M'$ by surgery on $X$. In that case one
applies a diffeomorphism of $X$ that is not in the identity component of
$\pi_0\Diff(X)$ and that is realized as an operator $\g$ on the vector
space $V_X$. The partition function of $M'$ can then be written as
\be
Z_{M'} = \<\Psi_{M_1},\g\Psi_{M_2}\>.
\ee

(4) If $M=[0,1]\times X$ and $N=S^1\times X$, we can compute the
dimension of the vector space $V_X$ as the partition function of $N$
(given the fact that this dimension is finite)
\be
Z_N = \Tr\F_M = \dim V_X.
\ee

\newsubsection{Topological field theory in two dimensions}
 
In this subsection we will restrict our investigations to two dimensions
where our manifolds $M$ are surfaces of genus $g$. This example has been
explained quite often \cite{witten-top,rhd-thesis,trieste}, but, with
the risk of repeating myself once too often, I explain it here since it
a necessary, though routine, part of our 2D route to string theory. For
a very thorough review of these and other aspects of two-dimensional TFT
see \cite{dubrovin}.

Since the only connected compact one-dimensional manifold is the circle
$S^1$, we have only one vector space to consider
\be
V = V_{S^1}
\ee
together with its dual. For convenience we will assume $V$ to be
finite dimensional. The data of a two-dimensional topological field
theory are now obtained by considering respectively the sphere with
one, two, and three holes.  Let us briefly run through the argument.

(1) Following remark (2) in the previous section,
the disk with an outgoing boundary
gives rise to a particular state
\be 
\insertfig{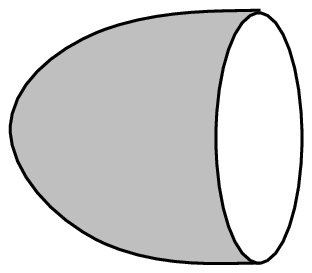}{1}\quad {\bf 1} \in V, 
\ee
that we will denote as the identity, for reasons that become obvious
in a moment. Similarly, the disk with one incoming boundary gives an
element of the dual vector space,\ie a linear functional
\be
\insertfig{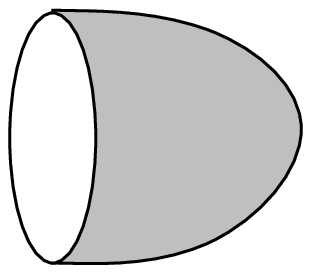}{1}\quad\<\.\>_0: V \ra \C.
\ee
According to (\ref{bloep})
the partition function of the two-sphere is then expressed as
\be
Z(S^2) = \<\id\>_0
\ee

(2) The cylinder with one incoming and one outgoing boundary is just the
identity map $\id: V \ra V$. If we choose two incoming boundaries, it
gives a bilinear map
\be
\insertfig{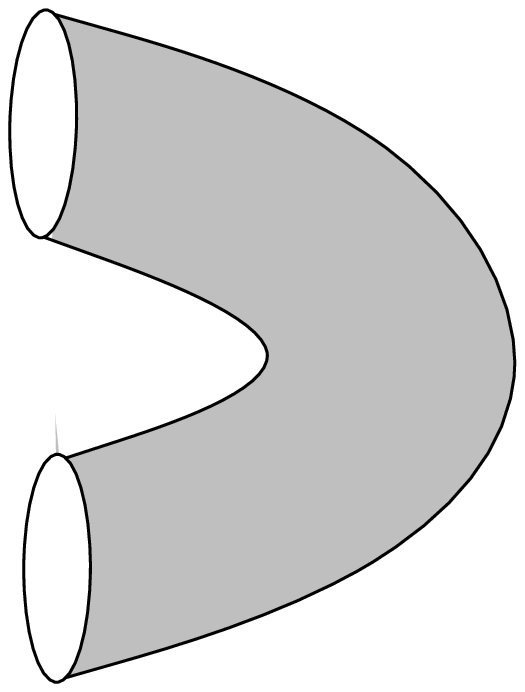}{1}\quad \eta : V \otimes V \ra \C,
\ee 
If we choose an explicit basis
$\f_i$ for $V$ (with $\f_0=\id$) we obtain the graded symmetric matrix
\be
\eta_{ij} = \eta(\f_i,\f_j)
\ee
By factorization, this inner product $\eta$ will be non-degenerate (but
not necessarily positive). 
It allows us to identify the incoming states
in $V$ with the outgoing states in the dual space $V^*$. The inverse
is given by the cylinder with two outgoing boundaries. We will write
$ \eta\inv(\f_i,\f_j) = \eta^{ij}$. It
is a simple exercise to check that $\h\.\h\inv=\h\.\inv\h=\id$.
(Draw the corresponding pictures.)

(3) Finally the pair of pants, or the sphere with three holes,
corresponds (again with the appropriate choice of orientations of
the boundaries) to a map
\be
\insertfig{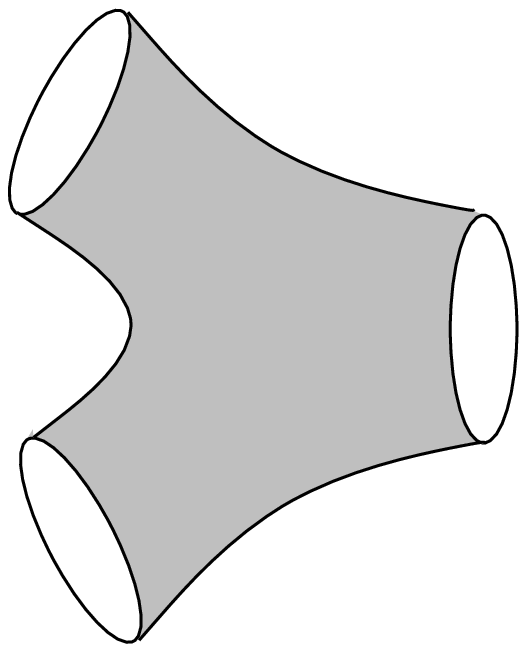}{1.3}\quad c : V \otimes V \ra V.
\ee 
If we introduce the notation
\be
 \a \cdot \b =c(\a,\b),
\ee
this makes $V$ into an algebra, the operator product or Verlinde algebra of the
topological field theory \cite{erik}.  
The multiplication $c$ allows us to identify states with
operators in a very simple way. It gives a map
\be
c: V \ra \End(V).
\ee
In a basis we will write
the matrix corresponding to $\f_i$ as $C_i$.
We also introduce the structure coefficients $c_{ij}{}^k$ in 
\be
\phi_i \cdot \phi_j = \sum_k c_{ij}{}^k \phi_k.
\ee 
Note that if we choose three incoming boundaries we get a map 
\be
\insertfig{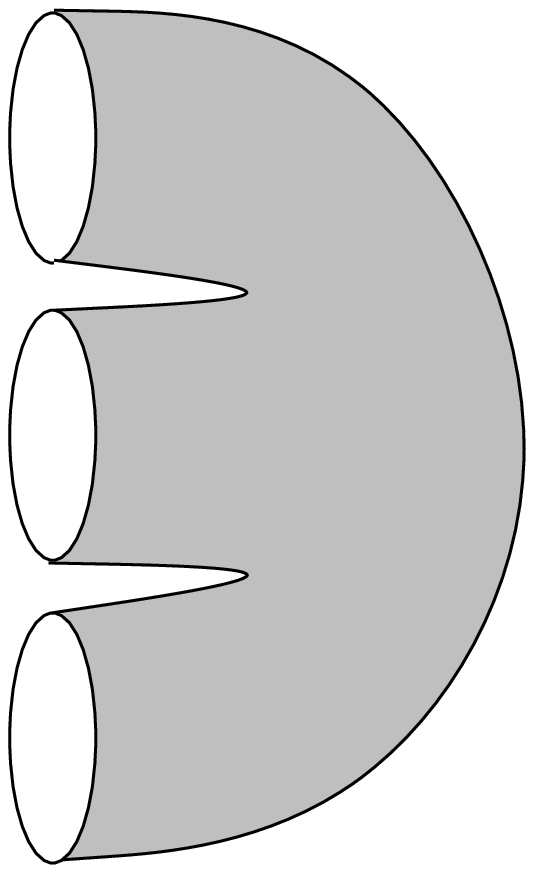}{1.1}\quad c: \Sym^3 V \ra \C,
\ee
which is characterized by the fully (graded) symmetric tensor
\be
c_{ijk} = c_{ij}{}^l\h_{lk}.
\ee
The inner product $\eta$ is obtained by inserting the identity
in this trilinear map, which gives in components 
\be
\h_{ij}=c_{ij0}.
\ee
 
All these data suffice to calculate any partition or correlation
function, since every surface can be reduced to a
collection of three-holed spheres, cylinders and disks, by cutting it often
enough. 
Of course, there are many
inequivalent ways to factorize a particular surface. The final answer should
however not depend on the particular choice of factorization. This gives
further constraints on the data $\eta$ and $c$. 

For instance, a simple
consequence of the symmetry of the 3-punctured sphere is the
compatibility of the metric $\eta$ with the multiplication
\be
\eta(\a \cdot \b,\g ) = \eta(\a, \b \cdot \g).
\ee
In a picture
$$
\insertfig{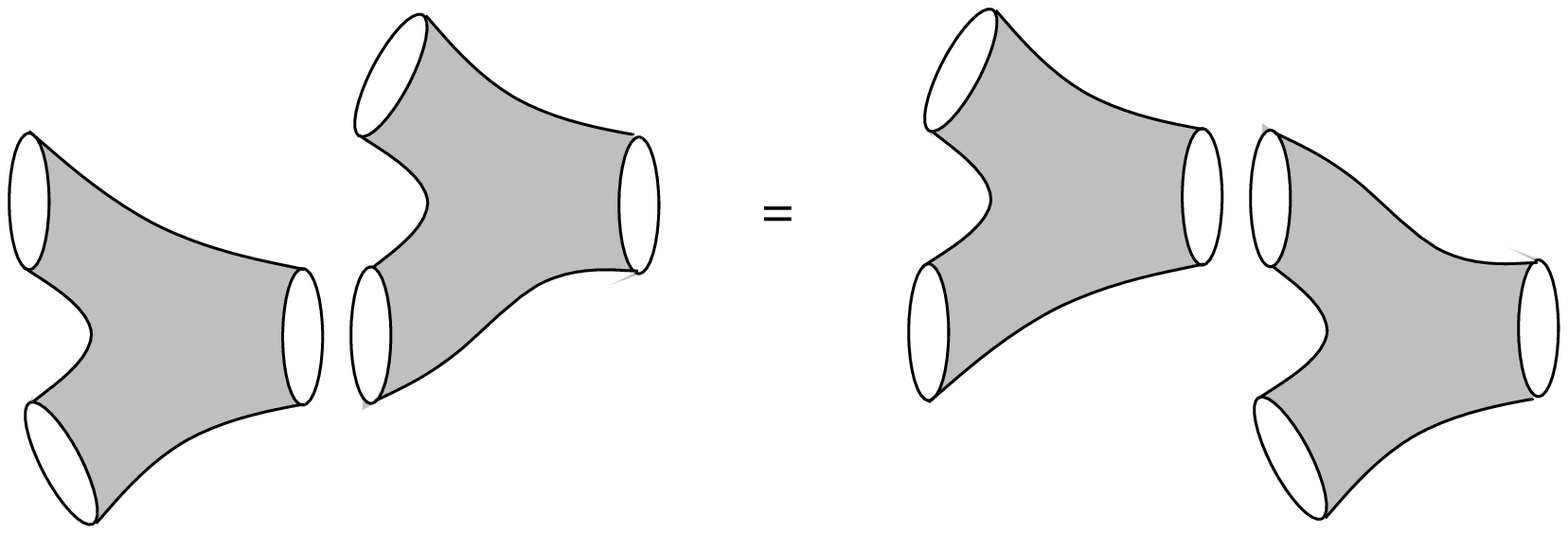}{7}
$$

Another relation expresses the inner product $\eta$ in terms of the
linear function $\<\.\>_0$ as
\be
\eta(\a,\b) = \< a\cdot\b\>_0
\ee

But most importantly,
if we consider the sphere with four holes, there are two inequivalent
ways of factorization
$$
\insertfig{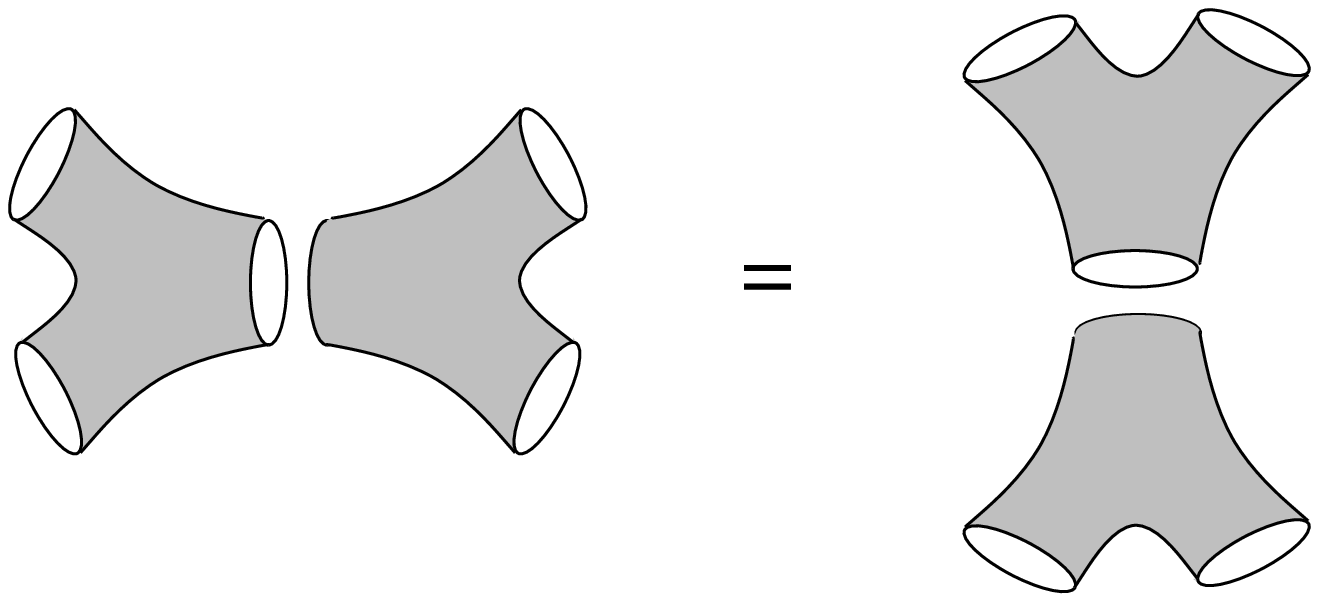}{7}
$$
which translates in associativity of the algebra
$V$ 
\be 
(\a \cdot \b) \cdot \g = \a \cdot (\b \cdot \g).
\ee

It can be easily checked that no further conditions will be found when
we consider more complicated surfaces. An associative algebra $V$ with
such an invariant inner product $\h$ is called a Frobenius algebra, and
this simple object captures a 2d TFT completely. (For more on 2D TFTs
see \cite{dubrovin}.)

With the concept of factorization, it is extremely easy to calculate
higher genus partition and correlation functions. In fact, we can introduce
an operator $H$ that creates an handle. It is defined as
the state associated to the torus with one puncture and has the
representation
\be
\insertfig{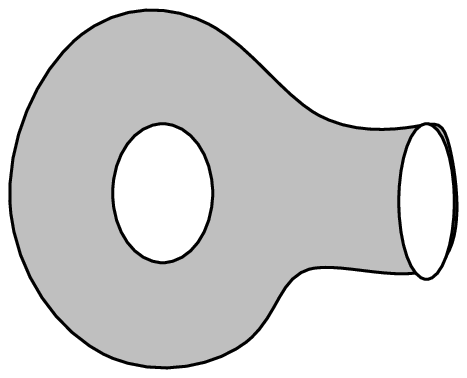}{1.7}\:\ \ H = \sum_{i,j} c_i{}^{ij} \phi_j.
\ee
In this fashion a genus $g$ partition function $Z_g$ can be written as
a genus zero correlation function
\be
Z_g = \Bl \, \underbrace{H \cdots H}\subg \, \Br_0 = \Tr H^{g-1}.
\ee

\newsubsection{Example ---  quantum cohomology}

Let us briefly discuss an important class of
examples of Frobenius algebras: classical and quantum cohomology. We
will see later how this is realized in (topological) sigma models.

We start with classical cohomology. Let $X$ be a compact orientable
manifold. We can take $V=H^*(X)$ with as multiplication the wedge
product of differential forms $\a\.\b = \a\wedge\b$, which is graded
commutative. On the states we now have an obvious inner product given by
the intersection form
\be
\eta(\a,\b)=\int_X \a\wedge\b.
\ee
Note that more generally the genus zero correlation functions are given
as
\be
\<\a_1\ldots\a_n\>_0 = \int_X \a_1\wedge
\cdots \wedge\a_n.
\ee

For example, if $X=\P^N$, the cohomology ring is simply generated by a
generator $x$ of degree two with the single relation
\be
x^{N+1}=0,
\ee
so $V=\C[x]/(x^{N+1})$, as an algebra.

It is clear that the cohomology ring $H^*(X)$ satisfies all the
relations of a two-dimensional topological field theory. We note that we
already represented $H^*(X)$ in supersymmetric quantum mechanics in
\S3.1. Here, however, we have a natural geometrical interpretation of
the ring structure (although the multiplication can also be defined in
quantum mechanics).

We really go beyond quantum mechanics if we consider quantum cohomology
or Gromov-Witten invariants.
For quantum cohomology we again take the vector space $V=H^*(X)$, but
now define a new, improved ``quantum'' product 
\cite{witten-sigma,witten-mirror,witten-top,kontsevich-manin}. Let $X$ now be a compact
K\"ahler manifold and consider holomorphic maps\footnote{One can also
take $X$ to be a symplectic manifold and study pseudo-holomorphic maps,
as introduced by Gromov to produce symplectic manifold invariants. This
is how the subject started in mathematics \cite{gromov}.} (sigma model
instantons) from the Riemann sphere into $X$
\be
x: \P^1 \ra X,\qquad \dbar x=0.
\ee
Let $\cN$ denote the moduli space of such ``stable'' holomorphic maps
(stability adds some special set of singular maps, making the moduli
space compact). Let furthermore $\w_1,\ldots,\w_n$ be a integer basis of
the Picard lattice $H^{1,1}(X)\cap H^2(X,\Z)$. 
For any map we define a multi-degree $d=(d_1,\ldots,d_n)$
by
\be
[x(\P^1)]=\sum d_i \w_i.
\ee
The moduli space $\cN$ will now decompose in components of different degree
$\cN_d$ (not necessarily irreducible)
\be
\cN = \bigcup_d \cN_d,
\ee
with in particular the constant maps
\be
\cN_0 \cong X.
\ee

It is not easy to describe these instanton spaces for the generic case.
However, there is a simple formula for the `virtual dimension' that is
defined as follows. (Here we consider maps $x:\Sigma\ra X$, with
$\Sigma$ a general Riemann surface of genus $g$.) Consider the tangent
space at a point $x\in\cN_d$. This tangent space $T_x\cN_d$ is by
definition given by the infinitesimal maps $\delta x$ satisfying
$\dbar\delta x = 0$, \ie
\be
\delta x \in H^0(\Sigma,x^*T_X).
\ee
($T_X$ denotes the holomorphic $(1,0)$ tangent bundle of $X$.) We also
have to consider the group $H^1(\Sigma,x^*T_X)$. The spaces $H^0$ and
$H^1$ are the kernel and cokernel of the $\dbar$ operator on $\Sigma$
twisted with the holomorphic vector bundle $x^*T_X$. The
Hirzebruch-Riemann-Roch theorem gives an expression for the difference
between the dimensions of these groups
\ba
\dim H^0 - \dim H^1 \is \int_\Sigma ch(x^*T_X)\. td(T_\Sigma)\nonu
\is n(1-g) - \int_\Sigma x^* c_1(X),
\ea
with $n=\dim_\C X$.
The RHS of this equation is known as the virtual dimension of the space
$\cN_d$. If $H^1$ vanishes, the virtual dimension equals the actual
dimension. We see that the virtual dimension is independent of the
homotopy of the map $x$ in the Calabi-Yau case, with $c_1(X)=0$,
\be
\dim H^0 - \dim H^1 = n(1-g).
\ee
In the particular case of
genus zero, $\cN$ has virtual dimension $n$, the same
as $X$.

We now choose parameters $t^i$ and define
the quantum cup product as
\be
\<\a_1\ldots\a_n\>_q = \sum_d q^d \int_{\cN_d} \v_i^*\a_1\wedge
\cdots\v_n^* \a_n,
\ee
where $q^d=\exp \sum {d_i t^i}$.
Here we pull-back the cohomology classes $\a_i$ on $X$ to the moduli space 
$\cN$ of holomorphic maps by first considering the ``universal
instanton''
\be
\F:\cN \times \P^1 \ra X,\qquad \F(x,z)=x(z),
\ee
and then choosing $n$ sections $s_i:\cN \ra \P^1$. We then define
$\v_i=\F\circ s_i$. One easily verifies that the answer does not 
depend on the various choices involved. This quantum product
defines again an associative algebra. The associativity condition can be
proved by a degeneration argument that we will sketch in more detail in 
\S8.9. In the limit $q\ra 0$ we recover the usual cohomology ring,
which is the contribution of the identity component $\cN_0$.
For example, in the case of projective space, $X=\P^N$, the relation
$x^{N+1}=0$ gets deformed to \cite{witten-top}
\be
x^{N+1} = q.
\ee
Many other cases have been worked out \cite{quantum-coh}.

For later use we note here that for a Calabi-Yau three-fold
the operator product coefficients $c_{ijk}$ of the elements $\w_i\in H^2(X)$
have an expansion \cite{candelas,aspinwall-morrison}
\be
c_{ijk} = \int_X \w_i\wedge\w_j\wedge\w_k + \sum_d
N_d {d_id_jd_k q^d \over 1-q^d},
\label{count}
\ee
where $N_d$ counts the number of rational curves of degree $d$ in $X$.

\newsection{Riemann surfaces and moduli}

We have seen that it is very simple to compute the partition function
of a two-dimensional topological field theory for a genus $g$ surface.
It is simply given as
\be
Z_g = \Tr H^{g-1},
\ee
with $H$ the handle operator. In string theory we also associate a
number $F_g$, the $g$ loop vacuum amplitude, to a topological surface of
genus $g$, but in a much more involved way. Instead we consider the
moduli space $\cM_g$ of Riemann surfaces and define the genus $g$
amplitude as
\be
F_g = \int_{\cM_g}\! Z_g,
\ee 
in terms of a particular volume form
\be
Z_g \in H^{top}(\overline{\cM}_g),
\ee
that is produced by a two-dimensional conformal field theory. (Here the
bar indicates the stable compactification, that we will discuss in
\S5.4.)

In terms of these higher-loop string amplitudes $F_g$ the {\it
space-time} partition function reads
\be
Z_{spacetime} \sim \exp {\sum_g \l^{2g-2} F_g}
\ee
with $\l$ the string coupling constant. (Unfortunately, these expansions
do not converge, so as it stands $Z_{spacetime}$ is only a formal
object.) Note that in the same notation
we can say that a 2d TFT, which by definition does not depend on the complex
structure on the surface, gives a partition function that is constant on
$\cM_g$ and thus can be regarded as an element of
\be
Z_g \in H^0(\overline{\cM}_g).
\ee

We now have to explain how quantum field theory leads to natural volume forms on
$\cM_g$. But before we do that, we spend the rest of this lecture
to make some
comments on Riemann surfaces and their moduli. This material is of
course completely standard mathematics, see {\it e.g.} \cite{riemann}.

\newsubsection{The moduli space of curves}

There are basically three different ways to think about Riemann
surfaces:

(1) As complex curves, \ie one-dimensional complex varieties where the
complex coordinates $z$ and $w$ in two patches are holomorphically
related, $w=w(z)$. Equivalently, we are given a complex structure on the
topological surface. Recall that a complex structure is a linear map
$J$, defined in the tangent space at each point, that satisfies $J^2=-1$
and the integrability condition $\nabla J=0$. The complex structure
allows us to split the complexified tangent space in holomorphic and
anti-holomorphic vectors, with eigenvalues $i$ and $-i$, and tells us
what the (local) analytic coordinate is. Equivalently, we are given a
$\dbar$ operator, and holomorphic functions are defined by $\dbar f =0$.

(2) As algebraic curves, \ie as the solutions of polynomial equations
\be
f(x,y)=0
\ee
in two complex variables $x,y\in\C$. (Strictly speaking, solutions of 
homogeneous equations $f(x,y,z)=0$ in complex projective space $\P^2$.)

(3) As surfaces with a conformal class of metrics. In two dimensions any
metric $g_{ab}$ defines a complex structure through the relation
\be
J_a{}^b = \sqrt g \e_{ac} g^{cb}
\ee
with $\e_{ab}$ the Levi-Civita symbol and $g=\det g_{ab}$. Furthermore,
all complex structures are obtained in this way. As we see, $J$ is
invariant under local rescaling (Weyl transformations) of the metric
$g_{ab} \ra \r(x) g_{ab}$, and so is only determined by the conformal
class of the metric. Locally we can choose coordinates $x^a$ such that
$g_{ab}=\r(x)\delta_{ab}$, and in these coordinates the complex structure
reduces to the usual identification of $\R^2=\C$, with analytic
coordinate $z=x^1+ix^2$.

The moduli space $\cM_g$ parametrizes all Riemann surfaces up to
equivalence. It is an orbifold space (even a quasi-projective algebraic
variety) of complex dimension
\be
\dim_\C \cM_g = \threeoptions 0 {$g=0$} 1 {$g=1$} {3g-3}{$g\geq 2$}
\ee

This can be proved using deformation theory of the $\dbar$ operator on
the surface $\Sigma$. We will assume $g\geq 2$ so that we are dealing
with the generic situation. The deformed operator is written as
$\dbar+\m\d$ with Beltrami differential $\m$, a section of $T_\Sigma
\otimes \Tbar_\Sigma^*$. There is no integrability condition for complex
curves. (By dimensional reasons, an obstruction would lie in
$H^2(T_\Sigma)$, which vanishes). Infinitesimal diffeomorphisms, which
lead to equivalent complex structures, are generated by vector fields
$\xi$ and induce the equivalence $\m \sim \m +\dbar\xi$. Therefore, the
tangent space at the point $\Sigma\in\cM_g$ is given by
\be
T_\Sigma\cM_g\cong H^1(T_\Sigma)
\ee
Using Serre duality one finds that $H^1(T_\Sigma)\cong H^0(K^2)^*$, with
$K=T^*_\Sigma$ the canonical line bundle on $\Sigma$. With Riemann-Roch one then
computes that 
\be
\dim H^0(K^2) - \dim H^0(T) = 3g -3.
\ee
For $g\geq 2$ the space $H^0(T)$ of holomorphic vector fields, \ie
infinitesimal automorphisms of the surface $\Sigma$, vanishes and
the space of quadratic differentials has the dimension
$\dim H^0(K^2) = 3g-3$, which then equals the dimension of the moduli
space. (For genus zero and one, we have $h^0(T)=3,\ 1$ respectively.) 

From the point of view of conformal metrics one starts with the space of
all metrics modulo diffeomorphism, and chooses a unique representative
in each conformal class. This can be conveniently done by requiring the
curvature $R$ to be constant. Taking into account the Gauss-Bonnet
theorem
\be
\int_\Sigma {d^2 z\over 2\pi} \sqrt g R = 2-2g,
\ee
$R$ can be normalized to be $1,0,-1$ for genus $g=0,1,\geq 2$. (For
genus one we also have to normalize $\int\! \sqrt g=1$.) We still have
to identify constant curvature metrics by the diffeomorphism group
$\Diff(\Sigma)$, which does not act freely. Therefore the moduli space
is not a smooth space; ``symmetric'' surfaces will be fixed points of
certain transformations, which makes $\cM_g$ into an orbifold. Actually,
one can do this identification in two steps, by first taking the
quotient by the identity component $\Diff(\Sigma)_0$ and subsequently
the mapping class group $\G_g$, which represents the global
diffeomorphisms and is defined by the exact sequence
\be
1\ra \Diff_0(\Sigma) \ra \Diff(\Sigma) \ra \G_g\ra 1.
\ee
 The first step is a smooth
operation that produces the so-called Teichm\"uller space $\cT_g \cong
\C^n$. However, the second step that
expresses the moduli space as the quotient
\be
\cM_g=\cT_g/\G_g,
\ee
can have fixed points, that correspond to surfaces with extra
automorphisms.

\newsubsection{Example ---  genus one}

A simple example is the case $g=1$: the two-torus $T^2$ or elliptic curve. 
Let us consider this moduli space from the three equivalent points of view.

(1) First we use the language of complex curves. A complex curve of genus
one, topologically a two-dimensional torus $T^2$, can be represented as
$\C/\L_\t$ with the lattice 
\be
\label{lat}
\L_\t = \Z\oplus\t\Z
\ee
and $\t$ an element of the
upper half plane $\H=\cT_1$ (defined by $\Im\t>0$). That is, we have
identifications
\be
z \sim z+1 \sim z+\t.
\ee
which produces a topological two-torus, as illustrated in
\figuurplus{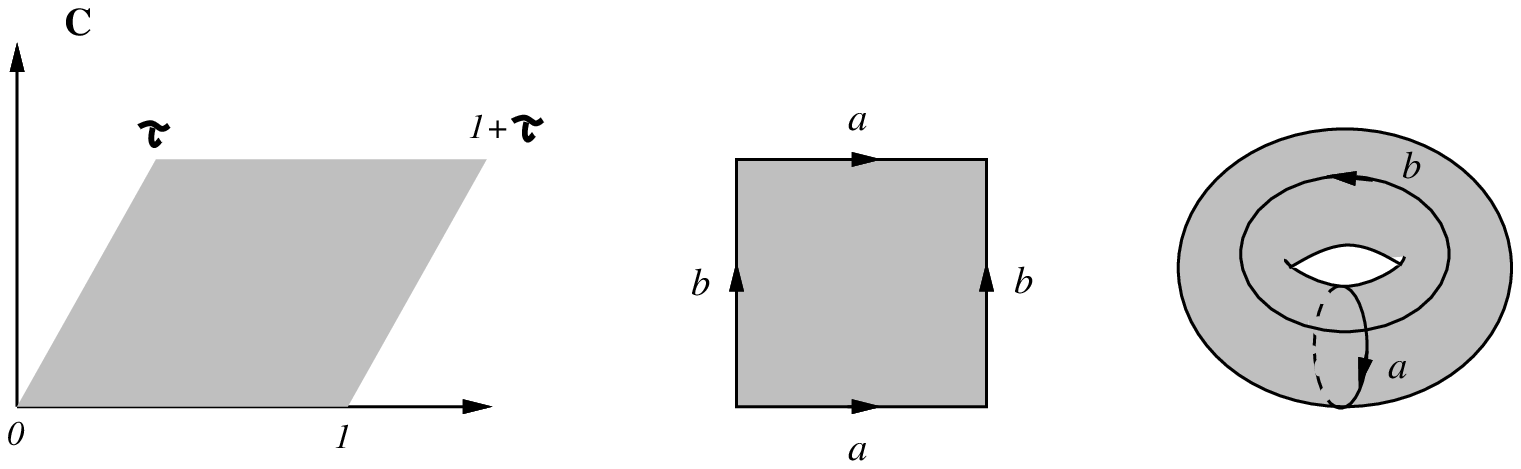}{12cm}{The torus or elliptic curve is obtained by
quotienting the complex plane by a two-dimensional lattice.}
Note that we used here the automorphisms of the torus to put one of the
two generators $e_1$, $e_2 \in\C$ of the rank two lattice $\L$ to $1$. A more
invariant expression for the modulus $\t$ would be as the quotient
$e_2/e_1$. Different basis choices for the lattice $\L_\t$, that are
related by elements in $SL(2,\Z)$, give the same elliptic curve. This
leads to a further identification of the modulus $\t$ by the modular
group $\G_1=PSL(2,\Z)$. This group acts on $\t$ by fractional linear
transformations 
\be
\t \ra {a\t +b \over c\t +d},\qquad
\twomatrixd abcd \in PSL(2,\Z).
\ee
It is generated by the transformations
\ba
T & : & \t \ra \t+1, \nonu
S & : & \t \ra -1/\t,
\ea
satisfying the relations $S^2=(ST)^3=1$. The action on the upper
half-plane is depicted in\ {\it fig.}
\addtocounter{fignum}{1}\figuurnum. \begin{figure}[t]\begin{center}
\leavevmode\hbox{\epsfxsize=15cm \epsffile{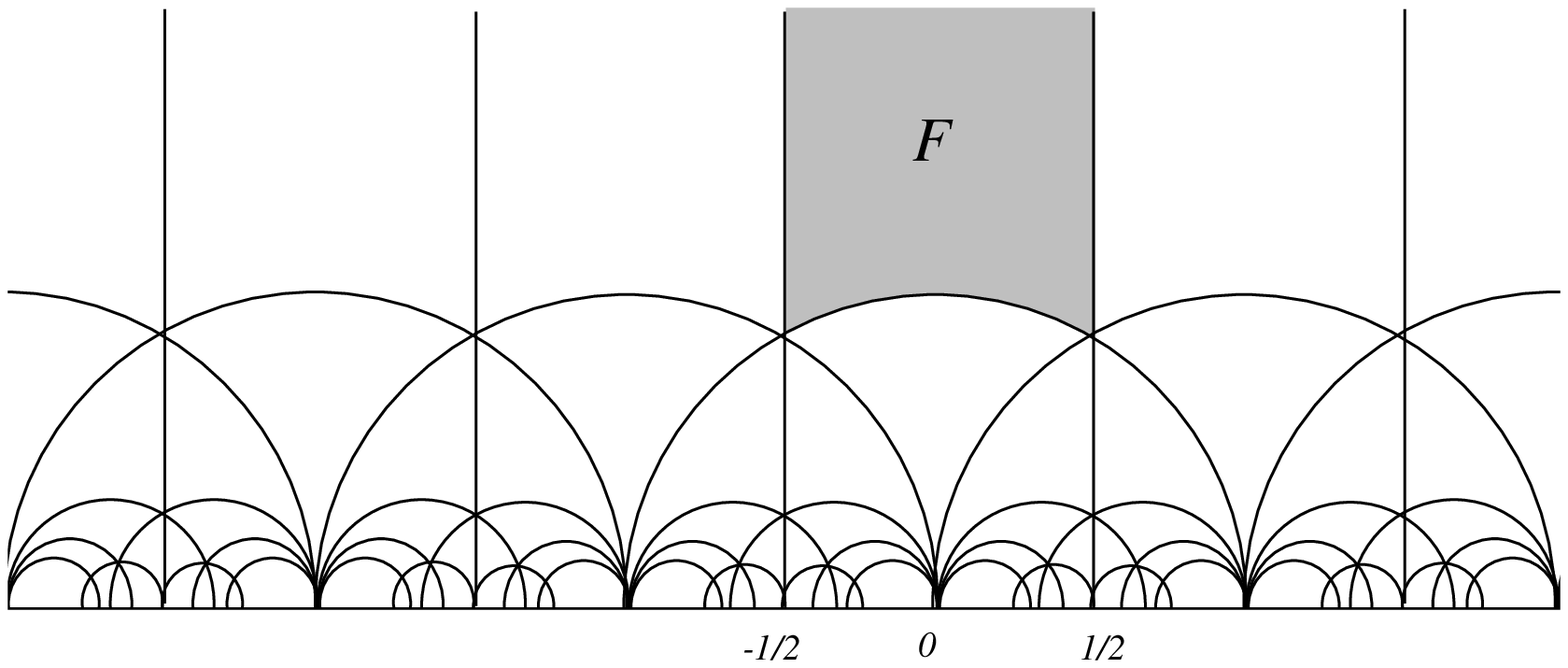}}\\[-24mm]
\parbox{10cm}{\small \bf Fig.\ \figuurnum: \it The action of the modular
$PSL(2,\Z)$ on the upper-half plane and a fundamental domain ${\cal
F}$.} \end{center} \end{figure}

The moduli space $\cM_1$ equals the quotient of the upper half-plane
$\H$ by the modular group $SL(2,\Z)$ and can be represented
by the well-known fundamental domain $\cF$, defined by restricting
$|\t|\geq 1,$ $|\Re\t|\leq \half$ and indicated in {\it fig.} 2.
Topologically we have $\cF \cong \C$. Note
that, once we compactify the moduli space by adding the point
$\t=i\infty$, the resulting ``stable '' moduli space $\overline{\cM}_1$
is topologically a (Riemann) sphere,
\be
\Mbar_1 \cong \P^1
\ee
It contains three orbifold singularities at $\t=i,$ $\t=e^{2\pi i/3}$
and $\t=i\infty$. These are respectively the fixed points of the
transformations $S,$ $ST$ and $T$ of order $2$, $3$ and $\infty$.
The point at infinity corresponds to a singular elliptic curve, as we
will see in a moment.

(2) From the point of algebraic curves, any genus one curve is
represented by the cubic equation (elliptic curve)
\be
y^2=x^3 + a x +b,
\label{ell}
\ee
for some constants $a,b\in \C$. However, this identification is not
unique (the moduli space is one-dimensional and here we have two 
parameters) and in order to express the
relation of the constants $a,b$ with the modulus $\t$ we first have to
introduce some facts about modular forms that we will need at various
places later in the lectures. 

A modular form of weight $k$
is a holomorphic function $f(\t)$ on the upper half plane satisfying
the condition
\be
f\left({a\t +b \over c\t +d}\right)=(c\t+d)^k f(\t)
\ee
and which is well-behaved at infinity, \ie has an expansion of the form
\be
f(\t) = \sum_{n\geq 0} a_nq^n,\qquad q=e^{2\pi i \t}.
\ee
Since the product of two modular forms of weight $k$ and $l$ gives
again a modular form, now of weight $k+l$, the space of modular forms is
a ring. As a ring it is generated
by the Eisenstein series $E_4(\t)$ and $E_6(\t)$ of weight $4$ and $6$
respectively. The normalized Eisenstein series $E_k(\t)$, for $k\geq
4$ and even, are defined as
\be
E_k(\t) = 1 -{2k\over B_k} \sum_{n>0} {n^{k-1} q^n\over 1 - q^n} =
\sum_{(m,n)=1} {1\over (m + n\t)^k}
\ee
There are two independent forms of weight $12$, namely $E_4^3$ and
$E_6^2$. Therefore, we can define a linear combination that vanishes at
$q=0$, the discriminant
\be
 \Delta = {E_4^3 - E_6^2 \over 1728} = \eta^{24}= q - 24 q^2 + \ldots
\ee 
where $\eta$ is Dedekind's eta function
\be
\eta(\t) = q^{1\over 24} \prod_{n>0} (1-q^n).
\ee
We need the discriminant $\Delta$ in our last definition, that of 
the modular $j$-function
\be
j(\t)={E_4^3\over \Delta} .
\ee
Since it is defined as a ratio of two modular forms both of weight 12,
it transforms with weight zero. However, since the moduli space
$\overline{\cM}_1\cong \P^1$ is compact, the only globally holomorphic
function is constant. Therefore $j$ has a pole at $q=0$,
\be
j= {1\over q}+ 744 + 196884 q + \ldots
\ee
In fact, one can proof that the map $j:\overline{\cM}_1 \ra \P^1$ is a
bijection, so the value of the $j$-function classifies the inequivalent
elliptic curves. (One often speaks about $\cM_1$ as the ``$j$-line.''
Also, all meromorphic modular functions are necessarily rational
expressions in the $j$-function, $j$ generates the function field on
$\cM_1$.)

We can now express the constants $a$ and $b$ in terms of the modulus
$\t$. One finds that $a= - {\pi^2\over3} E_4(\t)$ and $b= -{2\pi^6\over
27} E_6(\t)$ so that the $j$-value of the curve (\ref{ell}) is given by
\be
j(\t) = 1728 {4a^3\over 4a^3 +27 b^2}.
\ee
This tells us which elliptic curves are equivalent and which are not.

(3) Finally, from the point of view of conformal structures, we have to pick a
flat metric on $T^2$, which is of the general form
\be
ds^2 = \l \,dx^2 + \m \, dxdy + \n\, dy^2,
\ee
with constants $\l,\m,\n\in\R$.
Such a metric can always be written as
\be
ds^2 = \r |dx+\t dy|^2,\qquad g_{ab} = \r\twomatrixd 1 {\Re\t}{\Re\t}
{\t\taubar}
\ee
We then immediately make contact with the complex structure point of view.

\newsubsection{Surfaces with punctures}

We can also include marked points on the Riemann surface. If we fix $n$
different points $P_1,\ldots,P_n\in\Sigma_g$, the corresponding moduli space 
$\cM_{g,n}$ has dimension 
\be
\dim_\C \cM_{g,n} = 3g-3+n,
\ee
at least if $2g-2+n>0$, one extra dimension for each puncture. 

The simplest case is genus zero. We pick points
$z_1,\ldots,z_n\in\P^1$ with $z_i\neq z_j$ if $i\neq j$.
However, the moduli space $\cM_{0,n}$ is not simply $(\P^1)^n$ minus
diagonals, since
$\P^1$ has a non-trivial group of automorphisms
\be
\Aut(\P^1) = PGL(2,\C),
\ee
that we have to factor out. It acts by M\"obius transformations
on the coordinates $z_i$,
\be
z\ra {az+b\over cz+d},\qquad a,b,c,d\in\C,\ ad-bc\neq 0.
\ee
This quotient can be eliminated by fixing three points, say,
$\{z_1,z_2,z_3\}=\{0,1,\infty\}$. In this way we find for example that
\be
\cM_{0,3} \cong pt.
\ee
and
\be
\cM_{0,4} \cong \P^1-\{0,1,\infty\}.
\ee
Here we shouldn't confuse the Riemann surface and the moduli space!
We can compactify this space to $\Mbar_{0,4}\cong\P^1$ by adding back the
configurations $z=0,1,\infty$. However, it is better to think of adding
singular curves with a node and two points on each component, that is,
surfaces of the form
$$
\insertfig{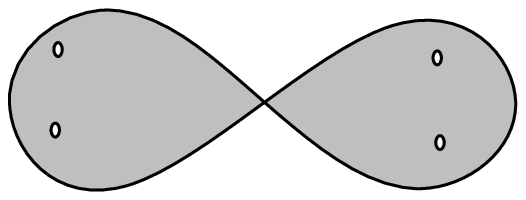}{4}
$$
Let us explain this in some more detail.

\newsubsection{The stable compactification}

The moduli space $\cM_{g,n}$ is non-compact because complex curves can
become degenerate. There exists a direct, intuitive interpretation of the
points at the boundary. There are basically two ways in which a
surface can degenerate. If we think in terms of a conformal class of
metrics, the surface can either form a node (equivalently, a long
neck) or two marked points can collide. The boundary of $\cM_{g,n}$
can be thought to lie at infinity. One would like to compactify the
moduli space by adding points at infinity, not unlike how one
compactifies the plane $\R^n$ to the $n$-dimensional sphere. In this
case the points at infinity represent particular singular Riemann
surfaces. The Knudsen-Deligne-Mumford or `stable' compactification
\cite{stable} tells us to add the following singular curves:

(1) The process in which two points $z_1$ and $z_2$ `collide', when the
difference $q=z_1-z_2$ tends to zero, can (after a coordinate
transformation $z \ra z/q$) alternatively be described as the process
in which a sphere, that contains $z_1$ and $z_2$ at fixed distance,
pinches off the surface by forming a neck of length $\log q$. These
two descriptions are fully equivalent, but the latter is actually more
in the spirit of conformal field theory, since we see in an obvious
way the operator product expansion emerge. The natural final configuration
is not simply the surface with $z_1=z_2$, but it consists of a
separate sphere containing the points $z_1$, $z_2$ and a third point
where the infinite long tube was attached, together with the original
surface with one marked point less.
$$
\insertfig{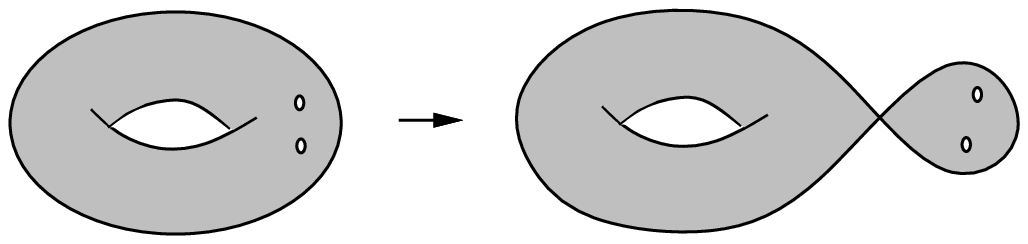}{8}
$$
In the stable compactification we add this configuration as limit point.
The crucial property of this compactification is that the points $z_i$
never are allowed to come together.

(2) If a cycle of non-trivial homology pinches, we replace the
surface by a surface with one handle less and two extra marked
points, the attachment points of the infinitely thin handle.
$$
\insertfig{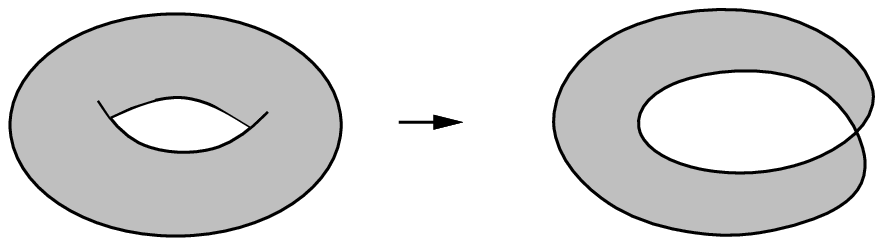}{7}
$$
So by this process we lower the genus by one.

(3) Finally, in case a dividing cycle pinches, the
resulting surface consists of two disconnected surfaces of genus $h$
and $g-h$, each having one extra puncture.
$$
\insertfig{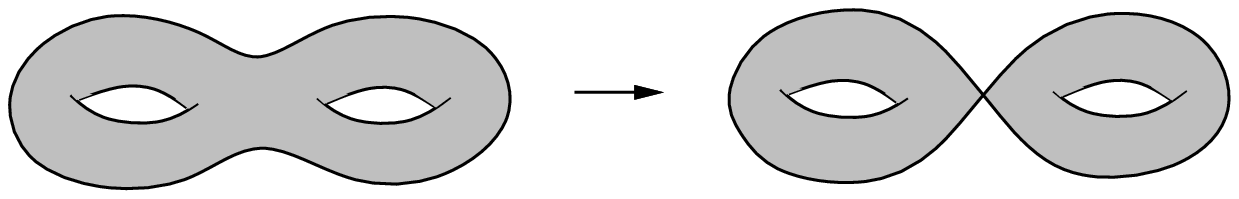}{9}
$$
It can be shown that this prescription makes $\cM_{g,n}$ into a compact
orbifold space $\overline\cM_{g,n}$ (and even a complex projective variety).

\newsection{Conformal field theory}

The next step in our hierarchy of two-dimensional field theories is
conformal field theory (CFT). There are two mathematically sound ways to
think about conformal field theories. Either in terms of representations
of the Virasoro algebra and operator algebras, or in terms of Riemann
surfaces. The algebraic and the geometric formalisms are equivalent, but
we will mainly focus on the latter.

\newsubsection{Algebraic approach}

We can think about CFTs in terms of a Hilbert space $\cH$ that carries a
representation of the Virasoro algebra $\Vir \oplus \overline{\Vir}$
generated by the local holomorphic and anti-holomorphic vector fields
\be
L_n = z^{n+1} \pp z,\qquad \Lbar_n = \zbar^{n+1} \pp \zbar
\ee
on the cylinder $\C^*$.
These  generators can be assembled in the (anti)holomorphic stress
tensor
\ba
T(z)\is\sum_n L_n z^{-n-2},\nonu
\Tbar(\zbar)\is\sum_n \Lbar_n \zbar^{-n-2}.
\ea
After quantization the Virasoro algebra $\Vir$ with central charge $c$ reads
\be
[L_n,L_m]=(n-m) L_{n+m} + {c\over 12} n(n^2-1) \delta_{n+m,0}.
\ee
To build a quantum field theory out of these representations,
one has to introduce additional
algebraic structure, in particular the operator algebra \cite{bpz}. 
This produces
in the end well-defined correlation functions
\be
\<\f_1(P_1)\cdots \f_n(P_n)\>_g,
\ee
where $\f_i \in \cH$ are states/operators and $P_i$ are points on a
genus $g$ Riemann surface $\Sigma$. Ignoring the anomaly for a moment,
these correlations functions are non-trivial functions on $\cM_{g,n}$.
Due to lack of time we will have to refer to the literature for a
further discussion of this ``algebraic'' point of view. (For a more
physical review see for example the excellent lectures by Ginsparg and
Cardy at the 1988 Les Houches school \cite{ginsparg,cardy} and the
reprint volume \cite{cft}. For a more mathematical exposition see \eg
\cite{fms,gawedzki}. )

\newsubsection{Functorial approach}

Alternatively, one can start with Segal's functorial axioms \cite{segal}
or what is known in the physics literature as the operator formalism
\cite{operator}. Hereto one has to consider a much bigger, ``dressed
up'' moduli space, denoted as $\cP_{g,n}$. It consists of Riemann
surfaces $\Sigma$ of genus $g$ and $n$ marked points $P_i\in\Sigma$
together with a choice of local coordinates $z_i$ around the punctures.
These local coordinates are chosen such that the point $P_i$ is given by
$z_i=0$. Of course, there is a projection map
\be
\cP_{g,n} \opra^\pi \cM_{g,n}
\ee
(with infinite-dimensional fibers) that simply forgets the information
about the local coordinate. The local coordinate not only allows one to
``cut a hole'' around the puncture by removing the disk $|z_i|<1$, it
also gives a parametrization of the resulting boundary $S^1$.

Generalizing the axioms of two-dimensional TFT, a CFT can now be 
regarded as a functor 
\be
\F: {\bf Riem} \ra {\bf Hilb}
\ee
from the category ${\bf Riem}$ of these Riemann surfaces with
parametrized boundaries to the category ${\bf Hilb}$ of Hilbert spaces.
(Here we completely ignore the notion of the central charge $c$. Strictly
speaking, everything here only holds for $c=0$, which is actually the
relevant case for string theory.)

That is, a CFT is map $\F$ that associates to each element in this extended moduli
space $\cP_{g,n}$  an element in the $n$-th tensor product of
the Hilbert space 
\be
\Sigma \in \cP_{g,n}\ \ \ \gives \ \ \ \F_\Sigma \in \cH^{\otimes n}.
\ee
(Note that for a Hilbert space we can identify $\cH$ with $\cH^*$ so
that we need not distinguish incoming and outgoing states.) If we compare
this point of view with the axioms of a TFT we see two important
differences: 

(1) The morphisms now depend on the choice of complex
structure;

(2) The vector space of states $\cH$ associated to the
circle is now infinite-dimensional and carries an hermitean inner
product.

The action of the Virasoro algebra is recovered by considering the
cylinder or annulus $\C^*$ which represents a map $\cH\ra\cH$, but we
will not go into this here, see \cite{segal}.

The CFT correlation functions are obtained by choosing vectors
$\f_1,\ldots,\f_n\in\cH$ and inserting them into the linear forms
$\F_\Sigma:\cH^{\otimes n}\ra\C$,
\be
\<\f_1(P_1)\cdots \f_n(P_n)\>_g = \F_\Sigma(\f_1,\ldots,\f_n)
\ee

The gluing property is more involved than in the case of a TFT, 
since it involves a local modulus $q$, comparable to $e^{-t}$ in the
case of quantum mechanics. We can glue two Riemann surfaces $\Sigma_1$
and $\Sigma_2$ at two punctures $P_1\in\Sigma_1$ and $P_2\in\Sigma_2$ by
identifying the local coordinates $z_1$ and $z_2$ as
\be
z_1z_2=q,\qquad q\in \C^*.
\ee
This is the inverse of the degeneration process we sketched in \S5.4.
The amplitudes of the new surface $\Sigma = \Sigma_1 \cup_q \Sigma_2$ are
now given by
\be
\F_\Sigma = \<\F_{\Sigma_1},q^{L_0}\qbar^{\Lbar_0}\F_{\Sigma_2}\>.
\ee
This corresponds geometrically to cutting unit disks $|z_i|\leq 1$ out
of the surfaces and inserting a cylinder of length $-\log |q|$ and
rotating this cylinder (``twisting'') around an angle $\arg q$. Note
that the dimensions of the moduli spaces work out, since we introduce a
new modulus $q$ in the gluing process.

The Virasoro algebra can be recovered in the following general way: if we
pick a non-single valued holomorphic vector field $\xi$ on the surface,
possibly with poles at the punctures, then we can associate to this a
deformation of the complex structure $\m=\dbar\xi$ (a quasi-conformal
transformation). Thereby we obtain a vector field on the {\it
moduli space} $\cP_{g,n}$. We denote the Lie derivative of this vector
field (as it acts on functions on the moduli space) as $\cL_\xi$. On the
other hand the Virasoro generators $L^{(i)}_n$ that act on the Hilbert space
$\cH$ at puncture $z_i=0$, are given in terms of the stress-tensor as
\be
L_n = \oint {dz \over 2\pi} z^{n+1} T(z).
\ee
More generally, we can define for an arbitrary local vector field $\xi$
\be
L_\xi = \oint {dz \over 2\pi i} \xi(z)T(z).
\ee
The relation between the functorial approach and the Virasoro algebra
is now given by the symbolic equation
\be
\left(\cL_\xi + \sum_i L^{(i)}_\xi\right)\F=0,
\ee
that tells us that a deformation of the moduli in $\cP_{g,n}$ can be
translated into an action of $\Vir$ on the state spaces at the punctures.

\newsubsection{Free bosons}

Let us put our feet back on the ground and briefly review the ur-CFTs
associated with free bosonic and fermionic fields, since we will need
these results later. A free boson, with action
\be
S= {1\over 2\pi} \int d^2\!z \, \d x \dbar x
\ee
can be conveniently discussed in terms of the spin one currents
$\d x$ and $\dbar x$. Because of the equation of motion, $\d\dbar x=0$,
these currents have a meromorphic expansion
\be
-i\d x = \sum_{n\in \Z} \a_n z^{-n-1}, \qquad \a_n^\dagger = \a_{-n}.
\ee
The standard free field quantization procedure gives rise to a Fock space
description of the Hilbert space. We have canonical commutation
relations
\be
[\a_n,\a_m]=n\delta_{n+m},
\ee
and the Fock states are created out of the ground states $|p\>$, with
$p\in\R$, defined by
\be
\a_0|p\>=p|p\>,\qquad
\a_n|p\>=0,\quad n> 0.
\ee
The bosonic Fock space, that we will denote as $\cB_p$,
is spanned by states of the form
\be
|\v\> = \a_{-n_1}\cdots \a_{-n_s}|p\> .
\ee
The operator-state correspondence
relates these states to the vertex operators
\be
\v(z) = \d^{n_1}x \cdots \d^{n_s}x e^{ipx(z)}.
\ee
The Fock space forms a $c=1$ representations of the Virasoro
algebra, generated by the stress-tensor
\be
T=-\half (\d x)^2.
\ee
The space is graded by conformal dimension,
the eigenvalues of the operator
\be
L_0 = \oint z T(z) = \half \a_0^2 + \sum_{n>0} \a_{-n}\a_n
\ee
and $U(1)$ charge or (space-time) momentum
\be
J_0=-i \oint  \d x = \a_0.
\ee
The degeneracies for fixed eigenvalues
can be read off from the character
\be
{\rm Tr}_{\strut \cB_p}\left(y^{J_0}q^{L_0+ c/24}\right) = 
{y^pq^{{1\over 2} p^2}\over \eta(q)}.
\label{B}
\ee
If we combine left-movers and right-movers, the full Hilbert space
is of the form
\be
\cH = \int\!\! dp\; \cB_p \otimes\overline{\cB}_p.
\ee
Its partition function reads
\ba
Z \is  {\rm Tr}_{\strut \cH} \left(q^{L_0+ {1\over 24}} 
\qbar^{\Lbar_0 + {1\over24}}\right)\nonu
\is \int dp \, {q^{{1\over 2}p^2} \qbar^{{1\over 2}p^2} \over |\eta|^2} \nonu
\is {1 \over \sqrt{\Im \tau} |\eta|^2}= {1\over \sqrt{\det'\Delta}}
\ea
with $\Delta$ the scalar Laplacian on $T^2$.

We can make a slight variation on this model by including a background
charge $Q$. This modifies the stress-tensor to
\be
T = -\half(\d x)^2 -iQ\d^2 x
\ee
and gives $c=1-3Q^2$. This shift produces
\be
L_0 \ra L_0 + \half Q J_0
\ee
and gives the vertex operator $e^{ipx}$ a conformal dimension $\half
p(p+Q)$. In the character (\ref{B}) this can be done by the ``twist'' $y
\ra y q^{Q/2}$.

\newsubsection{Free fermions}

Free two-dimensional chiral (Dirac) fermions $b,c$ have an action
\be
S = {1\over \pi}\int d^2\!z\, b\dbar c
\ee
where we can take the spins of $b$ and $c$ to be $\l$ and $1-\l$, with
$\l$ half-integer for physical fermions (that obey spin-statistics) and
$\l$ integer for ghost fields (that violate spin-statistics). The
equations of motion give $\dbar b=\dbar c=0$, so we have again mode
compositions of the form
\ba
b(z) \is \sum_{n\in\Z+\e} b_n z^{-n-\l}, \nonu
c(z) \is \sum_{n\in\Z+\e} c_n z^{-n-1 +\l}.
\ea
Here $\e=\half,0$ indicates the two possible spin structures on the
circle, usually referred to as Neveu-Schwarz (NS) and Ramond (R)
respectively.

Quantization starts from the canonical anti-commutation relations
\be
\{b_n,c_m\}=\delta_{n+m,0}, 
\ee
with
\be
\{b_n,b_m\}=\{c_n,c_m\}=0.
\ee
One defines a Fermi sea $|\m\>$ satisfying
\ba
b_n|\m\>=0, & &\quad n > \m-\l, \nonu
c_n|\m\>=0, & & \quad n \geq \l - \m.
\ea
and this gives rise to a fermionic Fock space,
written as $\cF^\e_{\m}$ and spanned by
states of the form
\be
b_{-m_1}\cdots b_{-m_r} c_{-n_1}\cdots c_{-n_s}|\m\>.
\ee
This will form a representation of the Virasoro algebra with central
charge
\be
c=1-3(1-2\l)^2,
\ee 
and stress tensor
\be
T =  -\l b\d c + (1-\l) \d b c.
\ee
We also have a $U(1)$ quantum number given by the eigenvalues
of the operator
\be
J_0 = - \oint  b(z)c(z)
\ee
With these conventions the modes $b_n$ carry charge $-1$ and $c_n$ carry
charge $+1$. The degeneracies for fixed eigenvalues of $L_0$ and
$J_0$ follow from the expansion of the character 
\be
{\rm Tr}_{\strut \cF^\e_\m}\left(q^{L_0+ c/24}y^{J_0}\right) = 
y^\m q^{\half \m(1+\m-2\l)}
\prod_{n \geq \l - \m}\left(1+yq^n\right) \prod_{m > \m-\l}
\left(1+y\inv q^m\right)
\ee

\newsection{Sigma models and T-duality}

We now want to discuss some examples of less trivial CFTs describing
strings moving on spaces with non-trivial topology. One of the
interesting phenomena that we meet here is that two manifolds that are
classically very different, turn out to be equivalent when considered as
a background in string theory. This is one of the manifestations of
duality. In its more sophisticated form this is known as mirror symmetry
\cite{greene-plesser,candelas,morrison-rat,mirror}, as discussed at
great length in the lectures of Brian Greene \cite{brian}. Here we will
mainly stick with the more tractable abelian dualities. These so-called
T-dualities are reviewed in great detail in \cite{giveon}. Actually, as
we will see, abelian dualities are basically infinite-dimensional
versions of the fact that under Fourier transformation the gaussian
behaves as 
\be 
e^{-ax^2} \ra e^{-x^2/a} .
\ee

\newsubsection{Two-dimensional sigma models}

In this lecture we discuss two-dimensional sigma models. Here the
fundamental fields are maps
\be
x:\Sigma \ra X,
\ee
where $\Sigma$ is a closed Riemann surface (compact without boundary)
and $X$ is {\it a priori} an arbitrary compact $n$-dimensional
Riemannian manifold. The corresponding quantum field theory is defined
through the path-integral over all such smooth maps
\be
Z = \int  \cD x \, e^{-S},
\ee
where the action $S$ is defined as follows. We choose a conformal
class of metrics on the surface $\Sigma$, so that we obtain a Hodge star
$*$, and pick coordinates $x^\mu$ and a metric $G_{\mu\nu}$ on $X$.
The action is then given by
\be
S = {1\over 4\pi\alpha'} \int_\Sigma
G_{\mu\nu}(x) dx^\m \wedge * dx^\n .
\ee
Here $\alpha'$ is the coupling constant of the sigma model that we
usually simply set to one by absorbing it into the metric $G$. Note that
weak coupling $\a'\ra 0$ (in which sigma model perturbation theory makes
sense and the classical action, in particular the target space metric
can be recovered) corresponds to large volume of the target space $X$.
It is well-known that the critical points of the action are given by the
so-called {\it harmonic maps} --- a beautiful subject in classical
differential geometry (see \eg\ \cite{harmonic}). 

The action can be
generalized by also picking a two-form $B=\half B_{\m\n}dx^\m \wedge dx^\n$
on $X$ and adding the term
\be
{i\over 4\pi} \int_\Sigma B_{\m\n} dx^\m \wedge  dx^\n
= {i\over 2\pi} \int_\Sigma x^*B
\ee
to the action.
Note that by Stokes' theorem this term is invariant under shifts $B \ra
B + d\L$. 

Actually, we have an even stronger quantum equivalence by
which we shift $B \ra B + 4\pi^2C$ with $C$ a closed two-form with 
quantized periods, 
\be 
C \in H^2(X,\Z),
\ee 
since the path-integral
picks  up a term
\be
Z \ra Z \. e^{2\pi i \int_\Sigma x^*C}=Z
\ee

When is such a sigma model conformal invariant? The action
only uses the Hodge star, so classically the theory only depends on 
the conformal class of the metric on $\Sigma$. This conformal invariance
is in general not preserved at the quantum level. It is ruined
by the implicit but very important metric dependence of the measure $\cD
x$ in the path-integral. This fact is reflected by a non-vanishing
beta-function, which for a sigma model is given by an 
expression in the
target space Riemann tensor \cite{beta}
\be
\beta_{\mu\nu} = R_{\mu\nu} + 
\alpha' \half R_{\mu\k\l\r}R_\nu{}^{\k\l\r}
+ O((\alpha')^2)
\label{beta}
\ee
So to first order in sigma model perturbation theory, conformal 
invariance implies that the target space
should provide a solution to the vacuum Einstein equations
\be
R_{\m\n}=0.
\ee
One similarly derives an equation for the background $B_{\m\n}$ field,
that reads to first order in $\a'$
\be
dB=0.
\ee
Together with the gauge equivalence that we discussed above this gives
us an element in the torus\footnote{Here and everywhere we ignore
torsion.}
\be
B \in H^2(X,\R)/ H^2(X,\Z).
\ee

Actually, for compact spaces, the full `elliptic' equations
$\beta_{\m\n}=0$ only allows for flat space solutions,
\be
R_{\m\n\l\r}=0
\ee
leaving only tori (and singular quotients) as possible candidates for a
conformal invariant sigma-model. This situation improves remarkably if
we also include fermionic degrees of freedom, as we will indicate in
\S7.7. (But see Brian Greene's lectures \cite{brian} for more details.)

\newsubsection{Toroidal models}

The simplest cases for conformal invariant sigma-models are tori, 
so let us assume that $X$ can be written as
\be
X=T^n=\R^n/2\pi \L
\ee
with $\L$ a rank $n$ lattice. The action of these toroidal models
is pure gaussian
\be
S = {1\over 2\pi}\int_\Sigma \left(G_{\m\n} + B_{\m\n}\right)\d x^\m
 \dbar x^\n .
\ee
Duality is in this context the statement that the sigma-model model is
insensitive (up to a change in normalization of the partition function)
to the interchange of the torus with the dual torus
\be
T^n \ \iff \ (T^n)^*.
\ee
This particular type duality transformation is  usually referred to as {\it
T-duality} \cite{giveon}. It interchanges target spaces with large
volumes and target spaces with small
volumes. From the point of view of sigma model perturbation theory it
relates strong and weak coupling
\be
\a' \ra 1/\a',
\ee
so we could have named it a {\it world-sheet} S-duality. But we prefer to
reserve the term S-duality only for space-time strong-weak coupling
transformation. Since a T-duality is an automorphism of the CFT and
acts for every given Riemann surface, it
is a perturbative symmetry from the point of string perturbation theory.

The simplest way to prove T-duality starts from the following
path-integral \cite{rocek} (we
drop the $B$-field for convenience)
\be
Z = \int \cD A\cD y\, \exp{-{1\over 2\pi} \int \left(
G_{\m\n}A^\m\wedge *A^\n  + A^\m\wedge dy_\m\right)}
\ee
with $A^\m$ (for fixed $\m=1,\ldots,n=\dim X$) a vector-valued
one-form on $\Sigma$ and $y_\m$ a set of scalars. 
On the one hand we can integrate out $y_\m$, which imposes the condition
\be
dA^\m=0 \gives A^\m=dx^\m,
\ee
at least locally. This gives us back the sigma model on the original
torus $T^n$. On the other hand we can integrate out the field $A^\m$
which gives the integral
\be
Z = \int\cD y\, \exp{-{1\over 2\pi} \int G^{\m\n}dy_\m\wedge *dy_\n}
\ee
with $G^{\m\n}$ the dual metric on the dual torus $\widehat{T}^n$, $G^{\m\n}
G_{\n\l} = \delta^\m{}_\l$. This produces the T-dual model.

Before we discuss the full T-duality group, let us first briefly review
some general facts about lattices, that will also come in useful later, when
we discuss four-dimensional geometry in \S9.1.

\newsubsection{Intermezzo --- lattices}

 A lattice $\G$ is by definition a $\Z$ vector
space,
\be
\G \cong \Z^n,
\ee
together with a non-degenerate, symmetric bilinear
form (inner product)
\be
Q: \G \times \G \ra \Z.
\ee
This bilinear form will have a signature $(p,q)$.
The dual lattice $\G^*$ is defined as the set
\be
\G^*=\{ x \in\R^n;\ Q(x,y)\in \Z,\ \forall y\in\G\}.
\ee
A lattice is self-dual, $\G^*=\G$, if and only if the inner product
is unimodular
\be
\det Q = \pm 1.
\ee
A lattice $\G$ is called even if
\be
x^2\in 2\Z,\qquad \forall x\in\G,
\ee
and odd otherwise. For even lattices the signature $\s=p-q$ satisfies
necessarily
\be
\s \equiv 0 \mod 8.
\ee

The classification of self-dual lattices now proceeds depending on
whether the intersection form is even or odd and whether it is
(positive or negative) definite or not \cite{lattices}. 
The various possibilities are
$$
\renewcommand{\arraystretch}{1.5}
\begin{tabular}{|c|c|c|}
\hline
\strut \qquad & odd & even \\ \hline
\strut indef & $p\id \oplus q (-\id)$ & $pH\oplus n
E_8$ \\ \hline
\strut def &$ n\id,$ \it exotic & \it exotic:\ $E_8,D_{16},\ldots$ 
\\ 
\hline
\end{tabular}
\renewcommand{\arraystretch}{1.0}
$$
Here $H$ denotes the two-dimensional hyperbolic lattice
\be
H =\twomatrix 0110 
\ee
and $E_8$ the root lattice of the Lie algebra of the same name.  The
most remarkable fact is that there exists (for given rank and
signature) a {\it unique} indefinite self-dual lattice --- a theorem
by Hasse and Minkowski. 
In the odd case it carries simply the diagonal form, so the lattice
equals the hypercube lattice. In the
even case, where we will denote the lattice as $\G^{p,q}$ with $p,q>0$,
$p-q=0$ (mod 8),
it is given by a sum of hyperbolic and $E_8$ lattices,
\be
\G^{p+8n,p} = pH \oplus n E_8.
\ee
 
Unfortunately, the situation in the definite case is much more
complicated, because there are many more possibilities apart from the
obvious cubic lattice $n\id$. These are usually referred to as exotic
lattices. Although the number of possibilities remains finite for
given rank, it grows rather dramatically: for rank $8,16,24,32,\ldots$
we find $1,2,24,\sim 10^7,\ldots$ inequivalent even definite self-dual
lattices. We will need these facts about lattices again if we consider
four-manifolds in \S9.1.

For any lattice $\G$ of positive signature $(n,0)$
we define the theta-function as
\be
\theta_\G(\t) = \sum_{v\in\G} q^{\half v^2},\qquad q=e^{2\pi i\t},\
\Im\t>0.
\ee
Poisson resummation gives that
\be
\theta_\G(-1/\t) = \left({i\over \t}\right)^{n/2} \theta_{\G^*}(\t)
\ee
So for $\G$ even and self-dual, $\th_\G(\t)$ is  a modular form of weight
$n/2$, \eg in the simplest case
\be
\th_{E_8}(\t) = E_4(\t).
\ee
 
\newsubsection{Spectrum and moduli of toroidal models}

Returning to our discussion of toroidal sigma models, we want
to show that the full duality symmetry group of the
sigma model with target space $T^n$ is given by
\be
G=O(n,n,\Z),
\ee
by which we denote the group of automorphisms of the lattice
\be
\G^{n,n}=\L^* \oplus \L,
\ee
with inner product
\be
p^2=2k\cdot m,\qquad p=(k,m),\ k\in\L^*, \ m\in\L.
\ee
This so-called Narain \cite{narain} lattice $\G^{n,n}$ is an even self-dual lattice of
signature $(n,n)$. As we have seen, it is unique up to isomorphism,
\be
\G^{n,n} \cong \underbrace{H\oplus \cdots \oplus H}_{n\ times}.
\ee

The action of the duality symmetry group $O(n,n,\Z)$ is particular clear
if we consider the Hilbert space of the toroidal model. It is again
given by a sum of tensor products of left-moving and right-moving Fock
spaces as in \S6.3, now however built on momentum states $p_L$ and $p_R$
that take value in the Narain lattice
\be
\cH = \bigoplus_{(p_L,p_R)\in\G^{n,n}} \!\!\!
\cB_{p_L} \otimes \cB_{p_R},
\ee
with
\be
p^\m_{L,R}= k^\m \pm m^\m + B^{\m\n} m_\n,\qquad 
k\in\L^*, \ m\in\L,
\ee
(Here indices are raised and lowered with the metric $G_{\m\n}$.)
Note that the choice of metric and $B$-field determines the split
of the momentum $p$,
\be
p=p_L \oplus p_R, \qquad p^2 = p_L^2 - p_R^2.
\ee
The one-loop partition function reads
\be
Z = {\rm Tr}_{\strut \cH}
\left(q^{L_0+{n\over 24}} \qbar^{\Lbar_0+{n\over 24}} 
\right)=
\sum_{(p_L,p_R)\in \G^{n,n}} {q^{\half p_L^2} \qbar^{\half p_R^2} \over
|\eta(q)|^{2n}}
\label{part}
\ee
One can prove quite straightforwardly that the partition function $Z$ is
invariant under modular transformations. Invariance under $T$ and $S$
transformations in $SL(2,\Z)$ is implied by the fact that $\G^{n,n}$ is
even and self-dual, respectively. Furthermore, the expression is by
inspection invariant under the action of $O(n,n,\Z)$ on the lattice,
since we sum over all elements. 

The ``classical'' moduli space of this CFT is parametrized by the
constant $n\times n$ matrix 
\be
G_{\m\n} + B_{\m\n},
\ee
\newcommand{\Gr}{{\rm Gr}}
and is given by the Grassmannian $\Gr^{n,n}$ of maximal positive subspaces
in $\R^{n,n}$ 
\ba
\Gr^{n,n} \is \{V\in \R^{n,n}, \dim V = n,\ (\.,\.)|_{V}>0\}\nonu
         & \cong & O(n,n;\R)/O(n;\R) \times O(n;\R).
\ea
The ``quantum'' moduli space is defined as the left-quotient of this
homogeneous space by the ``arithmetic subgroup'' $G=O(n,n,\Z)$
\be
\cM_{T^n} =G\backslash \Gr^{n,n}.
\ee
To see how this  Narain moduli space classifies toroidal 
sigma models, choose a  $V\in \Gr^{n,n}$, so that we can write
\be
\R^{n,n}= V \oplus V^\perp.
\ee
A vector $p\in\G^{n,n}$ can in this decomposition be written as
$p=(p_L,p_R)$ with $p_L\in V$, $p_R\in V^\perp$. The quadratic form
$p^2$ will be the difference of two positive definite
forms of rank $n$ 
\be
p^2 = p_L^2 - p_R^2.
\ee
So we recover the spectrum described above.
Note that inside the duality-group $G$ we have a ``geometrical''
subgroup
\be
SL(n,\Z) \subset O(n,n,\Z)
\ee
that represent the ``large'' diffeomorphisms of $T^n$.

\newsubsection{The two-torus}

A particular interesting case appears for $n=2$, it is described in
detail in \cite{copenhagen}. Here we are dealing with $X=T^2$,
a two-torus or
elliptic curve, which is the simplest example of a Calabi-Yau space. We
can represent the elliptic curve as 
\be
T^2 = \C/\Z\oplus \rho\Z,
\ee
with the modulus $\rho\in\H$, the upper half-plane, and naturally identified
modulo the action of the modular group $PSL(2,\Z)$. The K\"ahler form
and the $B$-field (which has here only one non-zero component) combine conveniently in
one complex (1,1) form
\be
\omega = \half (B+i g_{z\zbar}) dz\wedge d\zbar.
\ee
We can write this complexified K\"ahler form as
\be
\omega=\sigma \left({\pi\over \Im \rho}dz\wedge d\zbar\right),
\ee
where we introduced a second complex modulus $\sigma$ as
\be
\int_{T^2}\omega = 2\pi i \sigma
\ee
Note that the area is given by $\Im\sigma$ and should be positive, so $\sigma$ 
also takes value in the upper half-plane. The local moduli are therefore
\be
(\r,\s)\in \H \times \H
\ee
which indeed represents the Grassmannian $\Gr^{2,2}$.
The sigma model action reads in this parametrization, with complex
field $x=x^1+\r x^2$ that takes values
on the torus $\C/\L_\r$ (see (\ref{lat}) for notation)
\be
S = \int  {i\pi d^2\!z\over \Im\r} \left(\bar\s \d \xbar \dbar x-
\s \d x\dbar\xbar\right)
\ee

Now our claim is that there
is also a modular $PSL(2,\Z)$ group acting on the K\"ahler variable 
$\sigma$. In fact, one transformation,
the usual shift of the $B$-field, is easily seen to be represented by
\be
T: \sigma \ra \sigma + 1.
\ee
The transformation $S:\s \ra -1/\s$ relates small and volume, and
equals the T-duality that we proved in \S7.2.

One way to see the full duality symmetries is to notice that the Narain
lattice can be written as
\be 
\Gamma^{2,2} = \bigoplus_{i=1}^4 \Z e_i
\ee
with $e_i \in \C^2$ the vectors
\be
e_1 = (1,1),\quad e_2 = (\rho,\rho),\quad
e_3 = (\sigma,\overline\sigma), \quad e_4 = (\rho\sigma,
\rho\overline\sigma).
\ee
and where the signature $(2,2)$ inner product of a vector
$(p_L,p_R)$ is given by 
\be
p^2= {p_L^2 - p_R^2 \over \Im \rho\. \Im \sigma}.
\ee
From this explicit realization of the Narain lattice we can simply read
off the full quantum symmetry group $O(2,2,\Z)$. In fact, we have the
isomorphism
\be
O(2,2;\Z)=PSL(2,\Z)\times PSL(2,\Z) \lltimes \Z_2.
\ee
The two copies of $\G=PSL(2,\Z)$ act in the obvious way on the two
moduli $\r,\s$ and the $\Z_2$ interchanges them
\be
(\rho,\sigma) \ \lra \ (\sigma,\rho)
\ee
Note that this transformation interchanges the complex K\"ahler form $\rho$
and the complex structure $\sigma$. It  can be 
seen as the simplest realization of mirror symmetry.
It can be intuitively understood by picking a rectangular two-torus with
radii $R_1$ and $R_2$. We then find 
\be
\r = i{R_1\over R_2},\qquad \s = i{R_1 R_2}.
\ee
A duality transformation on one $S^1$ will send $R_1 \ra 1/R_1$ and thus
interchange $\r$ and $\s$. 

Concluding, the moduli space for a sigma model with target space $T^2$ is 
\be
\cM_{T^2} \cong \left(\P^1_\r\times \P^1_\s\right)/\Z_2
\ee

\newsubsection{Path-integral computation of the partition function}

It might be instructive to also compute the partition function
(\ref{part}) of a toroidal model directly from the path-integral
\cite{higher-genus,c=1}. We will
do the general case of a field 
\be
x: \Sigma \ra T^n\cong\R^n/2\pi \L
\ee
on  a genus $g$ surface $\Sigma$.

We first choose an homology basis $A^i,B_i$, $i=1,\ldots, g$,
of $H_1(\Sigma,\Z)$ as in
\figuurplus{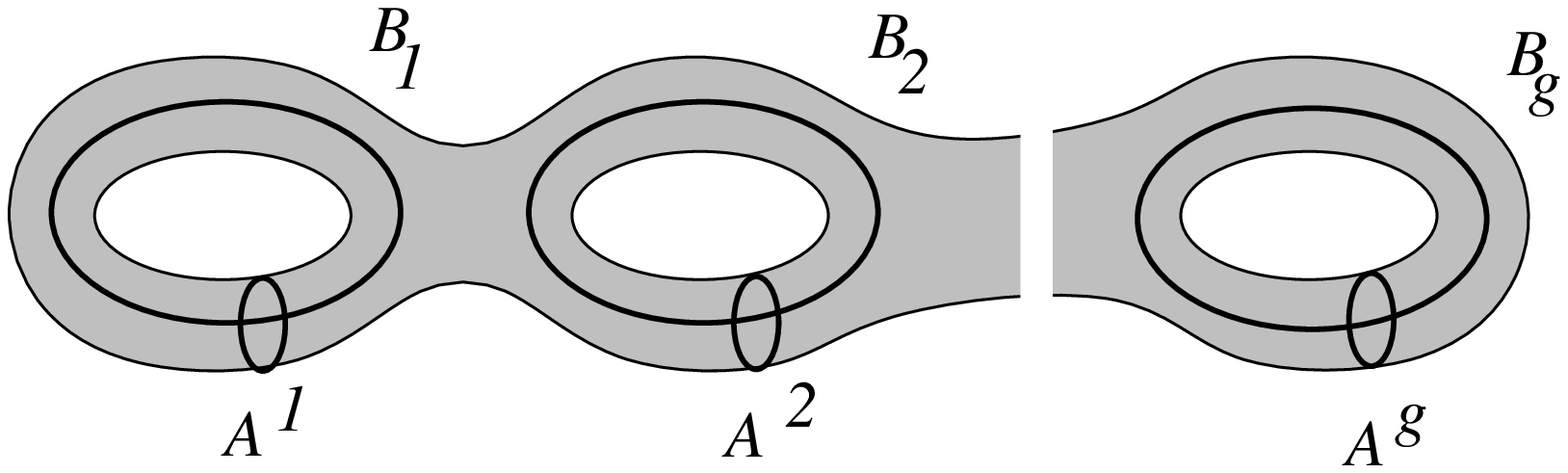}{8cm}{A Riemann surface with an homology basis.}
In terms of these cycles the one-form $dx$ has periods
\ba
{1\over 2\pi}\oint_{A^i} dx \is  m^i  \in \L, \nonu
{1\over 2\pi}\oint_{B_i} dx \is  n_i  \in \L .
\ea
Now we can write
\be
dx=\a + d y,
\ee
with $y$ a proper function $\Sigma_g \ra \R^n$ and $\a$ a harmonic
vector-valued one-form, $d\a=d^*\a=0$, with quantized periods 
\be
{\a \over 2\pi } \in H^1(\Sigma,\L). 
\ee
To compute the contribution to the action of this harmonic piece, the form
$\a$ must be decomposed in terms of the basis
$\w_i \in H^0(K_\Sigma)$ ($i=1,\ldots,g$) of holomorphic (1,0) one-forms on
$\Sigma$ and their conjugates. These satisfy canonical normalizations of their periods
\ba
\oint_{A^i} \w_j \is  \delta^i{}_j, \nonu
\oint_{B_i} \w_j \is \t_{ij}.
\ea
with $\t_{ij}$ the period matrix, a complex symmetric matrix with $\Im \t>0$. 
Note that these abelian differentials satisfy
\be
\int_\Sigma \w_i \wedge \overline\w_j = -2i\, \Im\t_{ij}.
\ee
If we now write
\be
\a = \l^i \w_i + \bar\l^i \bar\w_i,
\ee
we see that the periods are expressed as
\ba
m^i \is \l^i + \bar\l^i, \nonu
n_i \is  \t_{ij}\l^j + \bar\t_{ij}\bar\l^j,
\ea
so that the coefficients
 $\l^i$ can be solved in terms of the winding numbers $m^i,n_i\in\L$
as
\be
\l = i\pi (\Im\t)\inv (n-m\t).
\ee
If we insert this expression for $\a$ in the action, we obtain
the contribution
\be
S_{m,n} = \half \pi (n-m\bar\t)(\Im\t)\inv(n-m\t) + i\pi m\cdot B\cdot n.
\ee
So the partition function can be written as
\be
Z= Z_{qu} \sum_{m,n\in \L}e^{-S_{m,n}}.
\ee
Here $Z_{qu}$ is the contribution of the single-valued fields $y$, a
determinant that does not depend on the moduli of $T^n$ (apart from the
zero-mode). 

In order to reach the
canonical form (\ref{part}) we have to perform a further Poisson
resummation on the variables $n_i\in\L$, exchanging it for 
the dual variables $k^i\in \L^*$.
The soliton sum then takes the familiar form
\be
\sum_{m^i\in \L,\ k^i\in\L^*}
\exp\left(i\pi (k+m)\t(k+m) - i\pi (k-m)\t(k-m)\right)
\ee
in which we recognize the left-moving and right-moving momenta $p_L=k+m$
and $p_R=k-m$. If we now restrict to $g=1$ we obtain (\ref{part}).

This computation is particular interesting for the case of a two-torus.
If we compute the genus one partition function, we are dealing with maps
\be
x:T_\t^2 \ra T_{\r,\s}^2,
\ee
where the suffices indicate the moduli.
In the parametrization of \S7.5 the soliton sum reads
\be
\sum_{m,n\in\L_\r} \exp\left( -i\pi \s {|n-m\bar\t|^2 \over \Im\t\Im\r}
+ i\pi \bar\s {|n-m\t|^2 \over \Im\t\Im\r}\right)
\ee
were $m$ and $n$ are elements of the lattice $ \L_\r = \Z\oplus\r\Z.$ After
the Poisson resummation, we get the momentum sum 
\be
\sum_{k,m \in\L_\t} \exp\left(i\pi \t {|k-m\bar\s|^2 \over \Im\r\Im\s}
- i\pi \bar\t {|k-m\s|^2 \over \Im\r\Im\s}\right)
\ee
Here we note a remarkable triality permuting the {\it three} moduli
$\r,\s,\t$ \cite{copenhagen}.

\newsubsection{Supersymmetric sigma models and Calabi-Yau spaces}

If we add world-sheet supersymmetry, the set of conformal invariant
sigma models becomes more interesting. For $N=1$ supersymmetry not much
happens: only tori and orbifolds survive. However, for $N=2$
supersymmetry there is a much richer class, the Calabi-Yau spaces
\cite{calabi,yau}. 

Calabi-Yau manifolds, first introduced in the physics literature in
\cite{chsw}, are K\"ahler spaces of complex dimension $d$ for which the
holonomy, which generically lies in $U(d)$ for a K\"ahler space,
restricts to $SU(d)$. Calabi-Yau spaces can equivalently be defined by
the property that for fixed K\"ahler class they have a unique solution
to the vacuum Einstein equation
\be
R_{i\jbar} = 0,
\ee
\ie they allow for a Ricci-flat K\"ahler metric. Indeed, the usual
second order term in sigma model perturbation theory in the beta
function (\ref{beta}) disappears for $N=2$ models, as does a third order
term. So in the context of $N=2$ supersymmetry Ricci flat models stand a
change to define a CFT. It turns out that there is a nontrivial fourth
order correction, but this correction can always be compensated by
adding a piece to the metric, that keeps the cohomology of the K\"ahler
class invariant.

As conjectured by Calabi \cite{calabi} and proven by Yau \cite{yau},
a necessary and sufficient condition for Ricci flatness
is given by the condition that the canonical line bundle of $X$,
defined as
\be
K_X = \ext^d T^*_X
\ee
is trivial
\be
c_1(K_X) = c_1(X) = 0.
\ee
So we have in particular one global holomorphic section
$\W$ of $K_X$ (unique up to a scale factor).
That is, a CY space comes with a unique holomorphic
volume-form 
\be
\Omega = \Omega(x) dz^1 \wedge \ldots
\wedge dz^d
\ee
Therefore the space $H^{d,0}(X)$ is one-dimensional,
\be
h^{d,0} = 1.
\ee
If a manifold has precisely $SU(d)$ holonomy and not less (so that it is
a Calabi-Yau space in the strict sense) we further
have
\be
h^{1,0}=h^{2,0}=\ldots=h^{d-1,0}=0.
\ee
In dimension  $d\leq 3$ we have the following list of possible CY spaces 
$$
\renewcommand{\arraystretch}{1.5}
\begin{tabular}{|c|c|}
\hline
\strut $d=1$ & $T^2$ \\
\hline
\strut $d=2$ & $T^4,\ K3$ \\
\hline
\strut $d=3$ & $T^6,\ T^2\times K3, \ CY_3$ \\ 
\hline
\end{tabular}
\renewcommand{\arraystretch}{1.0}
$$ Here $CY_3$ denotes a simply connected three-fold; all tori have
trivial holonomy group and $K3$ is the unique compact two-dimensional
CY with holonomy $SU(2)$. We will not discuss supersymmetric sigma
models on CY spaces in much detail here, since they are covered in
detail in Brian Greene's lectures \cite{brian}. To be self-contained
in these notes, let us just make a few comments about their moduli
spaces, that we need later.

For Calabi-Yau three-folds we only understand the local structure of the
sigma-model moduli space in full detail. It is clear that we have a
unique CFT on $X$ given a complex structure and a K\"ahler class.
Deformations of complex structures are infinitesimally given by the
cohomology group $ H^1(T_X). $ Using the existence of the holomorphic
three-form one can show that these deformations are unobstructed
\cite{cy-period}.
Furthermore this cohomology group is isomorphic with $H^{1,2}(X)$. Thus
variations $\delta\sigma$ of the complex structure are parametrized by
\be
\delta\sigma \in H^{2,1}(X),
\ee
and there are $h^{2,1}$ of these complex structure deformation
parameters. The inequivalent K\"ahler deformations $\delta\rho$,
including the $B$ field) are
elements of the cohomology group
\be
\delta\rho \in H^{1,1}(X)=H^1(T^*_X).
\ee
These give rise to another $h^{1,1}$ complex 
moduli. Summarizing, the moduli space of CY sigma models takes the local form
\be
T_X\cM \cong H^{2,1}(X) \times H^{1,1}(X)
\ee
In the case of the two-torus we already saw the important role these
moduli-deformations $\delta\rho,\delta\sigma$ played, and that there
were strong relations between the two. Actually, in that case we could
simply interchange the two. Mirror symmetry is the generalization of
that phenomenon to higher dimensional manifolds \cite{mirror}.

\newsubsection{Calabi-Yau moduli space and special geometry}

Let us make a few further remarks about the moduli space of complex
structures of Calabi-Yau three-folds. Let $\cM$ be the moduli space of
inequivalent complex structures on the CY space $X$. If $h^{2,1} = g$,
then is $\cM$ a $g$ dimensional complex space. On $X$ we can pick a
holomorphic volume-form $\W$. Let $\cL$ denote the moduli space of pairs
$(X,\W)$. Since the choice of $\W$ is unique up to multiplication by
complex numbers, $\cL$ is a line bundle over $\cM$,
\be
\cL \opra^\pi \cM.
\ee
We can now consider the so-called period map that is defined
as follows. First we recall that $H^3(X,\C)=H^3(X,\Z)\otimes \C$, where 
the cohomology group $H^3(X,\Z)$ is the dual to the space $H_3(X,\Z)$
of three-cycles. By the intersection form 
\be
\eta(\a,\b)=\int_X \a\wedge\b,\qquad \a,\b\in H^3(X,\Z),
\ee
it is naturally a $2g+2$ integer dimensional symplectic vector space. It is a
topological object whose definition does not depend on the choice of
complex structure. 

If we choose a particular complex structure we get a
Hodge decomposition of $H^3(X,\C)$ in terms of the Dolbeault groups
\be
H^3(X,\C) \cong H^{3,0} \oplus H^{2,1} \oplus H^{1,2} \oplus H^{0,3}
\ee
with $\bar H^{p,q} = H^{q,p}$. The piece $H^{3,0}$ 
represents the space of holomorphic volume forms and is one-dimensional.
The period map now associates to the pair $(X,\W)$ the point in $H^3(X,\C)$
represented by $\W$. An important mathematical result is that this 
map
\be
\cL \ra H^3(X,\C)
\ee
is injective \cite{cy-period}. This gives a description of $\cL$ as a
cone in $H^3(X,\C)$. (It is a cone, since we can scale the holomorphic
three-form by multiplication by $\C$.)

Now the period map can be described in more detail
if we choose once and for all a basis in $H^3(X,\Z)$
by picking a canonical basis $A^i,B_i$ ($i=0,\ldots,g$) of homology
three-cycles. This gives a dual basis $\a_i,\b^i$ of integer
three-forms. On this basis we can decompose the holomorphic (3,0) form
as
\be
\W = \f^i \a_i + \cF_i \b^i
\ee
in terms of the periods
\ba
\oint_{A^i} \W \is \f^i,\nonu
\oint_{B_i} \W \is \cF_i.
\ea
Because of the properties of the period map, the components $\f^i$ may
be used as local coordinates on $\cL$, or, equivalently, as homogeneous
coordinates on $\cM$. The components $\cF_i$ become then functions of
the $\f^i$. We can now differentiate $\W$ with respect to these moduli
$\f^i$. The $g+1$ partial derivatives
\be
\w_i = \d_i\W
\ee
will give a basis of the space\footnote{Here we need the technical
assumption of Griffiths transversality: the first order variation of a
$(3,0)$ form gives at most a $(2,1)$ form \cite{griffiths}.}
\be
W=H^{3,0} \oplus H^{2,1}.
\ee
Note that we have
\be
H^3(X,\C) = W \oplus \bar W
\ee
and this is a complex polarization of the symplectic vector space, which
means concretely that
\be
\int_X \w_i \wedge \w_j = 0.
\label{pol}
\ee
The $g+1$ three-forms $\w_i$ are given in components as
\be
\w_i = \a_i + {\d \cF_j \over \d\f^i} \b^j.
\ee
We will use the notation
\be
\t_{ij} = {\d \cF_j \over \d\f^i}.
\ee
Now  we claim that 
\be
\t_{ij} = \t_{ji}.
\ee
This follows directly from
relation (\ref{pol}), by using the fact that for {\it any} two 3-form
$\F$ and $\Psi$
\be
\int_X \F\wedge\Psi = \sum_i \left(\oint_{A^i}\F \oint_{B_i}\Psi -
\oint_{A^i}\Psi \oint_{B_i}\F \right).
\ee
The LHS can only contribute if it is a form of total degree (3,3).
Various identities are obtained by inserting forms for which the LHS
vanishes. For instance, if we plug in $\w_i$ and $\w_j$ this is the
case, and we find the relation
\be
\t_{ij} = \oint_{B^j}\w_i =  \oint_{B^i}\w_j = \t_{ji}
\ee
Since $\t_{ij}=\d_i\cF_j$, 
this tells us that locally the period $\cF_i$ is the derivative of
a function on $\cL$, the prepotential $\cF(\f)$
\be
\cF_i = \d_i\cF.
\ee
Note that this function is homogeneous of degree two. This relation is
obtained easily by using the above trick again for the forms $\W$
and $\w_i$, which gives
\be
 0 = \int_X \W \wedge \w_i = \f^j \d_j\cF_i - \cF_j.
\ee
So, $\cF_i$ is homogeneous of degree 1, and therefore $\cF$ is homogeneous
of degree 2 (and therefore a section of the bundle $\cL^{\otimes 2}$.
Note that we can write
\be
\cF = \half \f^i\cF_i = \half \oint_{A^i}\W \oint_{B_i}\W.
\ee
Another useful identity is that
\be
\W = \f^i\w_i.
\ee
That is, in the local coordinates $\f^i$ on $W$, the three-form
$\W$ is given by the vector $\f$.

Finally, there is a beautiful formula for the natural Weil-Peterson
K\"ahler metric on the moduli space $\cM$
\cite{strominger-special}. (Note that such a metric
canonically exist, since the complex deformations can be written as
special metric deformations. The space of Riemannian metrics always
carries a natural metric itself.) In the case of a family of Calabi-Yau
spaces the metric is K\"ahler and given in terms of the K\"ahler
potential $K$, which can be written as
\be
e^{-K} = {i\over 2} \int_X \W \wedge \bar\W
\ee
or in terms of the special coordinates $\f^i$,
\be
e^{-K}=\Im(\bar\f^i\cF_i)=(\f,\bar\f).
\ee
Here we use the  inner product $(\.,\.)$ on the space $W$ defined by
\be
(x,y)=x^i(\Im \tau)_{ij} x^j.
\ee
Note that this inner product has signature $(g,1)$. 
The corresponding hermitian form  can be written as
\be
(x,\ybar)={i\over 2}\int x \wedge \ybar.
\ee
The corresponding K\"ahler (Weil-Peterson or Zamolodchikov) metric 
is given by
\be
G_{i\jbar}=\d_i\d_\jbar K,
\ee
so that for general vectors $x,y\in W$, with $x=x^i\w_i$, we have
\be
G(x,\ybar)=G_{i\jbar}x^i\ybar^\jbar=-{(x,\ybar)\over (\f,\bar\f)}+
{(x,\bar\f)(\f,\ybar)\over (\f,\bar\f)^2}.
\ee
Here we recall that the three-form $\W$ is expressed as $\W=\f^i\w_i$.  Since
$e^{-K}$ gives the metric on the line bundle $\cL$, almost by
definition the corresponding K\"ahler form $\w=\d\dbar K$ has the property 
\be
\w = 2\pi c_1(\cL).
\ee
Hence we have the quantization condition $\w/2\pi \in H^2(\cM,\Z)$.
(The technical term for such a K\"ahler metric that arises as the
Chern class of a line bundle is {\it restricted K\"ahler}.)
 
The K\"ahler metric is by inspection
degenerate in the direction of the vector $\f=\W$,
\be
G(x,\bar\f)=G(\f,\ybar)=0
\ee
In fact, here it is useful to introduce a slightly differently
normalized metric, the so-called $tt^*$-metric (see \S8.12) defined as
\be
g_{i\jbar}=e^{-K} G_{i\jbar}
\ee
In terms of this metric we have
\be
g(x,\ybar)=-(x,\ybar)+{(x,\zbar)(\f,\bar\f)\over (\f,\bar\f)}=-{i\over
2}\int x_\perp \wedge y_\perp
\ee
where $x_\perp$ is the $(2,1)$ part, satisfying $(x,\bar\f)=0$,
of the general 3-form $x\in W=H^{3,0}\oplus H^{2,1}$.
So, when we restrict to $H^{2,1}$, and thereby descend from $\cL$
down to $\cM$, we have
\be
g_{i\jbar}=-\Im\tau_{ij}.
\ee
More precisely, with $P_\perp$ the orthogonal projection on $H^{2,1}$,
\be
g=-P_\perp\. \Im\tau \. P_\perp
\ee
We now see that $g$ (and $G$) give a good metric on the moduli space
$\cM$.

This description of the geometry of the moduli space of complex
structures of a CY manifold in terms of the homogeneous function $\cF$
is known as {\it special geometry} \cite{special-ref}. Special geometry
was first discovered in the context of $N=2$ $D=4$ supergravity
\cite{special-geometry}. It was then shown that CY spaces give a
concrete realization \cite{strominger-special}. This is not a
coincidence, since the CY manifold can be used as a compactification in
string theory. We will see later in \S8.12 how the same structure also
emerges in CFT, or more precisely, families of $N=2$ string vacua. 

It is interesting to see what happens at a singularity in $\cM_X$. If
the Calabi-Yau space develops a node, a particular three-cycle, say
$A^0$, will shrink to zero volume. So the period
\be
\f^0 = \int_{A^0}\W \ra 0
\ee
will go to zero.
Now one can prove that if we make a monodromy at the singular locus
$\f^0=0$, \ie transform $\f^0 \ra e^{2\pi i}\f^0$,
there will be an action on the other cycles given by the
Picard-Lefshetz formula \cite{picard}
\be
C \ra C + \eta(C,A_0)C.
\ee
So the dual cycle $B_0$ will transform as
\be
B_0 \ra B_0 + A_0,
\ee
and therefore we have
\be
\d_0\cF = \int_{B_0} \W \ \ra\ \d_0\cF + \f^0.
\ee
From this we read off immediately that the singularity of $\cF$ around
$\f^0=0$ takes the form
\be
\cF \sim {1\over 4\pi} (\f^0)^2 \log \f^0.
\ee
We will make use of the result later.

\newsection{Perturbative string theory}

From our deliberations on conformal field theories we now move upwards
to the world-sheet formulation of perturbative (bosonic) string theory.
This a subject for a lecture series in its own, and fortunately there
are excellent set of lectures in the literature 
\cite{polch-leshouches,dhoker-phong,luest-theisen,oguri-tasi}
(as there are excellent text books \cite{gsw}). So we will mainly focus
on the more abstract mathematical aspects, that are usually not stressed
so much in an introductory course.

\newsubsection{Axioms for string vacuum}

From an axiomatic point of view a perturbative string vacuum consists of
three ingredients: a representation $\cH$ of the Virasoro algebra (actually
the classical Witt algebra), a BRST differential $Q$, and an anti-ghost
field $G$ that can be used to make volume forms on the moduli space
$\cM_{g,n}$. More precisely, we need (for a mathematical exposition
see \eg \cite{stas})

(1) A (necessarily non-unitary) 
CFT of central charge $c=0$. That is, we have Virasoro generators
$L_n,\Lbar_n$ acting on a Hilbert space $\cH$.
This Hilbert space is graded by ghost/fermion number $F$,
\be
\cH = \bigoplus_{n\in\Z} \cH^{(n)}.
\ee
Quite often we can introduce a bi-grading by
separately conserved left-moving and right-moving ghost charges
$F_L$ and $F_R$ with $F=F_L+F_R$.

(2) A BRST operator
\be
Q : \cH^{(n)}\ra \cH^{(n+1)}
\ee
satisfying $Q^2=0$. This makes the Hilbert space $\cH$ into a complex.

(3) Anti-ghosts $G,\Gbar$ 
that are primary spin two fields with a mode expansion 
\be
G(z) = \sum_k G_k z^{-k-2}
\ee
and similarly for $\Gbar(\zbar)$.
The generators
$G_n,\Gbar_n$ are odd and anticommuting, and satisfy the algebra
\be
\{Q,G_n\} = L_n,\qquad
\{Q,\Gbar_n\} = \Lbar_n,
\ee
\be
[L_n,G_m]=(n-m) G_{n+m}.
\ee

Given these three basic ingredients one defines the Hilbert space of
basic states as
\be
\cH_{basic}=\ker G_0^- \cap \ker L_0^-,
\label{basic}
\ee
with $G_0^\pm=G_0 \pm \Gbar_0$, \etc Here we note the important relation
\be
\{Q,G_0^-\}=L_0^-.
\ee
The space $V$ of physical states is finally defined as
the ``semi-relative'' cohomology
\be
V  = H_Q^*(\cH_{basic}).
\ee

This last definition needs some explaining. Suppose we have a
space $X$ that carries a circle action, say for convenience a free
action, so that we have a $S^1$ principal fiber bundle $\pi:X\ra B$ over
some base space $B$. Let $\xi$ be the vector field that generates the
$S^1$ action. Now suppose we want to compute the cohomology of $B$ in
terms of the cohomology of $X$. There is a general procedure to do this
for arbitrary quotient spaces called equivariant cohomology
\cite{equivariant-cohomology}, but here there is a much more simple
procedure. To this end we consider the following two operators that
act on the space $\W^*(X)$ of
differential forms on $X$: the inner product $\iota_\xi$
with the vector field
$\xi$, and the Lie derivative ${\cal L}_\xi$. These operators
satisfy Cartan's relation
\be
\{d,\iota_\xi\} = {\cal L}_\xi.
\ee
The differential forms on $B$ pulled-back to $X$ can now be
characterized by two properties: they are $S^1$ invariant, and thus
satisfy ${\cal L}_\xi\a=0$, and they have no components in the direction
of the fiber, $\iota_\xi\a=0$. This is the definition of a basic form,
\be
\W^*_{basic}(X) = \ker \iota_\xi \cap \ker {\cal L}_\xi.
\ee
The cohomology of $B$ can now be computed by working with these basic
forms. We see that to make an analogy with string theory, we have the
translation
\be
\W^*(X),\W^*_{basic}(X),H^*(B),d,\iota_\xi,{\cal L}_\xi \ \iff \ 
\cH^*,\cH^*_{basic},V,Q,G_0^-,L_0^-.
\ee
In string theory the relevant circle action is the rotation of the string
as implemented by the world-sheet momentum operator $L_0^-$.

With all this machinery in place, string amplitudes are defined by
choosing physical vectors $\f_1,\ldots, \f_n \in V$ and doing the following
integral over $\cM_{g,n}$
\be
A(\f_1,\ldots,\f_n) = \mathop{\int}_{\cM_{g,n}}
\<\f_1\cdots \f_n \prod_{I=1}^{3g-3+n}
\int_\Sigma \m_I G \int_\Sigma \mubar_I \Gbar \>.
\ee
Here $\m_I$ is a basis of Beltrami differentials spanning
$T_\Sigma\cM_{g,n}$. Since $G(z)$ is a section of $K^2$, the
combination $\m_I G$ is a $(1,1)$ form on the surface $\Sigma$
and so the integral makes sense. 
In particular we have the following form of the genus $g$ vacuum
amplitude
\be
F_g = \mathop{\int}_{\cM_{g,n}}
\<\prod_{I=1}^{3g-3}
\int_\Sigma \m_I G \int_\Sigma \mubar_I \Gbar \>.
\ee
By some formal manipulations this expression can be simplified for the
case of the one-loop amplitude, giving \cite{bcov1}
\be
F_1 = {1\over 2} \int_{\cF} {d^2\t \over \Im \t}
{\rm Tr}\left(F_LF_R (-1)^F q^{L_0} \qbar^{\Lbar_0}
\right).
\ee
(The factor $\half$ is because of the $\Z_2$ symmetry $z\ra -z$
that any elliptic curve allows.)

\newsubsection{Intermezzo --- twisting and supersymmetry}

Before we continue our discussion of string theory it is
useful to review briefly the concept of twisted supersymmetries and
non-local operators, since these concepts play an important role in the
following and are quite generally applicable, also in the
four-dimensional context in later lectures.

Supersymmetry transformations $Q$ are fermionic operations that square
into the translation operator. The usual supersymmetry algebra has the
form
\be
\{Q_\a,Q_\b\}=\g^\m_{\a\b}P_\m.
\ee
Here the supercharges $Q_\a$ are spinors on the space-time manifold,
and $P_\m$ is the momentum operator that generates translations. Now
generically there are no covariant constant spinors on a curved
manifold, so it is difficult to find manifolds that allow global
supersymmetries. (Just as it is generally speaking
not true that a manifold allows
isometries.) In fact, to find a global
supersymmetry that always exists, independent of the space-time
topology, one needs a {\it scalar} supersymmetry.
(There is always a covariant constant function, the constant
function.) That is, we are looking for a slightly
different supersymmetry algebra of the form
\be
\{Q,G_{\m}\}=P_\m,
\label{twist}
\ee
with $Q$ a scalar and $G_\m$ a vector, both odd and satisfying
\be
Q^2=0,\qquad \{G_\m,G_{\n}\}=0.
\ee
Since the ``BRST operator''
$Q$ squares to zero, for any QFT with such an operator
we can define in general the cohomology group
\be
V = H_Q(\cH)
\ee
of ``physical'' states annihilated by $Q$, modulo $Q$ exact states.
If $Q$ symmetry is unbroken, \ie if the vacuum satisfies $Q|vac\>=0$,
all expectation values of $Q$ commutators vanish, since
\be
\<[Q,\cO]\>=0.
\ee
(We write $\<\cdots\>=\<vac|\cdots|vac\>$.)
This implies that if a set of local
operators $\cO_1,\ldots,\cO_n$ all satisfy
\be
[Q,\cO_i(x)]=0,
\ee
their correlation function
\be
\<\cO_1(x_1)\cdots \cO_n(x_n)\>
\ee
is  constant. This is proven algebraically by observing that
\be
\Pp\cO {x^\m} =[P_\m,\cO]=[Q,\cO_\m^{(1)}],
\ee
where we define the one-form
\be
\cO_\m^{(1)}=[G_\m,\cO].
\ee
We thus find that\footnote{We write in the following $[\.,\.]$
for the graded (anti)commutator in the following.}
\be
\pp {x^\m} \<\cO_1(x_1)\cdots \cO_n(x_n)\>=
\<[Q,\cO_{1,\m}(x_1)\cdots \cO_n(x_n)]\>=0.
\ee

All this can be done in a more systematic way as follows
\cite{witten-tft}. In fact, regarding the $Q$ symmetry as a spin zero
supersymmetry is a very fruitful analogy. It is convenient to go to a
``superspace'' formulation of the theory where, in addition to the
space-time coordinates $x^{\mu}$ we have additional Grassmannian
coordinates $\theta ^{\mu}$, just as when we considered supersymmetric
quantum mechanics. Starting from a physical field $\cO(x)$, that we will
now denote as $\cO^{(0)}(x)$ to indicate that it is a zero-form, the
superfield $\cO(x,\theta)$ is defined as
\ba
\cO(x,\theta) \is e^{\theta^\mu G_\mu}  \cO(x)\nonu
\is \sum_k \cO^{(k)}_{\mu_1 \ldots \mu_k}(x)\theta^{\mu_1}
\cdots \theta^{\mu_k}.
\label{superfield}
\ea
Here the fields $\cO^{(k)}$ are generated from $\cO^{(0)}$ by repeated
application of $G_{\mu}$
\be
\cO^{(k)}_{\mu_1 \ldots \mu_k}(x) = [ G_{\mu_1},[ G_{\mu_2},\ldots
, [G_{\mu_k},\cO^{(0)}(x)] \ldots ] ].
\label{g}
\ee   
Since $\cO^{(k)}_{\mu_1 \ldots \mu_k}$ is antisymmetric in all its indices, it
represents a $k$-form, and we can write this relation as
\be
[G,\cO^{(k)}]= \cO^{(k+1)},\qquad G=G_\m dx^\m.
\ee
Since we have $[Q,G]=d$, these differential forms satisfy 
the important {\it descent equation} \cite{witten-tft}
\be
d \cO^{(k)} = [Q,\cO^{(k+1)}]. 
\label{descent}
\ee
We can draw two conclusions from this equation. First, it suggests a new class 
of {\it non-local} physical observables. If $C$ is a $k$-dimensional
closed submanifold, the descent
equation shows that
\be 
\cO_C = \int_{C} \cO^{(k)}(x)
\ee 
is a physical observable, since
\be
[Q,\cO_C] = \int_{C} d\cO^{(k-1)} 
= \int_{\partial C}  \cO^{(k-1)} = 0.
\ee 
Secondly, if $C$ and $C'$ represent the same class in $H_k(M)$, we have
$C-C'=\d B$ and
\be
\cO_C-\cO_{C'} =
\int_{B} d\cO^{(k)}= [Q,\int_{S} \cO^{(k+1)}].
\ee
So the physical observable depends only on the homology class of $C$.
That is, for each class in $H_k(M)$ and each element in $V=H_Q^*(\cH)$
we can construct a non-local operator
\be
C \in H_k(M) \gives \cO_C = \int_{C} \cO^{(k)}(x)
\ee

Why is all this relevant for string theory? As we have stressed, string
amplitudes are obtained by integrating over the moduli space. In
particular we have to integrate over the positions of the vertex
operators that create a certain string state in the push-forward
$\cM_{g,n} \ra \cM_{g,n-1}$. These vertex operators have to be $(1,1)$
forms on the Riemann surface. These volume forms are obtained through
the above procedure.

More precisely, we first observe that the translation operators are given by
\be
\pp z = L_{-1},\qquad \pp \zbar = \Lbar_{-1}.
\ee
Part of the algebra of the anti-ghosts tells us that
\be
\{Q,G_{-1}\}=L_{-1},\qquad
\{Q,\Gbar_{-1}\}=\Lbar_{-1}.
\ee
Therefore our scalar supersymmetry (\ref{twist}) is realized on the
world-sheet.

Suppose we now pick a physical state, that is a cohomology class
of $Q$. Such a state can be assumed to have conformal weight zero.
If not, we use the relation
\be
\{Q,G_0\}=L_0
\ee
to prove that it is $Q$ exact. Out of such a operator $\f$ of weight zero
we can now make the associated two-form as
\be
\f^{(2)}=[G_{-1},[\Gbar_{-1},\f]].
\ee
The integrated vertex operators are now given by the non-local
$Q$ closed expressions
\be
\f_\Sigma = \int_\Sigma \f^{(2)}.
\ee

\newsubsection{Example --- The critical bosonic string}

We now return to string theory by giving some simple examples.

In the bosonic string (see \eg \cite{fms}) 
the axioms are satisfied by starting with
26 bosonic free fields $x^\m$. These fields have stress-tensor
\be
T = -\half (\d x^\m)^2,
\ee 
which generates a $c=26$ representation of the Virasoro algebra. To this
we add the ghosts of the bosonic string, a $b,c$ system as described in
\S6.4 with $\l=2$. So $b(z)$ has spin two and $c(z)$ has spin $-1$ and the
central charge equals $c=-26$. The full Hilbert space is now given by
\be
\cH = \cH^{26\ bosons} \otimes \cH^{ghosts}.
\ee
It is graded by the ghost charge $F = - \oint bc$
 and carries the action of
the BRST charge
\be
Q = \oint \left( -\half c (\d x)^2 + c\d c b\right)
\ee
The anti-ghost is given by the $b$ field (which gives its name to $G$)
\be
G(z) = b(z),
\ee
which indeed has the defining property
\be
\{ Q,b(z)\} = T(z).
\ee
Physical fields typically take the form
\be
\f = c\cbar V,
\ee
with $V$ a primary field in the matter sector of conformal dimension
$(1,1)$. The most important fields are the states of the form
\be
V^{\m\n}_p = \d x^\m \dbar x^\n e^{ipx}\iff \a_{-1}^\m \bar\a_{-1}^\n
|p\>,
\ee 
which represent the Fourier modes
of the space-time fields $G_{\m\n}(x)$ and $B_{\m\n}(x)$.

\newsubsection{Example --- Twisted $N=2$ SCFT}

Another important example is the twisting of the $N=2$ superconformal algebra
in two dimensions, say with central charge $c=3d$. 
In that case we have the following currents: the
stress-energy tensor $T$ of spin 2, two supercurrents $G^\pm$ of spin
$3/2$, and a $U(1)$ current of spin 1. In these algebra states and fields are 
characterized by their conformal dimension $h$ and charges $q$
\be
L_0 |\phi\>= h |\phi\>,\qquad 
J_0 |\phi\> = q |\phi\>.
\ee
where $L_0$ and $J_0$ are the zero-modes of $T$ and $J$. 
Twisting amounts to redefining
the stress tensor as
\be
T \ra T + \half \d J,
\ee
which gives in particular $L_0 \ra L_0 + \half J_0$ and accordingly adds the
charge to the conformal dimensions
\be
h \ra h + \half q.
\ee
Since the supercurrent $G^\pm$ has $h=3/2$ and $q=\pm 1$, this gives
field $G=Q^+$ and $Q=Q^-$ of spin 2 and 1.
The modes $L_n,G_n,Q_n,J_n$ of
the four currents form a closed algebra, which is simple the $N=2$ superconformal
algebra written in a different basis
\begin{eqnarray}
\label{TCFTalg}
\lefteqn{
\begin{array}{ll} 
 [L_m,L_n]\; =\; (m-n)L_{m+n},\ \ \ \ & \ \ \ \ \
 [J_m,J_n]\; =\; d\cdot m \,\delta_{n+m,0},\\[4mm]
[L_m,G_n]\; =\; (m-n)G_{m+n},\ \ \ \ \ & \ \ \ \ \
 [J_m,G_n]\; =\; -G_{m+n},\\[4mm] 
[L_m,Q_n]\; =\; -n\,Q_{m+n},\ \ \ \ & \ \ \ \ \
 [J_m,Q_n]\; =\; Q_{m+n},
\end{array}}
\nonumber\\[4mm]
& \begin{array}{ll}
\ \ \{G_m,Q_n\} & \!\!\!=\; L_{m+n}+nJ_{m+n}+
{\textstyle{1\over 2}}d\cdot m(m\! +\! 1) \delta_{n+m,0}, \\[4mm]
\ \ \, [L_m,J_n]\, & \!\!\!=\; -n J_{m+n}
-{\textstyle {1\over 2}}d\cdot m(m\! +\! 1)\delta_{m+n,0}.
\end{array} &
\end{eqnarray}
So all the conditions of a string vacuum (and more) are satisfied.

We see in particular that out of this twisted $N=2$ we can produce a
two-dimensional topological field theory with Hilbert space
$V=H_Q(\cH)$. The ring $V$ is in that case known as the chiral ring
\cite{chiral-ring}.

\newsubsection{Example --- twisted minimal model}

As the simplest example of the twisting procedure
we consider one free boson $x$ with a background
charge $Q=-1/\sqrt 3$ so that the central charge vanishes $c=0$. This
is the twisted $k=1$ $N=2$ minimal model.  Now the
vertex operators
\be
V_p = e^{ipy/\sqrt 3}
\ee
have conformal dimension $h$ and ``ghost charge'' $q$ given by
\be
h = \half p(p-1),\qquad
q= \third p.
\ee
The BRST charge and the anti-ghost are defined as
\be
Q = \oint  V_3(z),\qquad G(z) = V_{-3}(z).
\ee
This string theory has two physical states
\be
\f_0=\id,\qquad \f_1=V_1,
\ee
with $(h,q)=(0,0)$ and $(0,\third)$. The algebra of these fields is
trivial
\be
\f_1\.\f_1=0.
\ee
It is a nice exercise to compute
the corresponding two-forms
\be
\f_0^{(2)}=0,\qquad  \f_1^{(2)} = V_{-2}.
\ee
The only non-zero amplitude in this toy model is the
four-point function
\be
A(\f_1,\f_1,\f_1,\f_1) =
\int \! d^2\!z\, |(z-1)^{2\over3}z^{1\over3}|^2
\ee
which is some finite ratio of $\G$-functions.

\newsubsection{Example --- topological string}

In its simplest form the topological string is obtained by twisting a
$N=2$ sigma model on a Calabi-Yau $X$. (See \cite{trieste} for an
extensive review of topological strings and their role in non-critical
string theory and two-dimensional quantum gravity, and see
\cite{cargese} for their importance for intersection theory on the
moduli space of Riemann surfaces.)

For the simple case $X=\C$ (or $T^2$), the topological string is given
by two free scalar fields $x^1,x^2$ that can be combined in complex
fields $x=x_1+ix_2$ and $\xbar = x^1-ix^2$. The central charge $c=2$ of
this system is matched by a $\l=1$ $(b,c)$ system with central charge
$c=-2$. The total stress-tensor is given as
\be
T = -\d x\d\xbar - b\d c
\ee
The BRST transformation rules are then taken as
\ba
Qx \is c\nonu
Q\xbar \is 0 \nonu
Qc \is 0 \nonu
Qb \is \d x
\ea
The anti-ghost is $G  = b\d x$. Typical physical fields are
$e^{ip\xbar}$, $c e^{ip\xbar}.$

\newsubsection{Functorial definition}

After all these concrete examples, we return into the clouds and
 our abstract discussion of perturbative string theory.
In order to give a more functorial definition of a string vacuum, using
the language of categories, we have to make two generalizations beyond CFT
\cite{segal-string,getzler}

(1) The Hilbert space $\cH$ is a complex, with differential $Q$. Its
cohomology we will denote again as $V$.

(2)  The amplitudes $\F_\Sigma$ are {\it differential forms} on the extended
moduli space. So, we have maps
\be
\F_\Sigma : \cH^{\otimes n} \ra \W^*(\cP_{g,n})
\ee

This structure is sometimes also called a {\it cohomological field theory}
\cite{witten-ctft,kontsevich-manin}.

If one wishes, one can introduce two new categories, the category ${\bf
Comp}$ of Hilbert space complexes and equivariant maps, and the category
${\bf TRiem}$ of so-called topological Riemann surfaces \cite{trieste}.
The definition of these objects is such that their moduli space is given
by the $3g-3|3g-3$ dimensional superspace
\be
\widehat{\cM}_{g,n} = \Pi T\cM_{g,n}
\ee
so that functions on the moduli space become differential forms on
$\cM_{g,n}$.
With all this in place, a perturbative string vacuum is a functor
\be
\F : {\bf TRiem} \ra {\bf Comp}.
\ee
We see that we have made the progression
$$
TFT \gives CFT \gives Strings 
$$
and at the same time our amplitudes or morphisms became objects in
bigger and bigger spaces
$$
H^0(\cP_{g,n}) \gives \W^0(\cP_{g,n}) \gives \W^*(\cP_{g,n}).
$$

In string perturbation theory we 
have two differentials: the supercharge or BRST operator $Q$ that
acts on the Hilbert space $\cH$ and the exterior differential $d$ on the moduli space.
Perturbative string theory is now abstractly defined by the 
extra relation
\be
\left(d+Q\right)\F_\Sigma=0.
\ee
The relation with the anti-ghost field
\be
G(z) = \sum_n G_n z^{-n-2}
\ee
is as follows. The amplitudes are differential forms on moduli space with 
a degree determined by the total ghost charge. If we
have a form of degree $p$, we can pick Beltrami differentials
$\m_1,\ldots,\m_p \in T_\Sigma \cP_{g,n} \cong H^1(T_\Sigma)$ and find the
equivalence
\ba
& & \<\prod_i \f_1(P_i) \prod_{I=1}^p \left(\int_\Sigma \m_I G \int_\Sigma 
\bar\m_I \bar G \right) \>\nonu
& & \quad \qquad \qquad 
= \<\F_\Sigma(\f_1,\ldots,\f_n),\bigwedge_I 
\left(\m_I \wedge \bar\m_I \right)\>.
\ea
If we now choose $\f_i \in V = H^*_Q(\cH_0)$ then the condition
\be
d\F_\Sigma(\f_1,\ldots,\f_n)=0
\ee
tells us that $\F_\Sigma$ actually is a map
\be
\F_\Sigma: V^{\otimes n} \ra H^*(\Mbar_{g,n}).
\ee

Here we use that the fibers in the projection $\cP_{g,n} \ra \cM_{g,n}$
do not contribute to the $Q$ cohomology \cite{getzler}. This is
guaranteed by the restriction to basic states, as in (\ref{basic}).
Indeed, the only non-trivial part in the infinite-dimensional fiber of
the projection is the circle action obtained by rotating the local
coordinate $z\ra e^{i\th} z$. This rotation is implemented by the
operator $L_0^-$. This explains the analogy with computing the
cohomology of the base manifold of a circle bundle that we mentioned in
\S8.1.

The final $g$ loop
string amplitudes, that computes scattering of string states, are then
expressed by a further integral over $\cM_{g,n}$ that picks out the top
component
\be
A(\f_1,\ldots,\f_n)=\int_{\cM_{g,n}}
\F_\Sigma(\f_1,\ldots,\f_n).
\ee

\newsubsection{Tree-level amplitudes}

This structure becomes more manageable in genus zero.
As a simple example consider the sphere with three holes. Since we have
seen that $\cM_{0,3}=pt$, the three point function is simply given by
a complex number
\be
\<\f_1(z_1)\f_2(z_2)\f_3(z_3)\>_0 = \F_{0,3}(\f_1,\f_2,\f_2) \in \C.
\ee
If we pick a basis $\f_i \in V$ this gives us the structure coefficients
of a TFT
\be
c_{ijk} = \F_{0,3}(\f_i,\f_j,\f_k) .
\ee
So we see that we actually went through a little loop,
\be
TFT \gives CFT \gives String \gives TFT.
\ee
For the more general case of an $n$-point function in genus zero, we
obtain a cohomology class on $\bar\cM_{0,n}$. In string theory we are
particularly interested in the forms of top degree. We will define the
$n$-point genus zero string amplitude as
\ba
c_{i_1\cdots i_n} \is \int_{\cM_{0,n}} \F_{0,n}(\f_{i_1},\ldots,\f_{i_n}) \nonu
\is \< \f_{i_1}(0) \f_{i_2}(1)\f_{i_3}(\infty)  \prod_{k=4}^n
\int \f_{i_k}^{(2)} \>
\ea
with $\f^{(2)}=[G_{-1},[\Gbar_{-1},\f]]$, a $(1,1)$ form on the surface.

As written above this amplitude has not the manifest permutation 
symmetry of the indices $i_1,\ldots,i_n$. That this symmetry is still
present we can show by considering again the automorphisms of $\P^1$.
To this end we recall that the non-local operators can be
combined into one superfield (\ref{superfield}) on the
topological Riemann sphere $\Pi T\P^1$,
\be
\f(z,\zbar,\theta,\thetabar) = \phi^{(0)} + \phi^{(1,0)}\theta
+ \phi^{(0,1)}\thetabar + \phi^{(2)}\theta\, \thetabar.
\ee
Here we suppressed the $(z,\zbar)$ dependence on the RHS. 
Let us now consider a
correlation function on the sphere of the form
\be
\Bl \prod_{i=1}^n \int \f_i(z,\zbar,\theta,\thetabar)
\Br\subzero.
\ee
This expression is not well-defined, since it is invariant under a
fermionic extension of the $PGL(2,\C)$ symmetry, generated by operators
$L_0$, $L_1$, $L_{-1}$ and $G_0$, $G_1$, $G_{-1}$ and their complex
conjugates. We have to factor out the infinite volume of this group in
order to obtain a finite answer. These symmetries correspond to the
super-M\"obius transformations
\be
z \rightarrow {az+b\over cz+d},\qquad \theta \rightarrow 
{\theta + \alpha z^2+\beta z+\gamma\over (cz+d)^2}.
\ee
This extended $PGL(2,C)$ symmetry can be used to fix three of the
$z$-coordinates, say $z_1,z_2,z_3$, at $0$, $1$, and $\infty$, and put
three of the $\theta$-coordinates to zero. We choose these
anti-commuting coordinates to be $\theta_1,\theta_2,\theta_3$. If we
recall that
\be
\int d^2\theta\cdot\f = \phi^{(2)},\qquad
\left. \f \right|_{\strut \theta=0} = \phi^{(0)},
\ee
then we see that after gauge fixing we are left with a correlation
function of the form
\be
\label{Widentity}
\Bl\phi^{(0)}_{i_1}\phi^{(0)}_{i_2}\phi^{(0)}_{i_3} \int\!
\phi^{(2)}_{i_4}\ldots \int\phi^{(2)}_{i_s}\Br\subzero.
\ee
Since we started from an expression that was explicitly symmetric in
{\it all} indices $i_1,\ldots,i_s$, this correlator also has this
permutational symmetry. That is, it does not matter which three
operators we represent as zero-forms. The generalized $PGL(2,\C)$
invariance tells us that we can interchange a zero and a two-form. 

\newsubsection{Families of string vacua}
 
Out of the generating function of $n$-point functions
we can construct a family of TFTs labeled by coordinates $t\in V^*$
by the definition
\be
c_{ijk}(t)= \< \phi_i\phi_j\phi_k\exp \int t^k\phi_k^{(2)}\>.
\ee
Physically one can think of deforming the two-dimensional action by
\be
\delta S = \int t^n \f_n^{(2)}
\ee
We claim that this defines a family of multiplications on the vector space 
$V$. Note that in this way we derive the higher $n$-point functions by taking derivatives
at $t=0$.
In terms of the coefficients $c_{ijk}(t)$ the permutation symmetry of
(\ref{Widentity}) gives the important integrability condition
\be
\d_i c_{jkl}  =c_{ijkl}=\d_j c_{ikl}.
\label{4pt}
\ee
We stress that this is an additional condition imposed on the family of
algebras $c_{ijk}(t)$, and a consequence of the topological invariance.
We can integrate this relation three times, at least locally, to find
that the three-point functions are actually the third derivatives of a
function $\cF(t)$, the so-called free energy or {\it prepotential}
\be 
\label{dddf} 
c_{ijk}(t) = \d_i\d_j\d_k \cF(t) .
\ee 
Symbolically, $\cF(t)$ is defined as 
\be 
\cF(t) = \< \exp \int t_i\f_i^{(2)}\>.
\label{free} 
\ee 

For example, comparing with our discussion of quantum cohomology, 
formula (\ref{count}) in \S4.3, we see
that in the case of a sigma model on a Calabi-Yau three-fold
we have the following expression for $\cF$
\be
\cF(t) = \int_X {t^3\over 3!}  + {\chi\over 2} \zeta(3)
+ \sum_{rational\ C} Li_3\left(e^{2\pi t\.C}\right).
\ee
with $t\in H^{1,1}$, $Li_3(x)=\sum_{k>0} x^k/k^3$, $\zeta(3)=Li_3(1)$,
and $\chi$ the Euler number of $X$. The sum is over all rational
curves $C\cong \P^1$ in $X$. (The constant term is
determined by mirror symmetry, or by a degeneration argument.)

As a corollary to the result (\ref{4pt}), consider the special identity
operator $\phi_0 = {\bf 1}$. There is no corresponding two-form, since $G$
commutes with the identity. So the coupling coefficient $t_0$ does not
exist, there is no modulus associated to ${\bf 1}\in V$. This fact, combined
with integrability relation (\ref{4pt}), shows that 
\be 
0 = \d_0 c_{ijk} =\d_i c_{0jk}=\d_i\eta_{jk}.
\ee 
So we have shown that the metric or two-point function $\eta$ is
independent of the deformation parameters $t_i$.

We have claimed that also for $t\neq 0$ the coefficients $c_{ijk}(t)$
define a Frobenius algebra. That is, the generating function $\cF(t)$
satisfies the so-called WDVV equation
\cite{witten-top,top-string,trieste,dubrovin,kontsevich-manin}
\be
\d_i\d_j\d_m\cF \eta^{mn} \d_n\d_k\d_l\cF
=\d_i\d_k\d_m\cF \eta^{mn} \d_n\d_j\d_l\cF.
\ee
This gives an infinite set of relations on the expansion
coefficients $c_{i_1\ldots i_k}$ of $\cF$.

There is a nice mathematical way to prove this last equation starting
from Segal's axioms \cite{kontsevich-manin}. Keel \cite{keel}
has shown that the cohomology of $\Mbar_{0,n}$
is generated by divisors (codimension one subvarieties) $D_S$ defined as
follows. Pick a subset $S\subset \{1,\ldots,n\}$ with $2 \leq \#S\leq
n-2$. $D_S$ is now defined as the set of all singular rational curves
with a single node, such that the punctures $P_i$, $i\in S$ lie all on
one component, whereas the punctures $P_j$, $j\not\in S$, all lie on the
second component:
$$
\insertfig{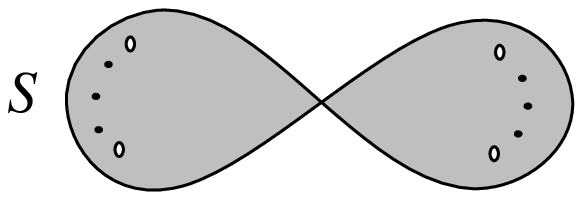}{4.5}
$$
Clearly these divisors lie in the compactification divisor, the
``boundary'' of $\cM_{0,n}$, since they represented singular curves. Now
Keel has shown that these divisors generate all cohomology (as a ring)
with a single constraint that is best expressed by a picture
\be
\sum_{S;\ i,j\in S,\ k,l \not\in S} \insertfig{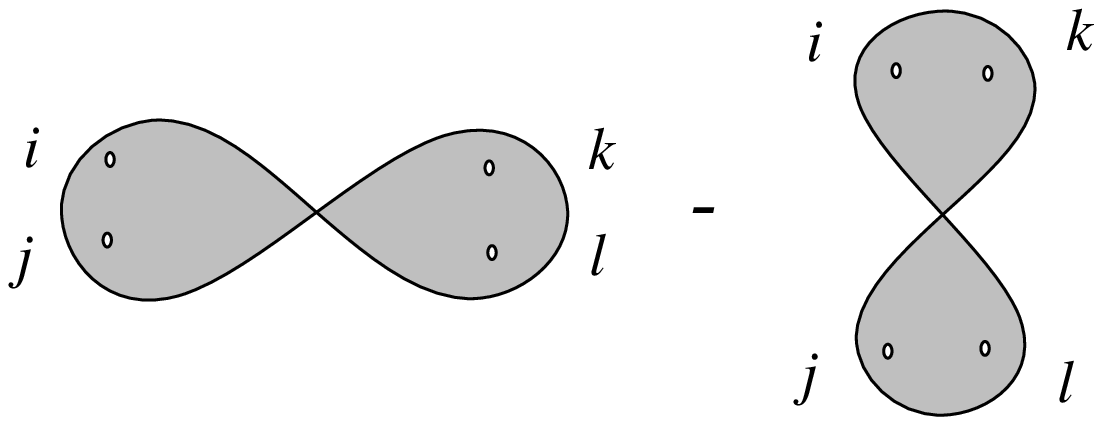}{6}=0
\ee
This directly implies the associativity condition on $c_{ijk}$.

\newsubsection{The Gauss-Manin connection}

Let us quickly summarize the most important ingredients of the world-sheet
discussion of string vacua and the associated topological field theory.
It consists of a moduli space $\cM$ of string vacua with over it
a vector bundle $V$ carrying a bilinear form $\eta$ together with an
associative multiplication on each fiber. In a concrete basis $\f_i\in V$ 
this multiplication is given in terms 
of the structure coefficients 
\be
c_{ijk}(t) = \<\f_i\f_j\f_k\>_0
\ee
Here $t^i$ are the (special) coordinates on the moduli space $\cM$ of TFTs.

More precisely, there are five important conditions that we can distinguish.

(1) The parameters $t^i$ are coordinates on $\cM$, \ie the vector
fields $D_i=\d/\d t_i$ commute
\be
[D_i,D_j]=0.
\ee
Moreover, the vector field $D_0$, corresponding to deformations by the identity
operator vanishes.

(2) The coefficients $c_{ijk}$ satisfy the integrability condition
\be
D_ic_{jkl}=D_jc_{ikl},
\ee
(symmetry of the 4-point function) that allows us to write them locally as
\be
c_{ijk}=D_iD_jD_k\cF,
\ee
in terms of the prepotential $\cF(t)$. 

(3) The algebra is associative for all values of the moduli,
\be
c_{ij}{}^n c_{nkl} = c_{ik}{}^n c_{njl}.
\ee

(4) The bilinear form $\eta_{ij}=c_{0ij}$ is conserved,
\be
D_i\eta_{jk}=0,
\ee
which is equivalent to the statement that $D_0c_{ijk}=0$ using (2).

(5) The algebra is compatible with $\eta$
\be
c_{ijk}\eta^{jk}=c_{ikj}\eta^{kj},
\ee
which is equivalent to the complete symmetry of $c_{ijk}$.

A more succinct description of these conditions uses the matrices
$C_i$ defined by the natural map $c:V \ra \End(V)$, with matrix elements
\be
\left(C_i\right)_j{}^k=c_{ij}{}^k=c_{ijl}\eta^{lk}.
\ee
With these matrices  the integrability condition reads
\be
[C_i,D_j]=[C_j,D_i],
\ee
whereas the associativity reads
\be
[C_i,C_j] = 0.
\ee
In fact, if we define the ``Gauss-Manin connection'' \cite{dubrovin}
\be
\nabla_i=D_i -\l C_i,
\ee
with spectral parameter $\l$, then the relations (1),(2) and (3) can be
summarized in the statement that 
\be
[\nabla_i,\nabla_j]=0,
\ee
for all values of $\l$. The compatibility of $\nabla$ with $\eta$
\be
\eta(\nabla\a,\b)=\eta(\a,\nabla\b)
\ee
gives the relations (4) $D_0\eta=0$, and (5) $\eta(C_i\a,\b) i=
\eta(\a,C_i\b)$. Why such a connection should exist will become clear in
the later lectures when we consider the integral structure behind all this.

\newsubsection{Anti-holomorphic dependence and special geometry}

Up to now we only discussed the holomorphic coordinates $t^i$, which are
analytical coordinates on the moduli space $\cM$ of string vacua. In
order to discuss the full geometry of $\cM$ we need also the dependence
on the complex-conjugate variables $t^\ibar$. This geometrical structure
is called special geometry, and we have seen an example when we
discussed the moduli space of Calabi-Yau three-folds in \S7.7.
The structure has been worked in the general context of twisted $N=2$
models by Bershadsky, Cecotti, Ooguri and Vafa. These matters are
explained in full detail in the beautiful papers \cite{tt*,bcov1,bcov2}, 
so we only summarize the most important results without proofs. This special
geometry of moduli space of string vacua is only present for string vacua that give rise to space-time
theories with $N=2$ supersymmetry.

We have seen that the vector space $V$ had a natural bilinear form (over
$\C$)
\be
\eta_{ij}=\eta(\f_i,\f_j).
\ee
However, in these models there is also an hermitian form
\be
g_{i\jbar} = g(\f_i,\f_\jbar).
\ee
Here the complex conjugate fields $\f_\ibar$ are defined in terms of
some anti-linear map
\be
\f_\ibar= M_\ibar{}^i\f_i,
\ee
satisfying $MM^*=-1$. 

In our deformation problem we now have to allow 
for anti-holomorphic dependence on the variables $t^\ibar$. The 
corresponding deformations are described by local operators
\be
\delta S = \int_\Sigma t^\ibar \f^{(2)}_\ibar  ,
\ee
with
\be
\f^{(2)}_\ibar = [Q_{-1},\{\Qbar_{-1},\f_\ibar\}].
\ee
Since this field it explicitly $Q$-exact, its insertions will lead
to total derivatives on $\cM_{g,n}$, which can only give contributions
at the compactification divisor. Indeed, naively one would have ignored
these boundary terms and would have concluded that all amplitudes
are holomorphic. If one is less naively, one has to investigate in a
precise way the behavior
on singular curves, and in this way one discovers that certain
objects do obtain an antiholomorphic dependence. 
This has been done in \cite{tt*,bcov1,bcov2}, and we now give
the results.

The hermitian form $g$ leads to a K\"ahler metric on the moduli space
$\cM$. In fact, the canonical metric is the the so-called
Zamolodchikov \cite{zam} or Weil-Peterson metric
\be
G_{i\jbar} = e^K g_{i\jbar}
\ee
with
\be
e^{-K} = g(\f_0,\f_{\bar0}) = \<0|0\>.
\label{K}
\ee
(Here $\f_0$ denotes the identity operator.)
It is normalized so that  $G_{0\bar0}=1$. This metric appears in the two-point
functions
\be
\<\f_i(z) \f_\jbar(w) \> = {G_{i\jbar} \over |z-w|^2}.
\ee
It is a K\"ahler metric 
\be
G_{i\jbar} =\d_i\d_\jbar K,
\ee
with K\"ahler form $K$ given in (\ref{K}).
Apart from the derivative $D_i$ and the operator product matrices $C_i$,
we now have conjugate objects $D_\ibar$ and $C_\ibar$. These
satisfy the relations
\be
[D_\ibar,C_j]=0,\qquad [D_i,D_\jbar=0].
\ee
These relations express the fact that the three-point function 
$c_{ijk}$ is a holomorphic object
\be
D_\ibar c_{ijk}=0
\ee
However there is an additional important relation that tells us
how the curvature of the K\"ahler metric is related to the associative
algebra, the so-called $tt^*$ relation \cite{tt*},
\be
[D_i,D_\ibar] + [C_i,C_\ibar] = 0
\ee
In terms of the Zamolodchikov metric $G_{i\jbar}$ this gives the 
so-called special geometry or special K\"ahler relation
\be
R_{i\jbar k}{}^l = G_{k\jbar}\delta_i{}^l
+ G_{i\jbar}\delta_k{}^l - e^{2K} c_{ikn} G^{n\bar n} c_{\jbar \bar
m\bar n} G^{l\bar m}
\ee

Let us note that relations on $D_i,D_\ibar,C_i,C_\ibar$ can be
summarized in terms of the Gauss-Manin connection as
\be
[\nabla_i,\nabla_j]=[\nabla_\ibar,\nabla_\jbar]=
[\nabla_i,\nabla_\jbar]=0
\ee
So apparently there is a natural flat connection on the moduli space
of string vacua, that preserves the bilinear form $\eta$. The reasons
why this connection exist will become apparent when we take a space-time
point of view. There we will uncover an integer structure on $V$, related
to electric and magnetic charges in the four-dimensional space-time.

The metric $G$ satisfies a second important quantization condition: the
associated K\"ahler form of type $(1,1)$,
\be
\w = {i\over 2} G_{i\jbar} dz^i \wedge dz^\jbar,
\ee
(which is by definition closed, $d\w=0$) has integer periods
\be
\w/2\pi \in H^2(\cM,\Z)
\ee
In the language of geometric quantization, this means that the phase
space $\cM$ can be quantized. There exists a line bundle $\cL$
over $\cM$ with first Chern class
\be
c_1(\cL)= \w/2\pi.
\ee
The K\"ahler form can thus be realized as the curvature
of the $U(1)$ connection on $\cL$.

\newsubsection{Local special geometry}

The above structure we also met in \S7.8, when we discussed the moduli
of Calabi-Yau manifolds, and is called {\it local special geometry} or
{\it special K\"ahler geometry} \cite{special-ref}. The formal
definition is as follows:

{\bf Definition:} a {\it local special geometry} $(\cM,V,\W)$ is 
defined by the following ingredients:

(1) A complex, K\"ahler manifold $\cM$, the moduli space, of 
complex dimension
\be
\dim \cM=g.
\ee

(2) A holomorphic, flat $Sp(2g+2,\Z)$ vector bundle $V\ra \cM$. We
denote the symplectic form as $\eta$. In the case of the Calabi-Yau
moduli space, this was the canonical bundle with fiber $H_3(X,\Z)$ and
$\eta$ was given by the intersection form on these three-cycles. The
flat connection is written as $\nabla,\bar\nabla$.  It satisfies
$d\eta(\a,\b)=\eta(\nabla\a,\b) + \eta(\a,\nabla\b)$.

(3) A holomorphic section $\W$ of $V$, 
satisfying the conditions
\be
\eta(\W, \nabla\W)=0,\qquad
\eta(\W,\overline\W)>0
\ee

(4) the K\"ahler form on the moduli space is given by
\be
e^{-K} =\eta(\W,\overline\W)
\ee

To establish the correspondence between the description of the
Calabi-Yau moduli space in \S7.8 and the deformations of $N=2$ string
vacua in the previous subsection, note that
the holomorphic section $\W$ determines a line bundle $\cL\ra \cM$. In
fact, in the CY case we can choose a polarization
\be
V = V^{3,0} \oplus  V^{2,1} \oplus V^{1,2} \oplus V^{0,3} 
\ee
where $\bar V^{p,q}=V^{q,p}$ and $V^{3,0}\cong \cL$ is the line bundle
generated by $\W$ and $W=V^{3,0} \oplus V^{2,1}$ is the subspace
generated by $\nabla\W$. Note that if $\a\in V^{p,q}$ and $\b\in
V^{p',q'}$ then the polarization satisfies the condition that
\be
\eta(\a,\b)\neq 0 \gives p+p'=q+q'=3
\ee
The description in terms of a local coordinate $\f^i$ follows exactly
the discussion in \S7.8. First we choose a (local) canonical integer
basis $\a_i,\b^i$ of $V$ using the flat connection $\nabla$. We then
expand the section $\W$ in terms of this basis in the familiar
form\footnote{Here, in order to make contact with \S7.8, we indicate
the special local coordinates as $\f^i$ instead of $t^i$.}
\be
\W = \f^i\a_i + \cF_i\b^i.
\ee
Condition (3) now leads to the integrability relation
\be
\cF_i = {\d \cF \over \d \f^i},
\ee
which defines quite generally the prepotential $\cF$ as a section of
$\cL^{\otimes 2}$. Finally, the all important three-point functions
$c_{ijk}$ are derived as
\be
c_{ijk} = -\eta(\W,\nabla_i\nabla_j\nabla_k\W)
\ee
or in local coordinates,
\be
c_{ijk} = {\d^3 \cF \over \d\f^i \d\f^j \d\f^k}
\ee
In the case of a Calabi-Yau model, the local parameters correspond to
the deformations of the complex structure in $H^{2,1}(X)$.  This then
gives the so-called Type B chiral ring \cite{witten-mirror}
\be
c_{ijk} = - \int_X \W \wedge \d_i\d_j\d_k\W 
\ee

This ring can be motivated as follows.
We can define the cohomology groups
\be
H^{-p,q}(X) = H^q_\dbar(Y,\ext^p(T^{(1,0)}X))
\ee
That is, we consider objects which are of the form
\be
\alpha(x) = \alpha^{i_1\ldots i_p}{}_{\jbar_1\ldots \jbar_q}
\pp {x^{i_1}}\wedge \ldots \wedge \pp {x^{i_p} }
dx^{\jbar_1}\wedge \ldots \wedge dx^{\jbar_q}.
\ee
Note that compared with the usual Dolbeault groups,
we replaced antisymmetric products of the
holomorphic {\it cotangent bundle} with antisymmetric products
of the holomorphic {\it tangent bundle}, which is essentially the
operation of mirror symmetry.
The unique holomorphic $3$-form $\Omega$ can be used to
give an isomorphism
\be
\Omega:\ext^p(T^{(1,0)}X) \ra\ext^{3-p}(T^{(1,0)}X)^*,
\ee
which also gives an isomorphism
\be
H^{-p,q}(X) \cong H^{3-p,q}(X).
\ee
The space
\be
H^{-*,*}(X) = \bigoplus_{0 \leq p,q \leq 3} H^{-p,q}(X)
\ee
acquires an obvious ring structure using the wedge product,
both on forms and vector fields. Furthermore, since
\be
H^{-3,3}(X) \cong H^{0,3}(X) = \C ,
\ee
we have an integration formula, and the structure
coefficients of the algebra can again be expressed
in terms of intersection numbers. By definition this
ring does depend crucially on the choice of complex
structure --- we use vector fields for the holomorphic
indices and forms for the anti-holomorphic indices.
There is however no dependence on the K\"ahler class.
If we use the isomorphism $H^{-1,1}\cong H^{2,1}$ and choose
an explicit basis for $H^{2,1}$ we reproduce the above formula.

\newsection{Gauge theories and S-duality}

We leave our discussion of the world-sheet formulation of string theory,
and now turn to the space-time description, where we will try to
reinterpret some of the mathematical structures that we found in the
previous lectures. Hereto we first discuss in the following lecture
S-duality in four-dimensional abelian gauge theories. However, before we
do this, it might be useful to review some four-dimensional geometry.
(The following sections closely follow \cite{four}.)

\newsubsection{Introduction to four-dimensional geometry}

Four-dimensional geometry is a very rich and difficult subject
\cite{freed,donaldson-kronheimer,friedman-morgan}. Since
we have been living exclusively in two dimensions in the previous lectures, 
we might be tempted to think too simple about topology in other dimensions.  For
example, the classification problem of four-manifolds is very much
different in nature from the classification of surfaces. Compact,
orientable surfaces are topologically classified by an integer, their
genus, 
$$
\insertfig{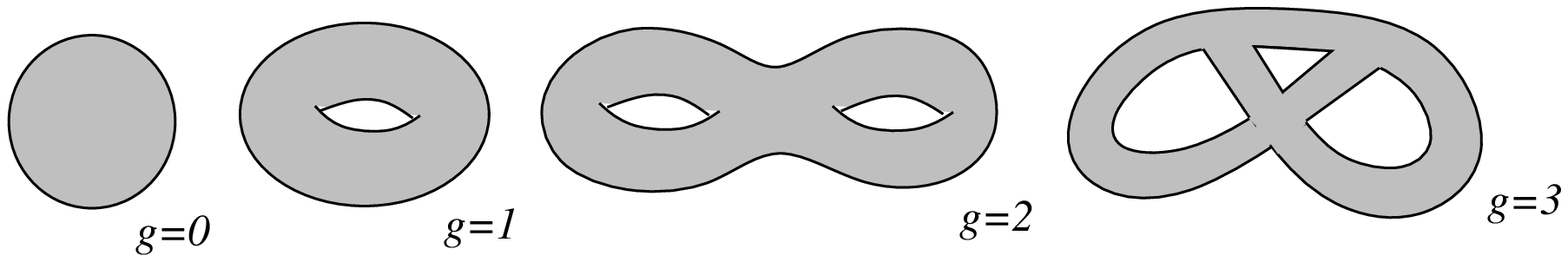}{11} \cdots
$$
That the situation is not that simple in four dimensions we can
already see by considering the most important invariant of any
manifold, the fundamental group $\pi_1$.

It is a classical theorem in topology that any finitely representable
group (\ie a group that can be represented by a finite number of
generators satisfying a finite number of relations --- not a very severe
restriction) can appear as the fundamental group of a four dimensional
manifold. There is a surgery algorithm that constructs the required
manifold starting from the generators and the relations. Markov's
solution of the word problem shows that the question whether two
finitely representable groups are actually isomorphic is undecidable.
That is, there is no computing algorithm that can decide this question
within a guaranteed finite time. This theorem has therefore rather
dramatic consequences for the classification problem of manifolds in
dimensions $d \geq 4$. There is simply not a conceivable `list' of
four-manifolds. This fundamental problem can however easily be
circumvented by assuming that the four-manifold $M$ is simply connected
\be \pi_1(M)=0. \ee This we will often assume in the following.
 
After the fundamental group, the next important invariant is the second
cohomology group $H^2(M)$. To understand better the role played by the
second cohomology of a four-manifold, it might be instructive to
consider first the analogous situation for two-dimensional surfaces, if
only because it is so much more easily visualized.
 
For any two-dimensional surface $\Sigma$ we can consider the first homology
group $H_1(\Sigma)$ of homology cycles or equivalently the dual cohomology
group $H^1(\Sigma)$ of cocycles. We can think of such a cocycle physically
as a flat abelian gauge field $A$, $F=dA=0$.  The pairing between a
cycle $C\in H_1$ and a cocycle $A\in H^1$ is the Wilson line
\be
\oint_C A.
\ee
For a genus $g$ surface $H^1$ has rank $b_1=2g$. It is naturally a
symplectic space by the intersection form
\be
Q(\a,\b)=\int_\Sigma \a\wedge \b.
\ee
$Q$ is an integer, unimodular {\it anti-symmetric} form. By a standard
theorem in symplectic linear algebra there exists a canonical symplectic
base $\a_1,\ldots,\a_g,\b^1,\ldots,\b^g \in H^2(\Sigma,\Z)$, satisfying
\be
Q(\a_i,\b^j)=\delta_i{}^j,
\ee
with all other intersections vanishing. The dual homology basis
$A^i,B_i$ we described in \S7.6 and looks like \fig 4.
If we have a orientation-preserving diffeomorphism on the surface, it
will act on the first cohomology by a symplectic transformation. That
is, there is a map (actually a surjection)
\be
\Diff^+(\Sigma) \ra Sp(2g,\Z).
\ee
 
In four dimensions the analogue object to consider is the second
cohomology group $H^2(M)$. A physical picture to keep in mind is that,
instead of considering Wilson lines, we now look at magnetic fluxes
$F$ (satisfying the Bianchi identity $dF=0$) through a two-cycle 
$\Sigma$ (a linear combination of surfaces)
\be
\int_\Sigma F.
\ee
In four dimensions we have an analogous intersection form
\be
Q: H^2\times H^2 \ra \R
\ee
defined by
\be
Q(\a,\b)=\int_M \a \wedge \b.
\ee
In the four-dimensional case $Q$ is
{\it symmetric} and non-degenerate. Over $\R$
such a form can be diagonalized and thus is easily classified by its
rank $b_2 = \dim H^2$ and signature $(b_2^+,b_2^-)$ (the number of positive respectively
negative eigenvalues). The signature $\s(M)$ of the 4-manifold $M$ is
defined as the difference of positive and negative eigenvalues of $Q$
\be
\s=b_2^+ - b_2^-.
\ee
It has a local expression due to Hirzebruch's signature theorem
\be
\s(M)=\int_M \third p_1= -{1\over 24\pi^2}\int_M \Tr R\wedge R.
\ee
The signature $\s$ and the Euler character $\chi$ are the two
classical invariants of a four-manifold. 
 
However, there is more sophisticated information hidden in the
intersection matrix $Q$ since it is actually defined on the lattice
\be
\G=H^2(M,\Z).
\ee
We can think of the elements of this lattice as
{\it quantized} fluxes, in the sense that
\be
F\in \G \gives \int_\Sigma F \in\Z
\ee
for all surfaces $\Sigma$. By Poincar\'e duality the lattice $\G$
is self-dual, so we can use the classification results of \S7.3.

It might be helpful to list some concrete examples of well-known
4-manifolds and their intersection forms
$$
\renewcommand{\arraystretch}{1.5}
\begin{tabular}{|c|c|}
\hline
\strut $M$ & $Q$ \\ \hline\hline
\strut  $\C\P^1$ & $ \id$ \\ \hline
\strut $\overline{\C\P^1}$ & $ -\id$ \\ \hline
\strut $S^2\times S^2$ & $H$ \\ \hline
\strut $T^4$ & $3\id \oplus 3(-\id)$ \\ \hline
\strut $K3$ & $3 H \oplus 2(-E_8)$ \\ \hline
\end{tabular}
\renewcommand{\arraystretch}{1.0}
$$
Herea bar indicates a complex manifold with opposite orientation.
(Every complex manifolds comes with a preferred orientation.)
 
Just as in the case of Riemann surfaces we can study the
representation of the diffeomorphism group on $\G$. Since this
action should preserve the intersection form, there is a homomorphism
of the group of orientation preserving diffeomorphisms $\Diff^+(M)$
into the arithmetic group $O(Q,\Z)=\Aut(\G)$, that leaves the lattice fixed,
\be
\Diff^+(M) \ra O(Q,\Z).
\ee
The  question is whether this map is actually onto (as was the
case for the modular group for a Riemann surface). This is an important issue for Donaldson theory, where we construct
differential invariants out of classes in $\G$. Since a manifold
invariant should by definition be invariant under diffeomorphisms,
these Donaldson polynomials will necessarily be constructed out of
tensors on $H^2$ that are invariant under the action of the image of
$\Diff^+$. In case that this image is the full group $O(Q,\Z)$,
standard invariant theory teaches us that the only invariants are
tensors constructed out of the intersection form $Q$ itself.
 
There is one more set of characteristic classes that we have not mentioned
yet, the Stiefel-Whitney classes
\be
w_i \in H^i(M,\Z_2).
\ee
For the case of a compact, simply-connected orientable manifold we have
automatically $w_1=w_3=0$. This leaves $w_2$ as the only new
characteristic class. It is the obstruction to a spin structure. The
class $w_2$ plays an important role if we consider the lattice reduced
modulo 2, because of Wu's formula that states that for all $x\in \G$
\be
w_2\.x=x^2 \mod 2.
\ee
This implies that $Q$ is necessarily even if $w_2=0$ or equivalently
if $M$ is spin.

\newsubsection{The Lorentz group}
 
It might be convenient to also review some standard facts concerning
the Lorentz group $SO(3,1)$ or its Euclidean version 
$SO(4)$ and its representations.  Consider a
4-dimensional vector space $V\cong\R^4$ and the corresponding space of
two-forms $\L^2 \cong \R^6$. If we pick a volume form $e\in \L^4
$, then we can define the intersection form
\be
q: \L^2 \times \L^2 \ra \R
\ee
by
\be
\a\wedge\b=q(\a,\b)e,\qquad \a,\b\in\L^2.
\ee
It is easily verified by explicit computation that the symmetric
bilinear form $q$ has signature $(3,3)$. The linear transformations
$SL(4,\R)$ preserve the volume form and induce linear maps on
$\L^2$ that preserve $q$. Thus we have a natural map
\be
SL(4,\R) \ra SO(3,3,\R).
\ee
This is actually a double cover of the identity component of
$SO(3,3)$.
 
The intersection form $q$ is not enough to decompose $\L^2$ into
maximal positive and negative subspaces
\be
\L^2 \cong \L^2_+\oplus \L^2_-.
\ee
In fact, there is a moduli space of such decompositions parametrized
by the grassmannian
\be
\Gr^{3,3}=SO(3,3,\R)/SO(3,\R)\times SO(3,\R).
\label{grass}
\ee
To find such an explicit decomposition we can pick a metric (positive
inner product) $g$ on $V$. This induces a metric on $\L^2$ (also
denoted by $g$) which decomposes $\L^2$ into orthogonal eigenspaces
of the Hodge $*$-operator, defined by
\be
g(\a,\b)=q(\a,*\b).
\ee
The $*$-operator squares to one
\be
*^2=1,
\ee
and the spaces $\L^2_\pm$ can now be defined as the eigenspaces with
$*=\pm$: the self-dual (SD) respectively anti-self-dual (ASD) forms.
The Hodge star only depends on the conformal class of the metric $g$,
which can be identified with the grassmannian (\ref{grass}).  The
choice of a metric $g$ gives thus rise to a natural homomorphism
\be
SO(4,\R) \ra SO(3)\times SO(3) = SO(3,3) \cap SO(6),
\ee
which is again a two-fold cover.
 
This local picture can be repeated globally by considering the tangent
bundle $TM$. There is a splitting of the 2-forms into SD and ASD
pieces
\be
\W^2(M) = \W^2_+(M) \oplus \W^2_-(M).
\ee
Descending to de Rahm cohomology, we find a similar decomposition
\be
H^2(M)=H^2_+(M)\oplus H^2_-(M)
\ee
with
\be
b^2_\pm=\dim H^2_\pm.
\ee
We also notice that (anti)self-dual cocycles are necessarily harmonic
\be
d\a=0,\ *\a=\pm \a \ \gives\  d^*\a=0.
\ee
 
Since we are also interested in fermions, we will in general consider
representations of the universal cover group of $SO(4)$
\be
Spin(4) \cong SU(2)_+\times SU(2)_-.
\ee
Its irreducible representations are labeled by the dimensions
$(\bn_+,\bn_-)$ with $\bn_\pm=1,2,\ldots$. In particular we have
\ban
\hbox{\it scalar}:& \psi & ({\bf 1},{\bf 1}),\\
\hbox{\it chiral spinor}:& \psi_\a &  ({\bf 2},{\bf 1}),\\
\hbox{\it anti-chiral spinor}:& \psi_{\dot\a}&  
({\bf 1},{\bf 2}),\\
\hbox{\it vector}:& \psi_{\a\dot\b}\sim\psi_\mu &  
({\bf 2},{\bf 2}),\\
\hbox{\it SD two-form}:& \psi_{\a\b}=\psi_{\b\a} \sim \psi_{\mu\nu}^+ &  
({\bf 3},{\bf 1}),\\
\hbox{\it ASD two-form}:& \quad\psi_{\dot\a\dot\b} = \psi_{\dot\b\dot\a} 
\sim \psi_{\mu\nu}^- \quad &  ({\bf 1},{\bf 3}).
\ean
As we already mentioned in \S9.2, the spinor
bundles, with ${\bf n_+} + {\bf n_-}$ odd,
only exist if the 4-manifold is spin $w_2=0$.
 
\newsubsection{Duality in Maxwell theory}

Our first example of a space-time theory which has a non-trivial
duality symmetry is Maxwell theory --- $U(1)$ gauge theory on
a four-manifold $M$. We pick a line bundle $L$ on $M$ and a connection
$A$ on $L$ with curvature $F$. The curvature satisfies the Bianchi identity
$dF=0$ and has integer periods around two-cycles $\Sigma\subset M$
\be
{1\over 2\pi} \int_\Sigma F \in \Z,
\ee
This gives of course the first Chern class 
\be
c_1(L)=\left[{F \over 2\pi}\right] \in H^2(M,\Z)
\ee
that classifies the line bundle $L$ topologically.

To consider the Maxwell equations we further need a metric on $M$, or more
precisely, a conformal structure or Hodge star $*$ (so the metric is
only defined up to local rescalings.) The equations of motion now read
$d^*F=$ or 
\be
d(*F)=0.
\ee
 Electric-magnetic duality or ``abelian S-duality''
is the observation that the equation $d^*F=0$ together with the Bianchi
identity 
\be
dF=0
\ee
are invariant under the transformation $F \lra *F$ that
interchanges the $E$ and $B$ field.

This duality is also present in the quantum theory, in fact it extends
to the group $SL(2,\Z)$.
This can be made more precise by considering the path-integral.
The Maxwell action can be concisely
written if we introduce for any complex variable 
\be
\t=\t_1+i\t_2 \in \H
\ee
the operator $\hat\t:\W^2(M) \ra \W^2(M)$ that acts on two-forms as
\be
\hat\t=\option{\t_1 +i\t_2*}{Euclidean signature $(4,0)$,}
{\t_1 + \t_2*}{Lorentzian signature $(3,1)$.}
\ee
(Note that in Lorentzian signature $*^2=-1$.) In terms of the usual coupling
constant $g$ and the theta angle $\theta$ we have
\be
\t = {\theta\over 2\pi} + {4\pi i\over g^2}.
\ee
The Maxwell action, including the theta-angle, now reads
\be
S = {1\over 4\pi} \int_M F \wedge \hat\t F.
\ee
The partition function is defined as the integral over the space of
connections $\cA$ (which includes a sum over all possible line
bundles $L$) modulo the action of the gauge group $\cG$
\be
Z=\mathop{\int}_{\cA/\cG} \cD A\, e^{iS}
\ee
We note immediately that under the theta-angle shift $\theta \ra \theta
+ 2\pi$, or equivalently under the transformation
\be
T:\ \t \ra \t + 1,
\ee
the partition transforms by the phase factor
\be
Z \ra Z \cdot  e^{2\pi i k},
\ee
with
\be
k = {1\over 8\pi^2} \int_M F\wedge F =  \int_M \half c^2_1 \in
\half \Z.
\ee
In case $M$ is spin, so that the intersection form  is even,  $k$ is an
integer, and the $T$ transformation is a quantum symmetry.

It is useful at this point 
to introduce the dual field strength, the two-form
\be
F_D = 2\pi i {\delta S \over \delta F} =\hat\t F.
\ee
In Maxwell theory (with a theta angle) we can then define electric and
magnetic charges $e,m \in \Z$ as
\ba
e \is {1\over 2\pi}  \int_{S^2}  F_D, \nonu 
m \is {1\over 2\pi}   \int_{S^2}  F  .
\ea
Here we choose a two-sphere $S^2$ surrounding a certain region
in a three-dimensional spatial slice. Under the T-duality we have the 
Witten effect \cite{witten-effect}: 
the field strength and dual field strength transform as
\be
T:\ \vectord{F_D} F \ra \vectord{F_D + F} F,
\ee
so that the charges transform as
\be
T:\ \vectord e m \ra \vectord{e+m} m.
\ee
In this way a monopole receives an electric charge equal to its magnetic
charge and becomes a dyon (an object with both electric and magnetic
charge).

We now claim that the full duality group of Maxwell theory is
\be
G = SL(2,\Z),
\ee
and that it acts on the coupling constant by fractional linear transformations
\be
\t \ra {a\t +b \over c\t+ d}.
\ee
In fact the vector $(F_D,F)$ transforms as a
doublet
\be
\vectord{F_D} F \ra \twomatrixd abcd \vectord{F_D} F.
\ee
To prove this duality we have to consider the second generator of
$SL(2,\Z)$, the well-known electric-magnetic S-duality
\be
S:\ \t \ra -{1\over \t},
\ee
that interchanges $F$ and $F_D$:
\be
S:\ \vectord{F_D} F \ra \vectord {-F}{F_D},
\ee
and consequently also the electric and magnetic charges
\be
T:\ \vectord e m \ra \vectord {-m} e.
\ee 
Note that this transformation squares to minus the identity (the parity
transform), $S^2=-1$. Since the coupling constant $g^2$ transforms into
$1/g^2$, $S$ relates strong and weak coupling (even though this is a
free theory, with no perturbation expansion!).

S-duality is derived in complete analogy with the proof of
T-duality in the toroidal sigma model in \S7.2. 
One starts with the path-integral
\be
Z = \int\cD F \cD A_D \, \exp\left(iS(F)+ {i\over 8\pi}\int_M F\wedge dA_D
\right)
\ee
with $F$ a two-form on $M$ and $A_D$ an abelian dual gauge field. The
manipulation is familiar from \S7.2. Integrating out $A_D$ forces the Bianchi
identity $dF=0$, which allows us to write $F$ locally as $dA$.
Integrating out $F$ produces the dual model, with $F_D=2\pi i \delta
S/\delta F=dA_D$, with dual coupling constant $-1/\t$.

\newsubsection{The partition function}

We now consider the computation of the Euclidean partition function,
following \cite{erik-ab,witten-ab}. The most interesting
part is the contribution of the zero-modes. The zero-modes are here the
first Chern classes
\be
p = [F/2\pi] \in H^2(M,\Z).
\ee
For classical field configurations $p$ will be an harmonic representative,
$dp=d^*p=0$. 
We can write a general two-form $F$ that satisfies the Bianchi identity
as
\be
F = 2\pi p + d\a,
\ee
with $\a$ a proper one-form.
Note that $\G=H^2(M,\Z)$ carries the intersection form
\be
p^2 = \int_M p \wedge p,
\ee 
which makes it into a self-dual lattice (even if $M$ is spin).
A choice of metric on $M$ gives a decomposition of two-forms in their
self-dual and anti-self-dual parts. We will write for the zero-mode piece
\be
p=p_L + p_R,
\ee
with $*p_L=p_L$ and $*p_R=-p_R$, so that
\be
p^2 = p_L^2 - p_R^2.
\ee
In this way the zero-mode contribution
to the action reads
\be
S=\pi(\tau p_L^2 - \taubar p_R^2)
\ee
The answer for the partition function now becomes a theta-function
\cite{erik-ab,witten-ab}
\be
Z = Z_{qu} \. \sum_{(p_L,p_R)\in \G} q^{\half p_L^2} \qbar^{\half p_R^2}
,\qquad q = e^{2\pi i \tau}
\ee
This generalized theta-function is identical to the zero-mode
contribution of the toroidal sigma-model with Narain lattice $\G$. The factor $Z_{qu}$ represents
the contribution from the oscillations and is a product of determinants.
By the usual arguments of string theory, the theta function is a
modular form (though not holomorphic) of weight $\half (b_+,b_-)$ if the
lattice $\G$ is even and self-dual, which is the case for spin
manifolds.

One way to understand this appearance of theta functions has been
suggested by Erik Verlinde \cite{erik-ab}. 
Consider a model in six dimensions on the
manifold $X^6 = M^4 \times T^2$, where the two-torus has modulus $\t$.
The dynamical field is a two-form $B$ with
three-form field strength $H=dB$ and action
\be
S = \int_X H \wedge *H.
\ee
Now demand that $H$ is self-dual and has a decomposition as $H=\sum_i 
F_i\wedge C_i$ with $F_i$ a two-form on $M^4$ and $C_i$ a one-form on
$T^2$. The self-duality on $X^6$ now tells us that the Hodge star on $M^4$
is correlated with the Hodge star on $T^2$. $H$ has then a
decomposition (with $dz=dx+\t dy$)
\be
H=F_+ dz + F_- d\zbar= F dx + F_D dy
\ee
(Note that $F=F_+ +F_-$ and $*F=F_+ - F_-$, so that $F_D=\t_1 F + i \t_2 *F
=\t F_+ + \bar\t F_-$.) If $(A,B)$ is a basis for $H_1(T^2)$ we can write
\be
F = \oint_A H,\qquad F_D  = \oint_B H
\ee
The partition function can now be computed either by first integrating
over $T^2$, which reproduces the gauge theory computation, or by first
integrating over $M^4$, which gives a sigma model with target space the
torus $H^2(M)/\G$. This Kaluza-Klein point of view will be substantially
upgraded if we go to string theory. 

\newsubsection{Higher rank groups}

We can make a simple generalization of Maxwell theory by considering an
abelian gauge group of rank $g$. In that case we have one-form gauge
fields $A^i$, $i=1,\ldots,g$, with corresponding curvatures $F^i=dA^i$.
We can now write an action with a matrix $\t_{ij}$ of coupling constants
and theta-angles
\be
S = {1\over 4\pi} \int_M F^i \wedge \hat\t_{ij} F^j
\ee
Note that we can take $\t_{ij}$ to be symmetric; the above action makes
only sense if also $\Im \t_{ij}>0$. So we are dealing with a period matrix
of an abelian variety of dimension $g$ (see \S11.6) 
for example the Jacobian of a
genus $g$ Riemann surface. The definitions of the dual field strengths
are now 
\be
F_{D,i}=\hat\t_{ij} F^j,
\ee
and we have a $2g$ dimensional vector of electric and magnetic charges
$e^i,m_i \in \Z$, $i=1,\ldots,g$,
\ba
e^i \is {1\over 2\pi} \int F_{D,i},
\nonu
m_i \is {1\over 2\pi} \int F^i.
\ea
The duality group is the rank $2g$ symplectic group
\be
G = Sp(2g,\Z),
\ee
that acts on the coupling constant matrix as $\t \ra (A\t+B)(C\t+D)\inv$.

\newsubsection{Dehn twists and monodromy}

This model can be obtained by considering the six-dimensional model of
\S9.4, now on the manifold $M^4 \times \Sigma_g$.  Note that from this
point of view the group $Sp(2g,\Z)$ is identified by (a quotient) of
the mapping class group of the surface $\Sigma_g$.  As such it is
generated by Dehn twists. A Dehn twist $D_C$ on a homology cycle $C$
is defined by cutting the surface on $C$ and gluing it back together
after a $2\pi$ rotation. The action on the homology cycles is given by
\be
D_C: C' \ra C' + \eta(C,C')C
\ee
where the symplectic form $\eta(\.,\.)$ denotes the intersection
product on the first homology group $H_1(\Sigma_g,\Z)$.

This Dehn twist can also be seen as a monodromy in the moduli space
$\cM_g$. If we pinch the cycle $C$ we obtain a double point. The
pinching process can be described by inserting first a tube connecting
the two marked points with modulus $q$, as in our discussion of CFT. We
write this modulus as $q=e^{- t + i\theta}$, so that the cycle $C$
shrinks to zero in the limit $t\ra\infty$. If we now fix $t$ and all
other moduli of the surface and follow the loop $\theta \ra \theta +
2\pi$ in $\cM_g$, the Dehn twist $D_C$ is implemented, see
\figuurplus{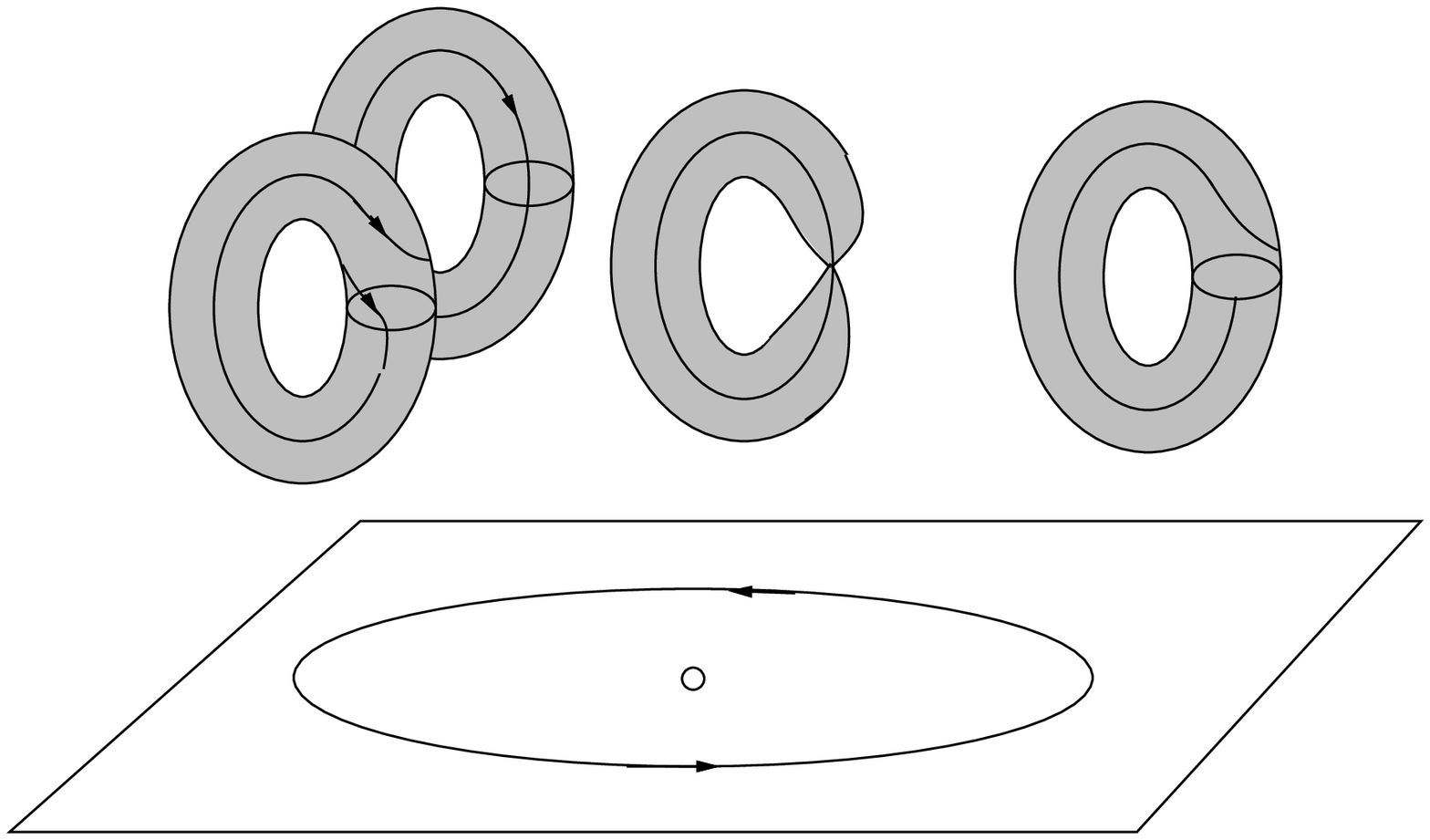}{9cm}{The action of a Dehn twist is implemented by
a monodromy in the moduli space.} 

This is a special case of what is called in generality Picard-Lefschetz
theory\footnote{Note that Picard-Lefschetz theory is the complex
analogon of Morse theory \cite{milnor}. In Morse theory we consider a
real function $f:M \ra \R$ and look at how the inverse image
$M_x=f\inv(x)$ varies as a function of $x$. The essential behaviour is
at the critical points $df=0$ where the topology of $M_x$ jumps, in a
controlled way determined by the signature of the Hessian of $f$. In the
complex case we have a complex function $f:M\ra \C$ and one can go {\it
around} a critical point, and study the corresponding transformation on
the homology of $f\inv(x)$.} \cite{picard}: modular/duality
transformations are related to vanishing cycles in the
``compactification'' $\Sigma_g$. This fact we will meet again when we
start studying Calabi-Yau compactifications in string theory. 

\newsection{Moduli spaces}

We have stressed the importance of the concept of a moduli space in the study
of quantum theories of fields and strings. There
are basically three ways in which moduli spaces enter in field theory or
string theory: (1) as special classical solutions, (2) as families of
QFTs and (3) as families of inequivalent vacua. A priori the three
ways are not related, however there are various deep relations between
these points of view that occur in string theory that we will use
to our advantage.

\newsubsection{Supersymmetric or BPS configurations}

The first occurrence of moduli spaces is as spaces of special
classical configurations of the (usually bosonic) fields $\f$, possibly
modulo gauge transformations. A good example is given by the
four-dimensional anti-self-duality (ASD) condition 
\be
F^+=F + *F=0,
\ee
that gives rise to the moduli space of instantons, defined as
the set of solutions to this equation modulo gauge equivalence
\cite{donaldson-kronheimer}. 

These configurations
typically arise as supersymmetric or BPS configurations. If we denote
the bosonic variables and fermionic variables generically as $\f$ and
$\psi$, then the supersymmetry variation $\delta\psi$ of the fermions reads
(at $\psi=0$)
\be
\delta \psi = G(\f)\e,
\ee
BPS configurations are characterized by the property that for a
particular subset of supersymmetry parameters $\e$ the RHS vanishes. 

For example, in $N=2$ supersymmetric gauge theories that we will
discuss in more detail in the next lectures, we have
a relation
\be
\delta\psi = F_{\m\n}\gamma^{\m\n}\e,
\ee
so that for a chiral spinor $\e$ that satisfies $\g_5\e=\e$, we find that ASD
configurations with $F^+=0$ are invariant under half the number of
supersymmetry variations.

These moduli spaces are often noncompact. In the case of ASD gauge
fields this is a familiar consequence of the fact that the instanton
equations allow solutions on flat space $\R^4$. Since the ASD equations
depend on the choice of Hodge $*$, they are invariant under conformal
transformations of the metric. So we can find a one-parameter family of
rescaled solutions $x\ra t\.x$. Since the instanton in the limit $t\ra
0$ becomes a $\delta$-function in the origin and therefore is no longer
a smooth solution to the ASD equation, the moduli space on $\R^4$ is
clearly non-compact. This argument can be repeated for general
four-manifolds. In the limit $t\ra 0$ we have almost point-like
instantons on $\R^4$. We can now cut out a little disk containing most
of the solution and ``graft'' this onto a point of $X$. Taubes has
proven that this can made into a solution for the ASD equation on $X$
\cite{taubes}. We see that therefore also on a general manifold
point-like instantons occur in the limit $t\ra 0$.

The moduli space also can have singularities. For the ASD equations
these singularities occur whenever the gauge group $\cG$ does not act
freely. This is the case if the holonomy group of the connection is
$U(1)$ instead of $SU(2)$. We can then restrict the gauge group
consistently to $U(1)$ and we are dealing with an abelian connection. So
the only singularities are the abelian instantons. In that case the
curvature is just a $\R$-valued 2-form $F\in H^2(X,\R)$. It must however
satisfy a quantization rule. The first Chern class of the $U(1)$ bundle
should be an integer cohomology class. If we now also impose the ASD
condition, we see that abelian instantons $*F=-F$ correspond one-to-one
with elements of the lattice
\be
H^2_- \cap H^2(X,\Z).
\ee
These singularities form in general not a serious problem. We should
remember that we are free to pick a metric. If we choose a generic
metric and thus a generic positioning of the subspace $H^2_-\subset
H^2$, the intersection with the integer lattice $H^2(X,\Z)$ will
typically be zero, unless $H_-^2$ has codimension zero (if $b^2_+=0$).
(Also the case $b^2_+=1$ has to be treated carefully.) This is a rather
general effect, quite often slight perturbations of our moduli problem
can make the singularities disappear.

\newsubsection{Localization in topological field theories}

One of the applications of the moduli spaces of the type discussed above
in QFT is that, under suitable circumstances, the path-integral can localize to
these configurations,
\be
Z =\int \cD\f\,e^{-S} \gives \int_{\cM} \cdots
\ee
Localization is a crucial ingredient in topological field theories
\cite{witten-tft,atiyah-jeffrey,mathai-quillen}, see the reviews
\cite{tft-review,blau,cmr}, that allows us to express partition and correlation
functions as integrals over finite dimensional moduli spaces instead
of over infinite field space. Of course, it is the unique structure of
topological models that allows such a drastic reduction in degrees of
freedom.
 
The mathematical idea of localization has rich applications. The most
familiar one is the calculation of the Euler characteristic of a
manifold $X$ \cite{bott-tu}. On the one hand it can be computed by
picking a Riemannian metric and integrating over $X$ the Euler density
$\Pf(R)$ that is constructed out of the Riemann curvature tensor. On the
other hand the Euler character can be computed by counting the number of
zeroes of a generic vector field on the manifold.
 
In quantum mechanics the localization principle is well-known as the
phenomenon of the Nicolai map \cite{nicolai}. This can be nicely
illustrated by a zero-dimensional example. Consider a polynomial $s(x)$
and the ``path-integral''
\be
Z = \Int_{-\infty}^\infty dx\,e^{-s^2/2}.
\ee
As it stands this integral has no interesting invariances and cannot
be computed in any simplified way. If instead we consider the modified
integral
\be
Z = \int dx\,e^{-s^2/2}\,{\d s\over\d x},
\ee
we can compute it without much effort, since it can be rewritten as
\be
Z=\int ds\, e^{-s^2/2} = \option 0 {if $s$ even} 1 {if $s$ odd}
\ee
So we see $Z$ computes a simple invariant of $s$: the degree of the
map $s:\R\ra\R$, which is either zero or one depending on whether $s$
is an even or odd polynomial. The partition function $Z$ is thus
invariant under all deformations of $s$ that leave the boundary
conditions at infinity invariant. 
 
Now that we have established the ``topological invariance'' of $Z$ we
can make use of this by deforming the action $s$ in such a way that
the localization becomes evident. To do this, we rescale $s \ra t\.s$
and take the limit $t\ra\infty$. In this limit the gaussian factor
will damp the integral for all values of $x$ except for the zeroes of
the function $s(x)$. Around these points we can perform a semi-classical
approximation. In this way the computation localizes to a finite number
of points. We compute $Z$ as sum over the zeroes of $s$ of a
factor $\pm 1$, very much in analogy with the Euler character,
\be
Z =\sum_{s(x)=0} {\rm sgn}\det(d s).
\ee
One way to express this localization is that the
semi-classical approximation gives an exact result.
 
A third way to compute the integral $Z$ will be a metaphor for the
actual computations in field theory. We start by
rewriting the factor $\d s/\d x$ as a fermionic gaussian integral. We
introduce two fermionic variables $\psi,\rho$ and rewrite $Z$ as
\be
Z =\int dxd\psi d\rho \, e^{-S},
\ee
with action
\be
S = \half s^2 + \rho \d s \psi.
\ee
This action $S$ has a BRST symmetry $Q$ given by
\be
Qx =\psi,\quad Q\psi = 0,\quad Q\rho = s(x),
\ee
As it stands this symmetry is not nilpotent. This property we obtain
if we introduce a further bosonic auxiliary field $H$ and write
\be
Z =\int dxdHd\psi d\rho \, e^{-S},
\ee
with
\be
S = isH + \half H^2+ \rho \d s \psi.
\ee
The symmetry is now extended as
\be
Q\rho=H,\qquad QH=0.
\ee
With this extra auxiliary field $H$ the BRST symmetry squares to zero
``off- shell''
\be
Q^2=0.
\ee

Now there is a simple localization argument for this BRST charge due
to Witten \cite{witten-N}. 
Our integral $Z$ is expressed as an integral over a $2|2$
dimensional superspace. On this space we have the action of the
fermionic symmetry $Q$ generated by a vector field $\xi$, that is,
$Q=\Lie \xi$. The odd vector field $\xi$ squares to zero, $\xi^2=\half
[\xi,\xi]=0$, a non-trivial property for an odd vector field. The
orbits of this group action are $0|1$ dimensional curves parametrized
by an odd coordinate $\theta$. Because of the fundamental identities
of grassmannian calculus
\ba
\int d\theta\, 1\is 0,\nonu
\int d\theta\, \theta\is 1,
\ea
the integral of a constant function, such as the BRST-invariant
action density $e^{-
S}$, along the orbit will automatically give zero. Therefore the only
non-vanishing contributions to the integral can come from the
zero-dimensional orbits, that is, the fixed points of $Q$ or
equivalently the zeroes of the vector field $\xi$.
 
\newsubsection{Quantization}

A second application is that moduli spaces can be quantized. Here we
consider a supersymmetric point particle moving on the moduli space
$\cM$. In this way there is a canonical way to associate a (graded)
vector space $V$ to the moduli problem as
\be
V = H^*(\cM).
\ee
The (super)dimension of this space can be computed using the Euler
character, that often also allows a representation as a TFT partition
function \cite{vafa-witten,btft,blau-thompson}
\be
\sdim V = \chi(\cM).
\ee
We will see an application of this in \S13 when we discuss the
quantization of D-branes.

\newsubsection{Families of QFTs}

The second way moduli spaces appear in our story is as
 families of quantum field 
theories. For example, in quantum mechanics we can consider a potential 
\be
V(x) = \sum t_n x^n
\ee
The coefficients $t_n$ now label a family of systems. We have seen in 
detail how the moduli space of two-dimensional TFTs and CFTs is a rich 
object. In field theory such deformation will be of the general form
\be
\delta S  = \int \cO
\ee
with $\cO$ some set of local operators. The Hilbert space of the quantum
field theory thus contains the tangent space to the moduli space $\cM$.

\newsubsection{Moduli spaces of vacua}

The third way moduli spaces occur is as spaces of inequivalent vacuum
states of a single QFT. Here we should be careful to distinguish the
classical moduli space $\cM_{cl}$ and the quantum moduli space
$\cM_{qu}$. For example, in the above case of a particle in a potential
$V(x)$, the classical vacua correspond to the absolute minima of $V$, so
that $\cM_{cl}$ might be very complicated; quantum mechanically we know
that tunneling will produce a unique vacuum state, so $\cM_{qu}=pt$.

The phenomena of inequivalent vacua in a QFT is strongly related the 
appearance of scalar fields. Scalar fields have the unique property that 
they can have expectation values compatible with Poincar\'e invariance. 
In fact, when we consider a QFT with scalar fields $\f$ on a noncompact 
spatial manifold, say $\R^n$, then we have to impose boundary conditions 
for the scalars
\be
\f(x) \mathop{\ra}^{|x|\ra \infty} \f_0.
\ee
The values $\f_0$ will minimalize the energy and thus take value in 
some moduli space
\be
\f_0\in\cM_{cl}.
\ee
(Stated otherwise, in the path-integral we have to separate out the constant 
modes $\f_0$, since they are not $L^2$.)

If we consider the same QFT on a {\it compact} manifold, the constant 
mode $\f_0$ becomes $L^2$ and the partition function will be obtained as 
an integral over $\cM$
\be
Z = \int_{\cM} d\f_0 \cdots
\ee

\newsection{Supersymmetric gauge theories}

A good class of examples of field theories with scalar fields and
non-trivial families of vacua are four-dimensional supersymmetric gauge
theories, to which we will now turn.

\newsubsection{Supersymmetric gauge theories} 

A general $D=4$ supersymmetric gauge theory with $N\leq 4$ supersymmetries
consists of the following field content: a gauge field $A_\m$ together with $N$ Majorana fermions
$\l_\a^I,\lambdabar_{\dot\a,I}$, $I=1,\ldots,N$, and $\half N(N-1)$
real scalar fields, conveniently described by an $N\times N$
anti-symmetric matrix $\f^{IJ}=-\f^{JI}$. All these fields take action
in the adjoint representation of the gauge group. The simplest
supersymmetric action is
of the form
\be
S = \int \Tr\left(F\wedge* F + i\lambdabar_I \Dslash\l^I + D\f_{IJ}\wedge *D\f^{IJ} +
\sum_{I,K} [\f^{IJ},\f_{JK}]^2
+ \l_I[\f^{IJ},\l_J]\right)
\ee
The supersymmetry transformations satisfy
\be
[Q^I_\a,\Qbar^J_{\dot\a}]= \delta^{IJ} \g^\m_{\a\dot\a} P_\m.
\ee
If we introduce a general linear combination
\be
\delta =  \eta_{\a,I} Q^I_\a + \etabar_{\dot\a,I} \Qbar^I_{\dot\a},
\ee
the transformations of the vector $A_\m$ read in terms of the
bispinor $A_{\a\dot\a}=A_\m \g^\m_{\a\dot\a} $
\be
\delta A_{\a\dot\a} = i \l^I_\a\etabar_{\dot\a,I} - i 
\eta_{\a,I} \lambdabar^I_{\dot\a}.
\ee
The fermions transform in the two chirality components of the field
strength
\ba
\delta \l^I_\a \is F_{\a\b}\eta^{\b,I} + i[\f^{IJ},\f_{JK}]\eta^K_\a +
D_{\a\dot\a}\f^{IJ} \etabar_J^{\dot a}, \nonu
\delta \lambdabar^I_{\dot\a} \is F_{\dot\a\dot\b}\etabar^{\dot\b,I}
-  i[\f^{IJ},\f_{JK}]\etabar^K_\a + D_{\a\dot\a}\f^{IJ} \eta_J^\a.
\ea
Finally, the scalar fields transform as
\be
\delta \f^{IJ}= \l^I_\a\eta^{\a,J} +
\lambdabar^I_{\dot\a}\etabar^{\adot,J}.
\ee
This model has an internal $U(N)$ R-symmetry, under which the charge $Q$
transforms as the fundamental representation ${\bf N}$ and $\Qbar$
transforms as the conjugate transformation $\overline{\bf N}$.
 
\newsubsection{Twisting and Donaldson theory}
 
We explained in section \S8.2 how the twisting procedure can produce a TFT
starting with a supersymmetric field theory. In four dimensions one needs
a model with at least $N=2$ supersymmetry \cite{witten-tft}. This has a
symmetry group
\be
Spin(4) \times U(2)_R,
\ee
where $Spin(4)$ is the double cover of the Lorentz group $SO(4)$ and
the internal $R$-symmetry $U(2)_R$ acts on the two supersymmetry
charges $Q^I$ which transform as a doublet. This group is (locally)
isomorphic to
\be
H=SU(2)_+\times SU(2)_- \times SU(2)_R \times U(1),
\ee
Irreducible representations of this group are labeled by $({\bf
n}_+,{\bf n}_-,{\bf n}_R,q)$, where $\bf n$ indicates the dimension of
the $SU(2)$ representation and $q$ indicates the $U(1)$ representation
$z \ra e^{iq\theta}z$. The twisting procedure replaces the component
$SU(2)_+$ by the diagonal subgroup of $SU(2)_+\times SU(2)_R$. That is,
the Lorentz spins are replaced as
\be
({\bf n}_+,{\bf n}_-) \gives
({\bf n}_+\otimes {\bf n}_R,{\bf n}_-).
\ee
We see that the supercharges $Q^I_\a,\Qbar^J_{\dot\a}$ transform under
$H$ as $({\bf 2},{\bf 1},{\bf 2},1)$ and $({\bf 1},{\bf 2},{\bf
2},1)$. Consequently, under the twisting procedure they decompose as
tensors of type $({\bf 1},{\bf 1}) \oplus ({\bf 3},{\bf 1})$ and $({\bf 2},
{\bf 2})$, that is, as a scalar and a self-dual two-form, and as a vector
\ba
Q^I_\a & \gives & Q, Q^+_{\m\n} , \nonu
\Qbar^J_{\dot\a} & \gives & G_\mu.
\ea
So we see that we obtain the required twisted supersymmetry algebra
(\ref{twist})
\be
\{Q,G_\mu\} = P_\mu.
\ee
Therefore, we are dealing with a topological field theory.

Note that on a general curved four-manifold the holonomy group will be
$SO(4)\cong SU(2)_+ \times SU(2)_-$, so the twisting procedure that
changes the Lorentz properties will be a measurable effect. The twisted
model is therefore physically speaking inequivalent to the untwisted
model. This is similarly true on a complex manifold with holonomy group
$U(2) = U(1)_+\times SU(2)_-$. However, on a hyperk\"ahler or Calabi-Yau
manifold, where the structure group of the tangent bundle is simply
$SU(2)_-$, the twisting is invisible. Twisting changes the $SU(2)_+$
representation, but that part of the frame bundle can be
trivialized\footnote{Stated otherwise, one does have (two) covariant
constant spinors on a hyperk\"ahler four-manifold. Unfortunately, in the
compact case, the only hyperk\"ahler spaces are $T^4$ and $K3$.}.

For an $N=2$ gauge theory the resulting topological field theory was
first introduced by Witten \cite{witten-tft,witten-sym}. It fits
perfectly in the general framework we sketched. Starting point is $N=2$
supersymmetric Yang-Mills theory. Its fundamental multiplet
$(A,\l^I,\phi)$ consists of a connection $A_\m$, two spinors
$(\l^I_\a,\l^I_\adot)$ $(I=1,2$) and a complex scalar field $\phi$, all
taking their values in the adjoint bundle. After twisting we recover the
following fields
\ba
A_\m & \gives & A_\m\nonu
\l^I_\a & \gives & \psi_\m \nonu
\l^I_\adot & \gives & \rho^+_{\m\n},\eta \nonu
\phi & \gives & \phi\nonu
\phibar & \gives &\phibar 
\ea
We recognize the fundamental topological multiplet $(A,\psi,\rho)$
together with an equivariant multiplet $(\eta,\phi,\phibar)$, that is
due to the fact that we quotient by the gauge group. The BRST
transformations can be derived from the $N=2$ supersymmetry algebra
that we gave before and read
\ba
\delta A_\mu \is \psi_\mu \nonu
\delta \psi_\mu \is D_\mu\phi \nonu
\delta \rho^+_{\mu\nu} \is F^+_{\mu\nu} \nonu
\delta \phibar \is \eta
\label{brst}
\ea
The ``section'' $s$ in the general localization setup of \S10.2
is thus identified as
\be
s(A)=F^+
\ee
and its zero locus is indeed the moduli space $\cM$ of ASD connections. 
 
\newsubsection{Observables}

In any TFT of cohomological type, the observables are the cohomology
classes of the BRST operator. The most important classes of the twisted
$N=2$ SYM model are constructed
out of the local operator
\be
\cO^{(0)} = {1\over 8\pi^2} \Tr \phi^2.
\ee
We construct its descendents by
\be
d\cO^{(i)} = \delta \cO^{(i+1)}.
\ee
As we reviewed in section \S8.2 any cycle  gives rise to a 
physical operator via the map
\be
C\in H_*(M) \ \gives \ \cO_C = \int_C \cO^{(i)}.
\ee
For twisted $N=2$ Yang-Mills this gives the following operators: 
(We will assume that $M$ is simply connected, so that only $H_0(M)$, $H_2(M)$
and $H_4(M)$ contribute.) Associated to
two-cycles $\Sigma \subset M$ we find
\be
\cO_\Sigma = 
\int_{\Sigma} \cO^{(2)} =  {1\over 4\pi^2}  
\int_{\Sigma} \Tr(\phi F + \psi^2).
\ee
For the top form we recover the chern class or instanton number
\be
\cO_M = \int_M \cO^{(4)} =  {1\over 8\pi^2} \int_{M} \Tr F\wedge F =
ch_2.
\ee
Note that by Poincar\'e duality,
the coupling constants to the operator $\int_C\cO^{(i)}$ naturally
takes value in $H^{4-i}(M)$. If we write
\be
\cO = \sum_i \cO^{(i)} \in \W^*(M),
\ee
and $t \in H^*(M)$, then a general coupling to the action reads
\be
\delta S = \int_M t\wedge \cO .
\ee
(The integral naturally picks out the forms of total degree four.)

Since the path-integral localizes to the moduli space $\cM$, the
(expectation values of the) operators have a direct interpretation as
cohomology classes on $\cM$. In fact, we get Donaldson's map
\cite{donaldson}
\be
\m: H_i(M)\cong H^{4-i}(M) \ \ra \ H^{4-i}(\cM).
\ee
The mathematical definition of this operation is roughly as follows.
There is a natural bundle on the product space $M \times \cM$. This
bundle has a second Chern class $\hat c_2$. We can now consider the
differential form $\a \wedge \hat c_2$, which is of degree $4+k$, and
integrate it over the fiber $X$
\be
\m(\a) = \int_X a\wedge \hat c_2.
\ee
We see that in the particular example of the identity $\id$ we have
\be
\m(\id) = \int_X c_2= n \in \Z.
\ee

The famous Donaldson polynomials are now defined in terms of the
generating functional
\be
D(v,\l)=\left\<
\exp\left(\int_M t \cO^{(4)} + v \wedge \cO^{(2)} + \l \cO^{(0)}\right)
\right\>,
\ee
with coupling constants
\be
t\in H^0(M),\ v\in H^2(M),\ \l\in H^4(M).
\ee
The generating function has an expansion
\be
D(v,\l)=\sum_{n\geq 0} e^{-nt} D_n(v,\l),
\ee
where $D_n(v,\l)$ is a finite sum of intersection products on $\cM_n$,
the component of $\cM$ of instanton charge $n$. However, it is a
nontrivial matter to make rigorous sense of these intersections, since
we have seen that the moduli space is non-compact. In fact, a tremendous
amount of work goes in proving that the heuristic definitions below
actually make sense in some precise way. It is a polynomial since only
contributions of degree $d=\dim\cM_n$ contribute (with $\deg(v)=2,$
$\deg(\l)=4.$) The main theorem of Donaldson states that the polynomials
$D_n$ are diffeomorphism invariants if $b_2^+>1$.

\newsubsection{Abelian models}

After this detour to TFT, let us now consider $N=2$ (untwisted) Maxwell
theory, with gauge group $G=U(1)^g$. We will write the gauge fields
again as $A^i=A_\m^i dx^\m$, $i=1,\ldots,g$. In that case it is
convenient to combine the two real scalars into one (rank $g$
vector-valued) complex scalar $\f(x)$. The most general action,
compatible with supersymmetry and at most second order in derivatives,
is given by
\be
S = {1\over 4\pi} \int \left(F^i \wedge \hat\t_{ij}(\f) F^j 
+ g_{i\jbar}(\f) \l^i_I \dslash \l^{\jbar,I}+ g_{i\jbar}(\f)
d\f^i\wedge *d\phibar^\jbar  \right),
\ee
where both the matrix of coupling constants $\t_{ij}$ and the
sigma-model metric $g_{i\jbar}$ are allowed to depend on $\f$.
Supersymmetry restricts this dependence in the following way: The
``period matrix'' $\tau_{ij}$ should be the second derivative of an
holomorphic function $\cF(\f)$ of $\f$, the prepotential,
\be
\t_{ij}={\d^2 \cF \over \d\f^i \d\f^j},\qquad {\d \cF \over
\d \phibar^i} =0.
\ee
Furthermore, the kinetic terms for the fermions and scalars are directly
related to this period matrix as
\be
g_{i\jbar} =\Im\t_{ij}.
\ee

Summarizing, the Lagrangian is essentially given by one complex
``function'' (it will turn out to be a section of a line bundle), the
prepotential $\cF(\f)$. This fact is more or less immediately clear if
one takes the superspace point of view.

In the case of a simple quadratic action, as we have been considering up
to now, this prepotential is given by 
\be
\cF = \half \t_{ij}\f^i\f^j.
\ee
Note that under S-duality $\t \ra -\t\inv$, the prepotential
transforms as under a Legendre transformation. This will be the
general behaviour. 

The important difference with the nonsupersymmetric case is that the
coupling constant $\t$ is now a function of the scalar fields, that take
some constant asymptotic value $\f\in\cM$. So, instead of a space
parametrizing a family of QFTs, we are dealing with one gauge theory
(with given prepotential) that has a family of vacua with varying
coupling constant $\t(\f)$.

In the quadratic case the abelian duality group
$Sp(2g,\Z)$ can be trivially extended to the fermions and scalars. We
simply have to define dual scalar field as
\be
\f_{Di} = \t_{ij} \f^i,
\ee
similarly as we defined the dual field strength.

The question is now if and how these duality transformations get modified for
general, not necessarily quadratic prepotential. For convenience we
mainly restrict to the case of one $U(1)$ factor, so put $g=1$. We claim that
we still obtain a duality group $SL(2,\Z)$ that acts by
linear fraction transformations on the coupling constant
$\t$, but that the transformation of the
scalar fields is more involved. In fact, we have to define the general
dual scalar fields as \cite{seiberg-witten,rigid-special}
\be
\f_{D,i} =  \Pp \cF {\f^i}.
\ee
Note that this implies that
\be
\t_{ij} = \d_i\d_j\cF = \Pp {\f_{D,i}} {\f^j}.
\ee
We now claim that the vector
\be
\vectord {\f_D} \f
\ee
transforms as a doublet under $SL(2,\Z)$. Indeed, to show that this
is consistent, remark that the transformation rule
\be
\vectord {\f_D} \f \ra \twomatrixd abcd \vectord {\f_D} \f 
\ee
implies that 
\be
\t = \Pp {\f_{D}} {\f} \ra {a\t + b \over c\t + d}.
\ee

Let us now see in detail how the $T$ and $S$ transformations that
generate $SL(2,\Z)$ act.
For $T$ we find
\be
T: {\f_D \choose \f} \ra {\f_D + \f \choose \f}
\ee
from which we recover the behaviour of the prepotential
\be
T:\ \cF \ra \cF+\half \f^2.
\ee
For the $S$-duality we find
\be
S: {\f_D \choose \f} \ra {-\f \choose \f_D}
\ee
so that $\cF(\f) \ra \cF_D(\f_D)$ with $\cF_D'=\f$. That is, $\cF_D$ is
the Legendre transform of $\cF$
\be
S:\ \cF \ra Legendre(\cF).
\ee

\newsubsection{Rigid special geometry}

\newcommand{\nablar}{{\overline{\nabla}}}

The question is now: on what space $\cM$ take the scalar fields their
values and what is the right definition of the prepotential $\cF(\f)$?

In fact, this is a good point to review the general structure of the
moduli space of vacua of a theory with global (rigid) $N=2$
supersymmetry --- the so-called rigid special geometry
\cite{rigid-special}. We will then show that it reduces in local
coordinates to the structure we have discussed above.

Local special geometry, as we discussed in \S\S7.8,8.11,8.12, arises in
theories with local (gauged) supersymmetry, such as string theory and
its low energy limit supergravity. Rigid special geometry is a property
of theories with global (non-gauged) supersymmetry, such as the
four-dimensional gauge theories that we discuss here. The two structures
are very closely related, as we will see from the following definition:

{\bf Definition:} a {\it rigid special geometry} $(\cM,V,\W)$ is 
defined by the following ingredients:

(1) A complex, K\"ahler manifold $\cM$ (the moduli space) of 
dimension
\be
\dim_\C \cM=g.
\ee

(2) A holomorphic, flat $Sp(2g,\Z)$ vector bundle $V\ra \cM$, \ie a
representation $\pi_1(\cM)\ra Sp(2g,\Z)$ that produces a lattice bundle
$\G$ such that $V=\G\otimes\C$ can be given a holomorphic structure. We
denote the symplectic form on the fiber as $\eta$. We further write
$\nabla,\nablar$ for the $(1,0)$ and $(0,1)$ pieces of the connection
with 
\be
\nabla^2=\nablar^2=[\nabla,\nablar]=0.
\ee
Compatibility with the symplectic form gives
\be
d\eta(\a,\b)=\eta(\nabla\a,\b) + \eta(\a,\nabla\b).
\ee

(3) A holomorphic section $\W$ of $V$, 
\be
\nablar\W=0,
\ee
satisfying the Lagrangian condition
\be
\eta(\nabla\W, \nabla\W) = 0,
\ee 
(so $\nabla\W$ spans a Lagrangian subspace in $V$)
and the positivity constraint
\be
-i \eta(\nabla_i\W,\bar\nabla_\jbar\W) > 0.
\ee

(4) Finally, the K\"ahler form on the moduli space is given by
\be
K = -{i\over 2} \eta(\W,\overline\W).
\ee

Note that the holomorphic section $\W$ gives a natural complex polarization
\be
V = V^{(1,0)} \oplus V^{(0,1)},
\ee
where $V^{(1,0)}$ is the image of $\nabla\W$.

To make contact with the previous formulas, notice that if we choose a
(local) canonical $\Z$-basis $\a_i,\b^i$ ($i=1,\ldots,g$) of $V$ with
$\eta(\a_i,\b^j)=\delta_i{}^j$, we can write the preferred section $\W$ as
\be
\W=\f^i\a_i + \cF_i\b^i.
\ee
Since $\W$ is a holomorphic section,  we can locally 
use the components $\f^i$ as analytic coordinates on $\cM$
(special coordinates), so that the conjugate components $\cF_i$ become
(holomorphic) functions of $\f^i$. We then have
\be
\nabla_i\W = \a_i + \d_i\cF_j\b^j,
\ee
The condition $\eta(\nabla\W,\nabla\W)$ now gives
\be
\d_i\cF_j=\d_j\cF_i,
\ee
which is an integrability condition that tells us that there exists a
local holomorphic function $\cF(\f)$ with
\be
\cF_i = \Pp \cF {\f^i}.
\ee
We now define the period matrix as 
\be
\t_{ij}=\d_i\d_j\cF.
\ee
Note that in these special coordinates the K\"ahler metric is given by
\be
g_{i\jbar}=\d_i\d_\jbar K= \Im\t_{ij}>0,
\ee
also by definition.
From this form one can derive that the Riemann tensor satisfies  the
following identity, that is often taken as the starting point of rigid
special geometry:
\be
R_{i\ibar j\jbar} = - c_{ijk} \cbar_{\ibar\jbar\kbar} g^{k\kbar}.
\ee
Here the ``three-point functions'' are defined as
\be
c_{ijk} = \d_i\d_j\d_k\cF.
\ee
We note that these can be written more covariantly as
\be
c_{ijk} =\eta(\nabla_i\nabla_j\W,\nabla_k\W)=
-\eta(\nabla_i\W,\nabla_j\nabla_k\W).
\ee

Also note that if we have the further relation $\eta(\W,\nabla\W)=0$,
the prepotential is of the quadratic form
\be
\cF = \half \t_{ij} \f^i\f^j,
\ee
with constant matrix $\t_{ij}$.

\newsubsection{Families of abelian varieties}

There is a natural geometric realization of rigid special geometry
using abelian varieties.

Consider an abelian variety $X$ of complex dimension $g$ (see
\eg\cite{griffiths-harris}). That is, $X$ is of the form $W/\G$ with
$W\cong \C^g$ a complex vector space and $\G\cong \Z^{2g}$ a rank $2g$
lattice. Topologically, $X$ is a real $2g$ dimensional torus, 
\be
X\cong T^{2g}.
\ee
Such a complex torus is called an
abelian variety if it can be embedded in projective space.
There is a simple criterion that determines whether that is the case. 

We have two natural coordinates to use on $W$ and $X$. First we can
consider the natural complex coordinates $z^1,\ldots,z^g$ coming from
the isomorphism $V \cong \C^g$. Secondly, there are real coordinates
$x^1,\ldots,x^{2g}$ from the isomorphism $W \cong \G\otimes \R$. 
Here we have chosen a basis $e_a$ in the lattice $\G$.

Consider the space $H^1(X,\C)$. This is a $2g$ dimensional complex
vector space with a natural decomposition
\be
H^1(X,\C) = H^{1,0} \oplus H^{0,1}
\ee
in terms of the one-form $dz^i$ of type (1,0) and the conjugate forms
$d\zbar^i$ of type (0,1). 
The dual space $H_1(X,\Z)$ is canonically isomorphic with the lattice
$\G$. We use the same symbol to indicate a basis $e_a$ of one-cycles in 
$X$. Given the natural pairing between homology and cohomology, we can
define a $2g \times g$ period matrix
\be
\oint_{e_a} dz^i = \pi_a{}^i
\ee
Since there is a natural dual basis of the space $H^1(X,\Z)$ given by
the $2g$ real one-forms $dx^a$, whose periods satisfy
\be
\oint_{e_a} dx^b = \delta_a{}^b,
\ee
the period matrix $\pi_a{}^i$ can be seen to relate 
the two natural coordinates as 
\be
z^i = x^a  \pi_a{}^i .
\ee

According to Kodaira's embedding theorem we now have to look for a
two-form $\h$ (the K\"ahler form)
 of type $(1,1)$ that is closed, integer and positive. Now
it is easy to write down an integer two-form in the real coordinates.
Any form of the type
\be
\h = \half \h_{ab} dx^a \wedge dx^b
\ee
with integer antisymmetric matrix $\h_{ab} = - \h_{ba}\in \Z$ will do.
We can rewrite this form in terms of the complex coordinates and now
demand that  the pieces of type (2,0) and (0,2) vanish, and the
piece of type (1,1) is positive. These conditions translate immediately
in terms of the period matrix $\pi$ as the famous Riemann conditions. In
fact, under these conditions we can always choose a basis $(a_i,b^i)$ of
$\G$ such that 
\be
\eta(a_i,b^j) = \delta_i{}^j,
\ee
and so that the periods are of the form
\ba
\oint_{a_i} dz^j \is \delta_i{}^j, \nonu
\oint_{b^i} dz^j \is \t^{ij}.
\ea
In terms of the period matrix $\t_{ij}$ we then find that the Riemann
conditions read
\be
\t_{ij}=\t_{ji},\qquad \Im\t>0.
\ee

In more invariant notation, the Riemann conditions tell us
the decomposition of the $2g$ dimensional vector space
$V = \G \otimes \C$, given by
\be
V \cong H^1(X,\C) = H^{1,0} \oplus H^{0,1},
\ee
is a polarization for the Hodge form $\h$. That is, 
\be
\eta(\a,\b)=0, 
\ee
if $\a,\b$ both in $V^{(1,0)}$ or $V^{(0,1)}$. We furthermore have the
positivity condition
\be
-i\eta(\a,\bar\a)>0
\ee
An important and famous class of examples of abelian varieties are 
the Jacobians of curves
\be
X = H^1(\Sigma_g,\C)/H^1(\Sigma,\Z)
\ee
Since abelian varieties are completely characterized by their period matrix
$\tau$, the moduli space $\cA_g$ of (principally polarized) abelian varieties
has a simple description as the quotient of a homogeneous space
\be
\cA_g \cong Sp(2g,\Z)\backslash Sp(2g,\R)/U(g)
\ee
Here $Sp(2g,\Z)$ is the naturally symmetry group of the lattice $\G$
endowed with the symplectic form $\eta$.
Over this moduli space we have a canonical flat, holomorphic $Sp(2g,\Z)$
vector bundle $V$ (the Hodge bundle) with fiber
\be
V_X = H^1(X,\C)
\ee

How do we now get a solution of special geometry out of all this? We are
clearly looking for a $g$ dimensional family of abelian varieties
parametrized by the moduli space $\cM$. That is, we are looking for a map
\be
\cM \ra \cA_g
\ee
with the property that the holomorphic tangent bundle to $\cM$ can be
identified with the space $H^{1,0}(X,\C)$. For more along this lines see
\eg \cite{donagi-witten}. 

\newsubsection{BPS states}

Any system with extended supersymmetry has a
special subspace of the full Hilbert space $\cH$, the so-called BPS
space 
\be
\cH_{BPS}\subset \cH,
\ee
that consists of ``small'' supermultiplets. These BPS states play
a fundamental role in understanding duality symmetries.

More precisely, suppose we
are dealing with some supersymmetry algebra with a set of $n$
supercharges $Q^\a$ ($n$ will always be even.) The Hilbert space will
decompose in irreducible representations of this algebra. Since the
supersymmetry algebra will be of the general form 
\be
\{Q^\a,Q^\b\}=\w^{\a\b}_i K^i,
\ee
with
\be
 [Q^\a,K^i]=0,\qquad [K^i,K^j]=0,
\ee
where the $K^i$ are some set of bosonic charges, consisting of the
translation operator $P_\m$ and some extra set of central charges, usually
denoted as $Z$. The symmetric bilinear forms $\w_i$ will always be
non-degenerate. Therefore, when we consider a  representation
where the operators $K^i$ have fixed generic eigenvalues $k^i$, so that
the total bilinear form 
\be
\w=\w_i k^i
\ee
is non-degenerate, we are essentially dealing with a representation
of a $n$-dimensional Clifford algebra.
The dimension of the representation will therefore be $2^{n/2}$.
However, for special values of the charges $k^i$, it might be the case
that the bilinear form $\w$ becomes accidentally degenerate. In that
case there are certain linear combinations of supercharges that
annihilate the representation. This representation is then called a BPS
representation. If it satisfies the conditions 
\be
\e_\a Q^\a |\bps\>=0,
\ee
for $m$ independent spinors $\e$, the rank of the Clifford algebra
will be $n-m$ and therefore the dimension of the 
representation will be $2^{(n-m)/2}$.

Let us now see what the role is of BPS states in models with global
$N=2$ supersymmetry. The four-dimensional
$N=2$ supersymmetry algebra is given as
\be
\{Q_\a^I,Q_{\dot\b,J}\}=\delta^{I}_{J} \g^\m_{\a\dot\b} P_\m,
\ee
However, we also have to consider the other commutators, which can take
the most general form
\ba
\{Q_\a^I,Q_{\b}^J\}=\e^{IJ}\e_{\a\b} Z,\nonu
\{\Qbar_{\dot\a,I},\Qbar_{\dot\b,J}\}=\e_{IJ}\e_{\dot\a\dot\b} \Zbar,
\ea
with complex central charge $Z\in\C$ satisfying
\be
[Z,Q]=[Z,\Qbar]=[Z,P]=0.
\ee 
For a general state with eigenvalues $P_\mu=M\delta_{\m,0}$ and $Z$, one easily
derives an upper bound for the mass \cite{olive-witten}
\be
M^2\geq |Z|^2.
\ee
In fact, states with $M^2=|Z|^2$ are precisely the small BPS 
representation (of dimension 2 instead of 4). 

The proof this statement follows a very general argument.
For a given multiplet
the BPS condition can be written as (we suppress the indices)
\be
\left(\e Q + \bar\e \Qbar\right)|\bps\> = 0 .
\ee 
This condition holds with fixed $\e,\ebar$ for all states in the
BPS multiplet. If we take the commutator with the supercharges, we derive
the conditions on the supersymmetry parameters of the form
\ba
\pslash \e + Z^\dagger \bar\e \is 0,\nonu 
Z \e + \pslash \bar\e        \is 0 .
\ea 
Combining these equations with the mass shell condition $p^2 =M^2$,
one deduces that $M^2$ coincides with the highest
eigenvalue of $ZZ^\dagger$ and $ Z^\dagger\!Z$,
with $\e$ and $\bar\e$ being the corresponding eigenvectors,
\ba
\label{ev}
(Z Z^\dagger)\e \is M^2 \e \nonu (Z^\dagger
\!Z )\ebar \is M^2 \ebar.
\ea 
This determines the BPS masses completely in terms of the central charge
$Z$. In the $N=2$ context we find that $Z^\dagger Z= Z Z^\dagger = |Z|^2
\id$, so that the above BPS mass formula is obtained immediately.

In the case of an $N=2$ abelian gauge theory with coupling constant
$\tau_{ij}$ the central charge $Z$ takes a simple form. If a state
carries charge $q=(e,m)$, with electric charges $e_i$ and magnetic charges 
$m^i$, the central charge $Z(q)$ takes the form \cite{olive-witten}
\be
Z(q) = \f^i(e_i - \t_{ij}m^j)
\ee
This result can be generalized to the case where one has non-trivial
moduli dependence $\tau_{ij}(\f)$.
Without proof, we mention that in that case the
general mass formula takes the following form (see \eg \cite{jeff})
\be
Z(q)=\eta(q,\W)=\f^i e_i - \cF_i m^i.
\label{bps-mass}
\ee
We note that this expression is explicitly $Sp(2g,\Z)$ invariant.

\newsubsection{Non-abelian $N=2$ gauge theory}

Our story has now come to the seminal work of Seiberg and Witten
\cite{seiberg-witten} on the non-abelian $N=2$ SYM model. This work has
been reviewed many times \cite{sw-review,jeff}, and we refer to these
references (and the original papers!) for more details.

So we turn to $N=2$ Yang-Mills theory with a non-abelian gauge group
$G$ and Lie algebra $g$. The fundamental multiplet is again of the form
\be
(\cA_\mu,\L_\a^I,\overline\L_{\dot\a}^I,\F),
\ee
where all fields take
value in the adjoint bundle $g$. The important difference with the
abelian case is that the action now contains a potential term of the
scalars $\F\in g\otimes \C$ of the form
\be
V(\F)=\int_M \Tr [\F,\F^\dagger]^2.
\ee
This potential vanishes if $\F$ is restricted to the complexified Cartan
torus $t\otimes \C$. In the case $G=SU(2)$, to which we will restrict
from now on, $\F$ is of the form
\be
\F = \twomatrixd {\f} 0 0 {-\f},\qquad \f \in\C.
\ee
Global action of $SU(2)$ gives the further identification 
\be
\f \lra -\f
\ee
(in the general case $\f\in t\otimes \C/W$, with $W$ the Weyl group), so that
the {\it classical} moduli space of vacua is given by (here we added the
point at infinity)
\be
\cM_{cl} \cong \P^1
\ee
with good local coordinate $u=\f^2=\half \Tr \F^2$. Note that we expect
singularities at $u=0$ and $u=\infty$.

What is quantum moduli space? This is the question Seiberg and Witten
posed and answered. We first notice that a question about the vacuum
structure refers to long distances, \ie infrared physics. So we are only
interested in the massless fields, whose correlation functions do not
decay exponentially in distance. Since the ``Higgs field'' $\F$ will in
general have a non-zero expectation value with $u=\half \<\Tr\F^2\>\neq
0$, by the Higgs phenomenon all fields will acquire masses except for the
component of the multiplet in the direction of $\F$. That is, we can
try to define a $U(1)$ gauge field
\be
A=\Tr(\widehat\F \cA),\qquad \widehat\F=\F/|\F|.
\ee
Of course, this definition does not make sense at the zeroes of $\F$.
Around such a zero the unit vector field $\widehat\F$ defines a possibly
non-trivial element of $\pi_2(S^2)$ with winding number $m$, which
represents an obstruction to deforming the zero away. These
configurations are characterized by the property that for a two-sphere
in $\R^3$ surrounding such a zero 
\be
\int_{S^2} dA = 2\pi m
\ee
and therefore the singularity carries possible magnetic charges. Indeed,
these are the famous 't Hooft-Polyakov monopoles \cite{monopole}.
In the long-distance limit we can think of them as point-like objects.
In order to correctly represent all $SU(2)$ gauge field configurations,
we therefore have to add a gas of these monopoles. This translates into
a second quantized monopole field. In fact, it is a long-standing
conjecture of 't Hooft that we can understand the vacuum structure of a
non-abelian gauge theory in terms of abelian gauge fields and magnetic
monopoles \cite{thooft-confinement}. This deep point of view has been
brilliantly confirmed by the work of Seiberg and Witten.
In general the monopoles will have
positive masses, so in the infra-red limit (where we
scale the metric on the volume of the four-manifold to infinity) it is
not necessary to include the monopole fields in the action.

As we sketched above, after integrating out the massive fields, we are
left with an effective abelian theory. Such a model is characterized (to
quadratic order in derivatives) by a prepotential $\cF(\f)$. The 
coupling constant $\t$ is given as its second derivative $\t=\cF''.$
We further have a $Sp(2,\Z)  \cong SL(2,\Z)$ vector bundle 
$V=\G\otimes \C$ with a holomorphic section
\be
\W={\f_{D} \choose \f},\qquad \f_D=\Pp\cF\f.
\ee
The electric and magnetic charges are, for given value of $\f$, given by
\be
q=\vectord{e}{ m}={1\over 2\pi} \int_{S^2} \vectord{F_D}{F}\in \G.
\ee
We now present the derivation of the prepotential $\cF$.

\newsubsection{The Seiberg-Witten solution}

Starting point for the SW solution is the behaviour of the prepotential
around the point $u=\infty$, where the leading contribution can be computed in perturbation
theory. The coupling constant $\t=\cF''$ has an expansion
\be
\t=\t_0 + {i\over \pi} \log (\f/\L) + \sum_n a_n (\L/\f)^{4n}
\ee
where the first term $\t_0$ is the classical contribution, the second
term represents the one-loop contribution (this is the only perturbative
correction), whereas the other terms indicate instanton corrections,
that are in general difficult to compute exactly. The logarithmic piece
gets us started, since we derive the local monodromy of the $SL(2,\Z)$
bundle around $\infty$. If $u \ra e^{2\pi i}u$ we see that
\be
\vectord {\f_D}\f =\vectord{\cF'}\f \ra \vectord{-\v_D+2\v}{-
\f}=\M_\infty \vectord {\f_D}\f.
\ee
With this monodromy 
\be
\M_\infty = \twomatrixd {-1} 2 0 {-1}
\ee
we can try to extend the vector bundle to $\P^1 - \{0,\infty\}$ by
having the opposite monodromy around the origin, where we expect our
second singularity. However, we also have to satisfy the positivity
constraint $\Im\t>0$. It turns out that this is impossible without
introducing a third singularity. Indeed, the SW solution assumes the
minimal amount of a total of three singularities, say at $u=\infty,1,$
and $-1$, with monodromy matrices
\ba
\M_1 \is \twomatrixd 1 0 {-2} 1,\nonu
\M_{-1} \is  \twomatrixd {-1} 2 {-2} 3.
\ea
These matrices generate the $SL(2,\Z)$ subgroup
\be
\G(2) = \{\g\in SL(2,\Z);\ \g\equiv \twomatrix 1 001\ ({\rm mod})\ 2\}.
\ee
The quantum moduli space is given by the Ansatz
\be
\cM_{qu} = \H/\G(2),
\ee
which is ``almost'' the moduli space of elliptic curves. In fact, $\cM$
parametrizes elliptic curves $\Sigma$ of the particular form
\be
y^2 = (x^2-1)(x-u).
\ee
The bundle $V$ is now the bundle with fiber over $\Sigma\in\cM$ given by
\be
V_\Sigma =H^1(\Sigma,\Z)\otimes\C.
\ee
So the charge lattice $\G$ is identified with the first homology group of
the elliptic curve $\G=H_1(\Sigma,\Z)$.

To such a surface we can associate a natural modulus $\t$ (of the elliptic curve)
that always satisfies the positivity condition $\Im\t>0$.
Note that the modulus can be expressed as a ratio of periods
\be
\t={ \oint_B \w\over \oint_A  \w} ,
\label{tau}
\ee
with $(A,B)$ a canonical basis of $H_1(\Sigma)$ and $\w=dz=dx+\t dy$ the
unique holomorphic one-form.

Now that we are given the bundle $V\ra\cM$, we should ask what  the
prepotential is. There is an obvious guess. Let's pick some one-form $\l$
on $\Sigma$. We can then consider the following Ansatz
\ba
\f \is \oint_A\l\nonu
\f_D \is \oint_B \l.
\ea
By definition this vector will transform correctly under $SL(2,\Z)$.
However, special geometry imposed the further constraint that
\be
\t = \Pp {\f_D}\f = { \oint_B \Pp \l u \over \oint_A  \Pp \l u}.
\ee
If we compare with (\ref{tau}) we see that necessarily
\be
\Pp \l u = c \w.
\ee
There is a (unique up to scalars) meromorphic one-form with this
property. This then completes the solution.

\newsubsection{Physical interpretation of the singularities}

The two new singularities in $\cM$ have a beautiful physical interpretation.
Recall the formula for the BPS formula
(\ref{bps-mass}) for a dyon with charges $(e,m)$
in a  model with $N=2$ supersymmetry:
\be
M = |Z| = |\f e - \f_D m|
\ee
For fixed charges this mass varies with the moduli. Now, if at some
point in the moduli space a linear combination $\f e - \f_D m$,
with $e,m\in\Z$ vanishes, this has important physical consequences. It
implies that the dyon with these particular charges will have zero mass.
But the SW solution was based on the premises that the abelian (super)
gauge field was the only massless field! Therefore, at these points in
$\cM$ we expect our formalism to break down. This breakdown will
manifest itself in a singularity in the solution. There are three such
points $u=1,- 1,\infty$. 

At $u=\infty$ the situation is clear: here the non-abelian
gauge fields with $(e,m)=(\pm1,0)$
become massless again, restoring the $SU(2)$ symmetry.

At $u=1$ we have a very different situation. Here the monodromy $\M_1$
forces the component $\f_D\ra 0$. Thus magnetic monopoles with
$(e,m)=(0,1)$ get zero mass at this point. In the 't Hooft philosophy we
had presumably taken into account the monopole fields, when we arrived
at our effective action encoded by the prepotential $\cF$. The fact that
these fields actually become massless declares this procedure after the
fact invalid. One should thus undo this integration procedure and
reintroduce the monopole fields in the path-integral. This removes the
singularity. It was simply the result of an overambitious simplification
of the dynamics.

It is very important that it is not a physical singularity. There is no
problem with having singularities in a moduli space that parametrizes a
family of classical field configurations. Neither is there a problem if
the moduli space parametrizes inequivalent quantum field theories. One
simply decides not to consider such a model. (A good example of such a
singularity is a sigma model on a singular target space.) However, one
cannot allow singularities in a moduli space that parametrizes
inequivalent vacua of a single theory. There is no way in which we can
prevent the theory of exploring the region in moduli space that contains
the singularity. (At least, if it is at finite distance, which is the
case in all situations that we will discuss.) There must be a physical
mechanism that ``explains'' the singularity, such as the massless
monopole in the SW point $u=1$.

One can reproduce the monodromy matrix $\M_1$ from this physical
argument --- an important consistency check on the solution. First we
use $S$-duality to go to a dual description in which we use the
expectation value of the dual scalar field $\f_D$ as the local coordinate
around $u=1$ and have a corresponding dual gauge field $A_D$. The
electric charges of this dual gauge field are the magnetic charges of the
original gauge field $A$, that we recover in the perturbative expansion
around $u=\infty$. In these new dual variables we have a dual
prepotential $\cF_D(\f_D)$ in which we can express $\f=\cF_D'$. We can
now make a {\it perturbative} calculation of the dual coupling constant
$\t_D=\cF_D''$ by considering the one-loop contribution of the nearly
massless monopoles. They give 
\be
\t_D \sim -{i\over \pi} \log (\f_D/\L)
\ee
from which we deduce that $\f \sim -{i\over \pi} \f_D\log \f_D$, so that 
the monodromy for $\f_D \ra e^{2\pi i} \f_D$ is given by
\be
\f \ra \f - 2 \f_D.
\ee
This gives the required monodromy matrix
\be
\M_1 = \twomatrixd 1 0 {-2} 1 .
\ee
In a similar fashion the massless dyon of charge $(1,1)$ produces the 
monodromy matrix $\M_{-1}$ around the second singularity $u=-1$.

There is a nice geometric interpretation of this monodromy in terms of
the elliptic curve. At the three singularities we have $\t\ra
i\infty$, so that we are dealing with a singular elliptic curve of
genus one. We can picture this singular curve by shrinking an homology
cycle to zero. For example, in the case $u=1$ the $B$-cycle shrinks to
zero, so that also
\be
\f_D = \oint_B \l \ra 0.
\ee
By doing a Dehn twist around this cycle, we recover the monodromy
matrix, precisely as in our discussion in section \S9.6.

\newsubsection{Implications for four-manifold invariants}

Somewhat as an aside, we can ask what the implications of the SW
solution are for four-dimensional topology \cite{witten-monopole}. 

We have seen how a twisted
version of $N=2$ super Yang-Mills with gauge group $SU(2)$ led to
Donaldson's manifold invariants by localization to the moduli space of
non-abelian instantons. In this argument, we applied more or less
semi-classical arguments. These arguments can be made valid if the gauge
coupling constant, that controls quantum corrections, is assumed to be
small. Thanks to the ``running'' of coupling constants in quantum gauge
theories, that makes the coupling constant a non-trivial function of the
scale, and thanks to asymptotic freedom, that tells us that the coupling
will actually decrease if we make this scale very small in non-abelian
$N=2$ SYM theory, we can conclude that this semi-classical
approximation is appropriate for small volumes of the four-manifold. But
the Donaldson invariants are topological invariants, they do not depend
on the choice of Riemannian metric (for $b^+>1$). We can therefore take
this metric to be as small as we want, and can thus correctly apply
semi-classical reasoning. This is the usual philosophy behind the
application of non-abelian gauge theory to four-dimensional differential
topology.

However, we can also take a completely contrary point of view and
consider instead the infrared limit of large volume\footnote{This point
of view is actually more intuitive. If we study topology we are
typically interested in global, large scale features. At least, this is
the colloquial use of the expression ``topological'' in physics.}. Then
we believe that only the massless degrees of freedom are relevant, since
all other fields have exponentially decaying interactions, which in the
IR limit become point-like. Of course, these massless fields are
described by an {\it effective} Lagrangian that takes correctly into
account the quantum loops of all the massive fields that we have
integrated out in the process. In this case, such an effective field
theory will be an abelian gauge theory of the type we have been
considering in \S11.4. 

If we now apply our localization formulas of \S10.2 to the BRST fixed
points of this effective abelian theory, we find that this model
localizes to the solutions of the abelian instanton equation
\be
F^+=0.
\ee
However, as we discussed already in section \S10.1, for an abelian gauge
field the curvature is necessarily quantized in integer fluxes,
$[F/2\pi]\in H^2(M,\Z)$. But $F$ is also ASD, so $F=F^-$. For a generic
metric and thus a generic positioning of the subspace $H^2_-\subset
H^2$, the intersection with the integer lattice $H^2(M,\Z)$ will only be
non-zero if $H^2_+=0$ \ie $b_2^+=0$, which one usually assumes not to be the
case. We conclude therefore that for the generic four-manifold and
generic metric the abelian BRST fixed point set is empty.

There is however one {\it caveat.} If we compute the partition function
on a {\it compact} four-manifold $M$, as we will certainly do in the
context of manifold invariants, there is no moduli space of disjunct
vacua, as labeled by the quantum moduli space $\cM_{qu}$. In fact, as we
already have explained in \S10.5, in this case one is forced to integrate
over the constant values of the scalar fields too. We thus have to
consider our model for all values of the modulus $u\in\P^1$
and integrate over $u$. Away from
the singularities this is not a problem, but at $u=\pm1$ we get a
modified equation, because we also have to take into account the
massless monopole and dyon fields respectively that we cannot ignore
in the IR limit.

What is the form of their contribution? If we use an $S$-dual
formulation, the massless fields are simply the field content of $N=2$
QED, \ie supersymmetric Maxwell theory with an additional massless
fermion field (a neutrino). Independently of the work of Seiberg and
Witten, one could have considered the twisted version of QED as a
generator of four-manifold invariants after Witten's paper
\cite{witten-tft} in 1988 showed how $N=2$ models can be twisted to
produce topological field theories. However, somehow the believe was
that only non-abelian Yang-Mills theories, that due to asymptotic
freedom stand a much better change to exist mathematically, would be
able to produce non-trivial differential geometry invariants! (Indeed,
it is a beautiful fact that both in physics and in mathematics
dimension four stands out as giving the richest phenomena.)

The extra $N=2$ matter multiplet in QED takes the form
$(M^I,\psi_\a,\psibar_{\dot\a})$, with $\psi$ the usual fermion
field and $M^I$ a doublet of two complex scalar fields (sneutrinos). The
twisting modifies the
spin of these fields as
\ba
M^I & \gives & M_\a \nonu
\psi_\a & \gives & \psi_\a \nonu
\psi_{\dot\a} & \gives & \rho_{\dot\a}
\ea
We obtain a commuting bosonic variable $M$ that transforms as a
spinor! The BRST transformations are
\ba
\delta M_\a \is \psi_\a, \nonu
\delta \rho_{\dot\a} \is (\dslash M)_{\dot\a}.
\ea
Finally, because of the coupling of the matter multiplet to the gauge
multiplet, the transformation for the field $\rho_{\mu\nu}^+$ in
(\ref{brst}) gets modified to
\be
\delta \rho_{\mu\nu}^+ = F_{\mu\nu}^+ - (\bar M M)_{\mu\nu}^+.
\ee
Looking at the transformation rules of the fields
$(\rho_{\mu\nu}^+,\rho_\adot)$ we see that in this case the section is
given by
\be
s=(F^+-(\Mbar M)^+,DM)
\ee
Its zero locus is the Seiberg-Witten moduli space described in 
\cite{witten-monopole} and discussed in the
lectures by Garcia. See also \cite{sw-math}.
 
\newsection{String vacua}

This finishes our introduction!

We now arrive at the last stretch of our journey --- the space-time
physics of string theory. The most important new ingredient is of course
the dynamical metric (together with its supersymmetric partners). String
theory is a theory of quantum gravity and string perturbation theory can
be used to compute weak-coupling quantum gravity effects. These
perturbative computations follow the pattern described in the first
half of the course. One picks a CFT describing the string vacuum around which
we can do our perturbative expansion in terms of Riemann surfaces. 
The expansion coefficients of the $n$-point scattering amplitudes
\be 
A \sim \sum_g A_g \l^{2g-2} 
\ee 
are then computed as integrals over the moduli space $\cM_{g,n}$. However,
this is only part of the total picture. 

First of all, the series expansion is only asymptotic, it does not
converge\footnote{See \cite{gross-periwal}. Indeed, one can estimate
that string perturbation theory generates terms of order $2g!$ at $g$
loops \cite{shenker}.}, so additional nonperturbative information is
needed to fix the final answer. These nonperturbative phenomena are not
necessarily completely ``stringy'' in nature. They can sometimes be
understood from a low-energy point of view, where we forget about the
infinite set of massive states. In fact, the issues that we will
discuss, such as dualities, moduli spaces and BPS states, can be even
understood without studying much dynamical issues in quantum gravity.

\newsubsection{Perturbative string theories}

At present we know of five different perturbative string theories, \ie
theories that allow a consistent perturbation expansion in terms of
Riemann surfaces, summarized in
\tabelplus{
\renewcommand{\arraystretch}{1.5}
\begin{tabular}{|c|c|c|c|c|}
\hline
\strut {\it perturbative} & {\it world-sheet} & {\it space-time} & 
{\it space-time} & {\it gauge group} \\[-2mm]
{\it vacuum} & {\it geometry} & {\it dimension} & 
{\it susy} & \\ \hline\hline
\strut I & unoriented, open & (9,1) & $N=1$ & $Spin(32)/\Z_2$ \\
\hline
\strut IIA & super, closed & (9,1)& $N=(1,1)$ & $U(1)$ \\
\hline
\strut IIB & super, closed & (9,1)& $N=(2,0)$ & ---  \\
\hline
\strut HO & heterotic, closed & (9,1) & $N=1$ & $Spin(32)/\Z_2$ \\
\hline
\strut HE & heterotic, closed & (9,1) & $N=1$ & $E_8\times E_8$ \\
\hline
\end{tabular}
\renewcommand{\arraystretch}{1.0}
}{The five known perturbative string theories.}
We apologize that we didn't clearly explained the precise world-sheet
formulations of these theories. We did explain the bosonic string, but
there are a few extra subtleties when we are dealing with super Riemann
surfaces that we didn't want to get into.

Recently, two more string vacua have been discovered. These theories,
called M-theory \cite{witten-duality,schwarz}
and F-theory \cite{vafa-f}
do not allow a perturbative formulation in
terms of surfaces. In fact, the geometrical objects seem to involve $(2,1)$
dimensional super membranes and $(2,2)$ signature objects that are even less
well understood. (See \cite{banks} for a conjectural definition
of M-theory.) For completeness we add these two new models in
\tabelplus{
\renewcommand{\arraystretch}{1.5}
\begin{tabular}{|c||c|c|c|c|}
\hline
\strut {\it perturbative} & {\it world-volume} & {\it space-time} & 
{\it space-time} & {\it gauge group} \\[-2mm]
{\it vacuum} & {\it geometry} & {\it dimension} & 
{\it susy} & \\ \hline\hline
\strut M & super membranes? & (10,1) & $N=1$ & --- \\
\hline
\strut F & SD three-branes? & (10,2) & $N=1$ & --- \\
\hline
\end{tabular}
\renewcommand{\arraystretch}{1.0}
}{The new string vacua M-theory and F-theory are less well-understood.}

The insight that all perturbative string theories are different
expansion of one theory is now known as {\it string duality.} The
precise way in which the different models are related is too involved to
be explained in this set of lectures (but see \eg
\cite{string-duality-reviews,polch-review}. It is one of the amazing new
insights following from string duality that these ``theories'' are all
expansions of one and the same theory around different points in the
moduli space of vacua. Indeed we have the beautiful picture of the
moduli space \figuurplus{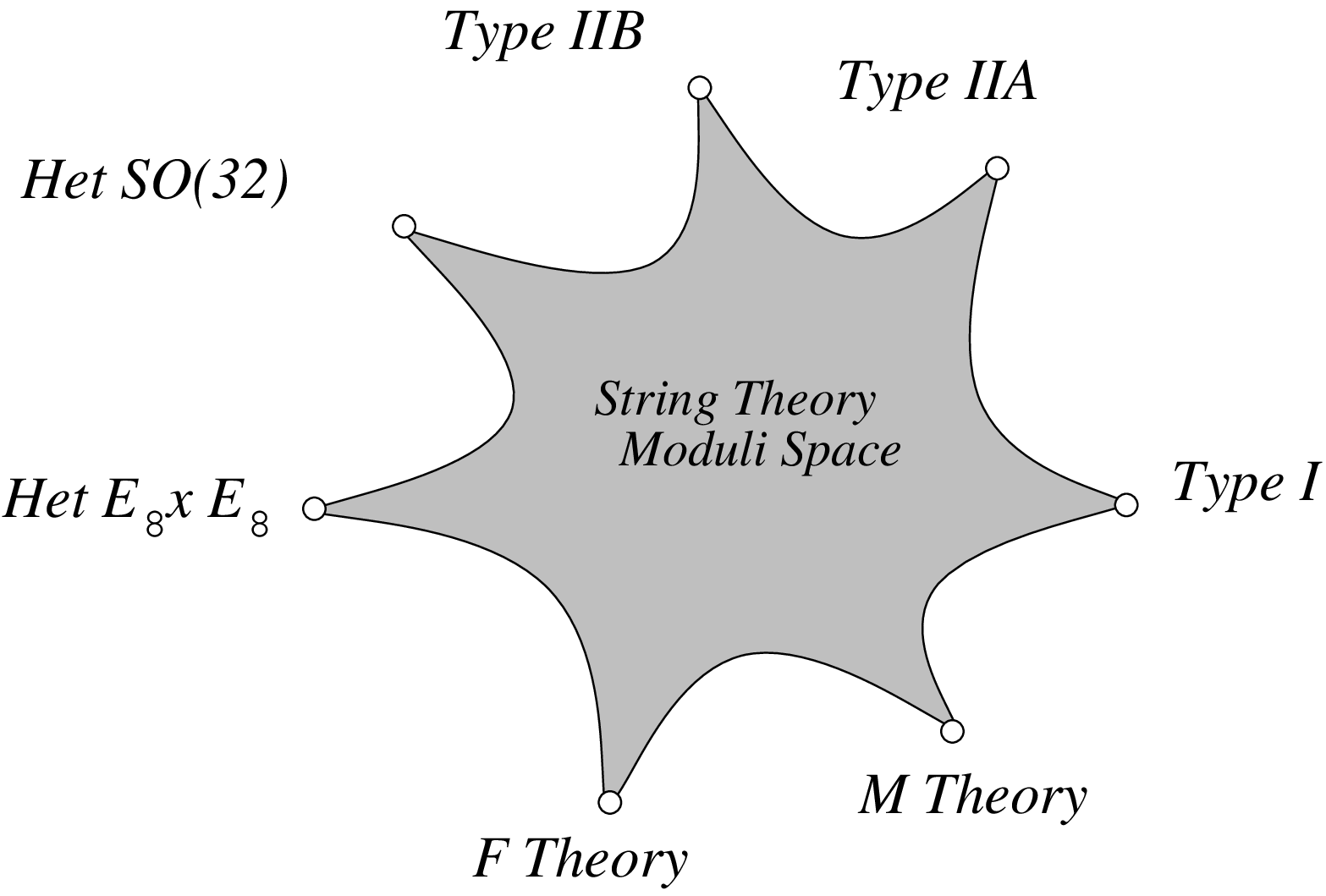}{8cm}{The moduli space of string vacua.
Various perturbative theories are recovered in expansions around the
cusps.} given in \cite{polch-review} where we have various ``cusps''
around which the amplitudes allow systematic geometric expansions.

We will say briefly a few words about the world-sheet formulation of the
perturbative string theories, although we do not have time to discuss
this in full detail (see \eg \cite{gsw} or more recent lecture notes
such as \cite{oguri-tasi}). Of course, all models follow roughly the line
discussed for the bosonic string in \S8.3. We will not discuss the open
string of Type I. According to the modern point of view it is by an
S-duality related to the heterotic string. For the remaining strings we
have the following left-moving and right-moving degrees of freedom
$$
\renewcommand{\arraystretch}{1.5}
\begin{tabular}{|c||c|c|c|}
\hline
\strut  \it string & \it Left & \it Right & \\
\hline\hline
\strut bosonic & $x^\m$ & $x^\m$ & $\m=1,\ldots,26$ \\
\hline
\strut IIA \& B & $x^\m$ & $x^\m$ & $\m=1,\ldots,10$ \\
\strut & $\psi^\m$ & $\psi^\m$ & \\ 
\hline
\strut heterotic & $x^\m$ & $x^\m$ & $\m=1,\ldots,10$ \\
\strut & $\psi^\m$ & $x^I$ & $I=1,\ldots,16$ \\ 
\hline
\end{tabular}
\renewcommand{\arraystretch}{1.0}
$$ 
We add the bosonic string for good measure, although in the end it
is inconsistent due to the existence of a tachyon.  Here the 16
internal right-moving bosonic fields $x^I$ of the heterotic string
take their value in the torus defined by modding out the lattice
$E_8\oplus E_8$ or $D_{16}$ giving the HE and HO theory respectively.

\newsubsection{IIA or IIB}

The distinction between the two type II theories has to do with
space-time chirality. It will be important in the following, so we
riefly review it. In the IIA theory we have supersymmetry generators
$Q_\a,Q_{\dot\a}$ of both 10-dimensional chiralities, whereas in the
type IIB theory the two supercharges $Q^1_\a,Q^2_\a$ carry the same
chirality. In terms of the world-sheet CFT description this distinction
is made in the way the GSO projection is defined. 

Recall that the 10 Majorana fermion fields $\psi^\m$ are most
conveniently quantized by combining them in pairs into 5 Dirac fermions
\be
\psi=\psi^1 +i\psi^2,\qquad \psi^*=\psi^1 - i\psi^2.
\ee
These fields have an expansion of the form
\be
\psi(z) = \sum_{n\in\Z+\e} \psi_n z^{-n-{1\over 2}},
\ee
where the spin structure $\e=0,\half$  is determined by the
NS or R boundary condition
\be
\psi(e^{2\pi i} z) = \option {\psi(z)}{$\e=\half$\ \ (NS)}
{-\psi(z)}{$\e=0$\ \ (R)}
\ee
In the NS sector we have a well-defined vacuum state with
\be
\psi_n |0\>=\psi^*_n |0\>=0,\qquad n>0.
\ee
The Fock space $\cF_{NS}$ built on this vacuum has a natural $\Z_2$
grading by the fermion number $J_0=0,1$ modulo 2,
\be
\cF_{NS} = \cF^0_{NS} \oplus \cF^1_{NS}
\ee
The GSO projection, that is an important ingredient in the definition of
the superstring, tells us to take the odd piece $\cF^1_{NS}$.

For the Ramond sector, the modes $\psi_n$ are integer valued. This means
that we have zero modes $\psi_0,\psi^*_0$ satisfying a Clifford algebra
\be
\{\psi_0,\psi_0^*\}=1
\ee
Therefore the vacuum $|\s_\pm\>$ will be two-fold degenerate. It is a spinor of
$SO(2)$, and it satisfies
\be
\psi_n |\s_\pm\>=\psi^*_n |\s_\pm\>=0,\qquad n>0.
\ee
The states $|\s_\pm\>$ have conformal dimension $h={1\over 8}$ and
Fermi number $\pm\half$. This can be seen by bosonization. Introduce a
scalar field $\v$ with $-i\d\v=\psi\psi^*$ so that 
\be
\psi = e^{i\v},\qquad \psi^* = e^{-i\v}.
\ee
The spin fields $\s_\pm$, that create the R ground states out of the 
NS vacuum now take the form
\be
\s_\pm = e^{\pm{i\over 2}\v}.
\ee
The Ramond Fock space is again graded
\be
\cF_{R} = \cF^+_{R} \oplus \cF^-_{R},
\ee
where the total Fermi number is $J_0 = \pm \half$ mod 2. The GSO
projection instructs us to take either $\cF^+_{R}$ or $\cF^-_{R}$. It is
hard to distinguish the two, since there is no intrinsic way to
determine the absolute sign of the fermion charge $J_0$. 

For the full multiplet of 10 fermion fields, the R ground states form a
rank $2^5=32$ spinor of $Spin(10)$. In fact, after bosonization they
take the form
\be
\s_\a = e^{i\a\cdot\v},\qquad 
\a =(\pm\half,\pm\half,\pm\half,\pm\half,\pm\half).
\ee
The spinor splits in two chiralities distinguished by
\be
\sum_i \a_i = \pm\half \ \rm mod\ \Z.
\ee
The GSO projection will project on one chirality. Again, which chirality
is a matter of taste, they will be permuted by a parity transformation.
But there will be a marked difference when we combine left-movers and
right-movers. The GSO projection is performed separately in both
sectors, and we have a two-fold ambiguity in each sector. Of the total of
four possibilities we can identify two by flipping the sign of the
overall fermi number $F=J_0+\Jbar_0$. 

The two remaining choices give
physically distinct theories, the type IIA and type IIB string theories.
As we mentioned already, they differ in the relative sign in the
chiralities of the two supercharges. In a formula we obtain the
following contribution to the Hilbert space
\ba
{\rm IIA:} & & \left(\cF_{NS}^1 \oplus \cF_R^+\right) \otimes
\left(\overline{\cF}_{NS}^1 \oplus \overline{\cF}_R^-\right),\nonu
{\rm IIB:} & & \left(\cF_{NS}^1 \oplus \cF_R^+\right) \otimes
\left(\overline{\cF}_{NS}^1 \oplus \overline{\cF}_R^+\right).
\ea
This gives four different sectors in the Hilbert space
\ba
& {\rm (NS,NS) :}  & \qquad \hbox{\it boson} \nonu
& {\rm (NS,R) :}  &  \qquad \hbox{\it fermion} \nonu
& {\rm (R,NS)  :} &  \qquad \hbox{\it fermion} \nonu
& {\rm (R,R) :}  &  \qquad \hbox{\it boson} \nonu
\ea
The (NS,NS) sector contains the familiar massless bosonic fields: the metric
$G_{\m\n}$, the two-form field $B_{\m\n}$ and the dilaton field $\F$,
all corresponding to states of the form
\be
\psi^\m_{-{1\over 2}} \psi^\n_{-{1\over 2}} |p\>.
\ee

The massless bosonic fields in the (R,R) sector have long been neglected, but
they play a crucial role in recent developments. By definition they are
bispinors. However, we can further decompose them in terms of irreducible
representations of $Spin(10)$. If $S^\pm$ denote the two chiral spinor
representations, and $\L^k$ the representation of $k$-forms, we have
\ba
{\rm IIA:} & & S^+ \otimes S^- \cong \L^0 \oplus \L^2 \oplus \L^4 \nonu
{\rm IIB:} & & S^+ \otimes S^+ \cong \L^1 \oplus \L^3 \oplus \L^5_+ 
\ea
Here $\L^5_+$ indicates the self-dual 5-forms.
These relations are derived in the familiar way, we write a $k$-form as a 
bispinor using the Dirac matrices
\ba
F_{\a\dot\b} \is  \sum_{k\ even}
F^{(k)}_{\m_1\ldots\m_k}\left(\g^{\m_1}\cdots
\g^{\m_k}\right)_{\a\dot\b}\nonu
F_{\a\b} \is  \sum_{k\ odd}
F^{(k)}_{\m_1\ldots\m_k}\left(\g^{\m_1}\cdots \g^{\m_k}\right)_{\a\b}
\ea
Here the total differential form
\be
F = \sum_{k\ odd/even} F^{(k)},\qquad
F^{(k)} \in \W^k(\R^{9,1})
\ee
satisfies the self-duality condition $ *F = F. $

The corresponding vertex operators couple to $k$-form field strengths.
This we can see by looking at the equation of motion. The bispinor
$F_{\a\b}$ satisfies two Dirac equations, one on the left and one
on the right.
In terms of differential forms this gives the equations
\be
dF=d^*F=0,
\ee
which we recognize as the Bianchi identity and the Maxwell equation.

\newsubsection{D-branes}

Summarizing, in the type II superstring we have a collection
of generalized gauge fields $A^{(k)}$, which are $k$-forms, with $k$ odd
and even in the IIA and IIB case respectively. 

Now we are familiar with one such generalized
gauge field, the two-form $B$ field that we find in the (NS,NS) sector.
The string carries a charge for this field, as can be seen from the
sima-model coupling
\be
\int d^2\!z\,\half B_{\m\n}(x)\d {x^\m} \dbar {x^\n} = 
\int_\Sigma B.
\ee
This minimal coupling is complete analogous to the way a particle with a one-dimensional
world-line $C$ couples with charge $q$ to a one-form gauge field $A$
\be
q \int d\t\, A_\m(x) \dot x^\m = q\int_C A.
\ee
So we see that the string has charge $+1$ with respect to $B$. In a
similar spirit the objects that would couple to a $p+1$ form gauge field
$A^{(p+1)}$ are $p$-branes (``pea brains''). 
These $p$-dimensional extended objects sweep out a $p+1$
dimensional world-volume $C_{p+1}$ if they propagate in time. The
coupling takes the form
\be
q\int_{C_{p+1}} A^{(p+1)}.
\ee
This coupling is invariant under gauge transformation  which shift $A$
by a $p$-form $\L$
\be
A \ra A  + d\L,
\ee
at least 
if the wave function of the $p$-brane transforms at the same time by
\be
\psi \ra e^{iq\int_{brane}\L} \psi,
\ee
which is the definition of charge $q$.

What are these mysterious extended objects that carry the charges of the
RR fields? They cannot be present in the perturbative string spectrum,
since we only find couplings to the field strength $F=dA$. (Since $F$ is
invariant under gauge transformations, the perturbative string states
are neutral.) String theory has therefore to be complemented with
non-perturbative states carrying the RR charges. These were found in a
brilliant proposal by Polchinsky and called D-branes \cite{polch}. These
D-branes are reviewed in great detail in a collection of excellent
lecture notes \cite{polch-review}, see also the lectures by Mike Douglas
at this school \cite{mike}. So we will skip their world-sheet
formulation. We cannot stress too much how important it is that for the
first time we have an exact description of non-perturbative solitons in
string theory.

\newsubsection{Compactification}

In string theory we will have to compactify the 10-dimensional
space-time manifold on a 6-dimensional compact space $X$. This space has
in general moduli $\f$ that take value in a moduli space $\cM_X$ that
parametrizes the inequivalent geometrical structures we can choose on
$X$. In perturbative string theory these moduli appear as the parameters
describing the family of sigma-models, or more general CFTs that can
appear as string vacua; the proper moduli space is therefore the ``quantum
moduli space'' that includes sigma-model quantum effects. It is
not necessary that the compactification fiber $X_x$ over the
4-dimensional space-time point $x\in \R^4$ is the same at each point; we
can allow small local variations. Stated otherwise, the moduli
$\f\in\cM_X$ become space-time scalar fields 
\be 
\f(x): M^4 \ra \cM_X .
\ee
Therefore we have the all-important relation 
\be 
\hbox{\it families of CFTs} \subset \hbox{\it vacua of string theory} .
\ee 
Here we write $\subset$
instead of $\cong$ since we realize that string theory can have moduli
such as the string coupling constant that are not present or difficult
to understand in string perturbation theory.

The amount of supersymmetry we expect in four dimensions will dictate
the combination of perturbative string plus compactification manifold.
We have gathered these data in
\tabelplus{
\renewcommand{\arraystretch}{1.5}
\begin{tabular}{|c||c|c|c|c|c|}
\hline
\strut  \it 4D susy & \it I/Het &  \it IIA &  \it IIB &  \it M &  \it F 
\\ \hline\hline
\strut $N=8$ & --- & $T^6$ & $T^6$ & $T^7$ & $T^8$ 
\\ \hline
\strut $N=4$ &  $T^6$ & $K3\times T^2$ & $K3\times T^2$ & $K3\times T^3$ & 
$K3\times T^4$ 
\\ \hline
\strut $N=2$ & $K3\times T^2$ & $CY_3$ & $\widetilde{CY}_3$ & 
$CY_3\times S^1$ & $\widetilde{CY}_3 \times T^2$
\\ \hline
\strut $N=1$ & $CY_3\cong\widetilde{CY}_3$ & \it orient. & \it orient .
& $G_2$ & $CY_4$
\\ \hline
\end{tabular}
\renewcommand{\arraystretch}{1.0}
}{Compactifications in string theory.}
A few scattered remarks about this table:
 The amount of supersymmetry in $4D$ is simply given by the
number of original supersymmetries in the uncompactified space-time
times the number of covariantly constant spinors on the compactification
space $X$. This latter number is determined by holonomy group of $X$. 
The holonomies are: trivial  for $T^n$, $SU(2)$ for $K3$, $SU(3)$ for
a Calabi-Yau three-fold $CY_3$, $G_2$ for a seven-dimensional manifold
with this exceptional holonomy, and $SU(4)$ for a Calabi-Yau four-fold
$CY_4$.

In this table all horizontal theories should give the same four
dimensional physics. There should therefore be relations among the
compactification data. In particular they should all have the same
moduli space. The two equivalent Calabi-Yau three-folds $CY_3$ and
$\widetilde {CY}_3$ are related by mirror symmetry.

In Type I or heterotic compactification we also have to pick
a gauge bundle $V\ra X$ over the compactification manifold $X$. This
gauge bundle should satisfy the topological constraints 
\be
c_1(V)=0,\quad c_2(V)=c_2(X).
\ee

We have studied in section \S\S8.9--12 the abstract structure of
string vacua. It corresponded to a family of 2d TFTs that satisfied
certain integrability conditions. We now want to make contact with the
space-time approach where they correspond with moduli spaces of
supersymmetric vacua. Indeed, we know that upon compactification we
obtain (in the low energy limit) a supergravity theory that contains a
supersymmetric sigma model with target space the moduli space $\cM_X$
of the compactification data $X$. This moduli space appears as the
expectation values of the scalar fields $\f^i$ and thus should be
interpreted as a moduli space of vacua of a single theory (string
theory respectively supergravity).

The scalar fields $\f^i$ will have an action that is highly
constrained by the requirements of supersymmetry. We have seen this in
detail for the scalar components of the so-called vector multiplets in
global $N=2$ supersymmetry in \S11.4 and this led directly to the
rigid special geometry for the moduli space $\cM$ as captured by the
prepotential $\cF$. The analogous discussion for local supersymmetry
can be done along the same lines and leads directly to local special
geometry that we met at various places. So we see that space-time
considerations give a direct interpretation of the geometry of the
moduli space of string vacua --- one of the themes of this lecture
series.

In this way we can reproduce the results from CFT obtained in \S7 and
\S8. However, there is one thing we can do from a space-time
perspective that is very difficult from the world-sheet point of view,
namely to explain the integer structure and the corresponding
canonical flat Gauss-Manin connection. Quite generally, the existence
of the integer structure is related to charge quantization and charge
lattices. Indeed, for a compactification on a Calabi-Yau $X$ we have
seen that the integer structure was related to the lattice
$H_3(X,\Z)$. This is the charge lattice for the the 4-form gauge field
of the type IIB string theory.  However, in perturbative string theory
there are no objects that carry charge with respect to this field. The
charged objects are 3-dimensional D-branes that can wrap around the
3-cycles of the Calabi-Yau, intrinsically non-perturbative objects
that are invisible from the CFT perspective. So the D-branes finally
complete our picture. In \S13 their role in determining the BPS
spectrum will be considered.

\newsubsection{Singularities revisited}

We have seen that the moduli spaces of string backgrounds might have
singularities. These singularities have often a completely
straightforward explanation. They simply parametrize the singular
geometries of the target space $X$. There is no reason why a sigma model
on such a singular space would have to make sense. However, once we
realize that the same moduli space is now used to label vacua of string
theory, we run into trouble. What will prevent the string of exploring
these singular points? And what happens at these points, where CFT and
thus string perturbation theory no longer makes sense?

Well, we have been at this point before. In the rigid special geometry
solution of Seiberg and Witten we also met singularities. At these
singularities the abelian gauge theory became infinite. But these
infinities had a good explanation. They were the result of integrating
out a nearly massless field, the monopole or dyon. If we had reinstated
this degree of freedom, the model would make perfect sense at these
points in moduli space.

It was the brilliant insight of Strominger \cite{strominger-sw}
that precisely the same thing
happens in string theory! For concreteness let's consider a
compactification of the Type IIB theory on a Calabi-Yau space $X$. 
At a typical singularity the space $X$ will develop a node. At this node
a three-cycle $C$ will shrink to zero, quite analogous to the shrinking of a
one-cycle in the case of the SW solution. Is there an accompanying
charged object that becomes  massless? Yes, Type IIB string theory
contains 3-branes
that can wrap around the cycle $C$. There mass will be given (at least
for BPS states) by the period of the holomorphic 3-form $\W$,
\be
M \sim \int_C \W .
\ee
So we see that in the singular limit $M\ra 0$, and we have precisely the
same situation. We have to introduce new degrees of freedom, describing
the quantum fluctuations of this new light soliton.

\newsubsection{String moduli spaces}

What is the general structure for moduli spaces of inequivalent string
vacua? In the low-energy limit string theory reduces to a particular
supergravity model, and the vacuum manifolds for theories with local
supersymmetry have been studied in great detail. In fact we always have
the following ingredients:

(1) A moduli space $\cM$ that carries
certain geometric structures (special K\"ahler, hyperk\"ahler, quaternionic) 
depending on the exact amount of supersymmetry. 

(2) A charge lattice $\G \cong \Z^r$ with rank $r$ equal to the
number of abelian gauge fields in the uncompactified dimensions\footnote{Here
the case $D=4$ is special since there $*F$ is also a two form and thus
an abelian gauge field gives rise to both electric and magnetic charges.
Consequently, in four dimensions the rank of the lattice is twice the
number of $U(1)$ gauge fields. Similarly, for $D=5$ there is one extra
generator.}. 

(3) A duality group $G$ that acts on $\G$. In many cases this
representation is actually irreducible. 

(4) A homomorphism $\pi_1(\cM) \ra G$, that is, a (necessarily flat) $\G$
bundle over the moduli space $\cM$. 

(5) A BPS spectrum $\cH_{BPS}\subset \cH$.

Let us make some comments on this last ingredient.
In supersymmetric theories the next step beyond the description of the
vacua is the spectrum of BPS states. For extended supersymmetry there is
a rich structure of such states. 
The BPS Hilbert space is graded by the charge lattice $\Gamma$ 
(as is the full space $\cH$)
\be
\cH_{BPS} = \bigoplus_{q\in \G} \cH^q.
\ee
The graded pieces $\cH^q$ have typically finite dimensions
\be
D(q) = \dim \cH^q.
\ee
In fact, it is interesting to consider the corresponding entropy, 
defined as
\be
S(q) = \log D(q).
\ee
Using relations with black hole physics, this entropy can be compared,
for large charges $q$, to the so-called Bekenstein-Hawking \cite{bh}
entropy 
$S_{BH}$ that can be computed by classical (super)gravitational methods,
\be
S(q) \mathop{\longrightarrow}^{q\ra\infty} S_{BH}(q).
\ee
This gives an indication of the growth of the degeneracy of BPS states.
In fact, in the interesting dimensions $d=4,5,6$ one finds that 
\be
S(q) \sim \threeoptions {2\pi |q|}{$d=6$} 
{2\pi |q|^{3/2}} {$d=5$} {2\pi |q|^4} {$d=4$}
\ee

\newsubsection{Example --- Type II on $T^6$}

As an example let us consider the compactification of the type
IIA or IIB theory on the six-torus $T^6$. In that
case one expects the following structure \cite{hull-townsend}: 

The ``classical'' moduli space is given by the homogeneous space
\be
\cM_{cl} = E_{7(7)}/SU(8).
\ee
Here $E_{7(7)}$ indicates the maximally noncompact real version of the Lie
group $E_7$, with 7 noncompact directions. 
The lattice $\Gamma$ has rank 56 and it forms an 
irreducible representation of the $U$-duality group
\be
G=E_{7(7)}(\Z)
\ee
a discrete group. The quantum moduli space is the quotient
\be
\cM = G\backslash \cM_{cl}
\ee
The flat vector bundle is the obvious one.

The degeneracies $D(q)$ of BPS states are expected to grow with entropy
\cite{e7}
\be
S(q) = 2\pi \sqrt{Q_4(q)},
\ee
where $Q_4$ is the unique quartic invariant of $E_7$ (that can
be used to {\it define} $E_7$). 
We see that this is already an incredibly rich structure.

\newsection{BPS states and D-branes}

We have seen that the next step beyond describing the string vacua is
the spectrum $\cH_{BPS}$ of BPS states. How are these BPS vector spaces
computed? This subject has seen a remarkable progress in the last year,
after the introduction of Polchinsky's D-branes. Seminal papers in the
field are, among others
\cite{polch,witten-bound,sen,vafa-gas,vafa-instanton,bsv}. There are two
cases that are relatively well-understood, the perturbative
states and the pure D-brane states, that we will now review.

\newsubsection{Perturbative string states}

Both in the type II and the heterotic string we can find BPS states in
the perturbative string spectrum. These are by far the easiest states to
understand. They are obtained by putting the right-handed bosonic and
fermionic oscillators in their ground state. This is most easily done in
the Green-Schwarz light-cone description of the physical Hilbert space
(see \cite{gsw}). 

Recall that in this description the spectrum is generated by the
following fields:

$8$ left-moving and $8$ right-moving bosonic fields $x^i(z)$ and
$x^i(\zbar)$ that transform as a vector ${\bf 8}_v$ under $SO(8)$; 

$8$ integer-moded left-moving fermionic fields $S^\a(z)$ that transform
as a spinor ${\bf 8}_s$ of $SO(8)$;

$8$ integer-moded right-moving fermions $S^{\dot\a}(\zbar)$ or
$S^\a(\zbar)$ that, depending on whether we are considering the type IIA
or IIB theory, transform as a (conjugated) spinor ${\bf 8}_c$
respectively ${\bf 8}_s$.

Note that the three representations ${\bf 8}_v$, ${\bf 8}_s$, ${\bf
8}_c$ of $SO(8)$ are related by triality (a necessary ingredient for
space-time supersymmetry) and all carry an invariant real inner
product.

These fields are quantized with the free action (written for the IIA
case)
\be
S  = {1\over \pi} 
\int d^2\!z \left(\half \d x^i \dbar x_i + S^\a\dbar S_\a +
S^{\dot\a}\d S_{\dot\a}\right).
\ee
 For future use we introduce the explicit mode
expansions 
\ba
\d x^i(z) \is \sum_k \a^i_k z^{-k-1},\nonu
S^\a(z) \is \sum_k S^\a_k z^{-k}.
\ea
Since the chiral fermions $S^\a$ are in the Ramond sector, their ground states 
form a representation of the Clifford algebra
\be
\{S^\a_0,S^\b_0\} = \delta^{\a\b}.
\ee
If $S^\a$ would have transformed in ${\bf 8}_v$, as is usually the case
for a Clifford algebra, the spinor representation would have been ${\bf
8}_s \oplus {\bf 8}_c$. Since $S^\a$ transforms here as ${\bf 8}_s$, one
finds (after applying triality) that the R ground states give the
representation ${\bf 8}_v \oplus {\bf 8}_c$. On the right-moving side,
we have a similar picture, possibly with the interchange of chirality.
In the GS formulation there is no GSO projection, therefore the ground
states have the form
\be
({\bf 8}_v \oplus {\bf 8}_c)\oplus
({\bf 8}_v \oplus {\bf 8}_{c,s})
\ee
The four terms in this product give the four sectors labeled
(NS,NS) through (R,R) in the covariant approach. In particular 
${\bf 8}_v\otimes {\bf 8}_v$ gives the light-cone modes of the
$G_{\m\n}$ and $B_{\m\n}$ field.

The other, massive physical states are produced by acting with the
creation operators $\a^i_{-n}$, $S^\a_{-n}$ on these ground states.
This gives a description of the Hilbert space tensor 
product of bosonic and fermionic Fock spaces both of rank 8.
These left-moving and right-moving Fock spaces have a gradation 
by the number operators $(N_L,N_R)$.

If we compactify on a torus $T^n$, we can fix momenta $(p_L,p_R)$ in the
Narain lattice $\G^{n,n}$. The physical Hilbert space is then defined as the
subspace satisfying the level-matching constraints
\be
\half p_L^2 + N_L = \half p_R^2 + N_R.
\ee
Note that this can be written in the signature $(n,n)$ inner product as
\be
\half p^2 =N_R-N_L.
\ee
The uncompactified momentum $k\in\R^{9-n,1}$ is fixed by the 
mass-shell condition
\be
\half k^2 + \half p_L^2 + N_L=0.
\ee
So for the type II string  compactified on $T^n$ the 
degeneracies $D(p)$ of physical states with momenta 
$p\in \G^{n,n}$ are given by
\be
D(p) = \sum_{n-m=p^2/2} d(n)d(m)
\ee
where the coefficients $d(k)$ are generated by 
\be
\sum_k d(k)q^k = 16 \prod_n \left( {1+q^n\over 1-q^n}\right)^8
\label{super}
\ee
which is the character of 8 fermionic and 8 bosonic Fock spaces
(the factor  $16$ counts the Ramond ground states).

For the heterotic string we tensor the left-moving supersymmetric 
spectrum-generating algebra we just described
with 24 right-moving transverse bosonic 
oscillators $x_R^I(\zbar)$. The compactified momenta $(p_L,p_R)$ now 
take value in the extended Narain lattice
\be
\G^{n,n+16}\cong \G^{n,n}\oplus (-E_8) \oplus (-E_8).
\ee
The right-moving degeneracies at oscillator level
$N_R$ are given by $c(N_R-1)$ with
\be
\sum c(k)q^k ={1\over {\displaystyle q \prod_n (1-q^n)^{24}}}
={1\over \eta(q)^{24}},
\label{eta}
\ee
the character of 24 bosonic Fock spaces.
Because of the bosonic intercept (which produces massless states
at level $N_R=1$) the level-matching now gives
\be
\half p_L^2 + N_L = \half p_R^2 + N_R -1.
\ee
Note that this can be written as
\be
\half p^2 =N_R-N_L -1.
\ee
So the full spectrum of perturbative heterotic string states with
$p\in\G^{n,n+16}$ takes the
form
\be
D(p) = \sum_{n-m=p^2/2} d(n)c(m).
\ee

\newsubsection{Perturbative BPS states}

For BPS states we simply put the left-moving supersymmetric oscillators 
in their ground states. That is we impose the conditions
\be
\a^i_n |\bps\>=S^\a_n|\bps\>=0, \qquad n>0.
\ee
Why is this a BPS state? To understand this we have to consider the
supersymmetry algebra. In the light-cone description of the physical
states, the (left-moving) supersymmetry generators are given by
(we put $p_+=1$)
\ba
Q^\a \is S_0^\a,\nonu
Q^{\dot\a} \is \oint \d x^i \g_i^{\dot\a \b} S_\b.
\ea
We can decompose $Q^{\dot\a}$ in zero-mode piece and a non-zero mode 
$Q^{\dot\a}_+$ piece as 
\be
Q^{\dot\a}=p^i\g_i^{{\dot\a} b}S_{0,\b} + Q^{{\dot\a}}_+.
\ee
On ground states $Q^{{\dot\a}}_+$ will act trivially, by definition, 
\be
Q^{{\dot\a}}_+|\bps\>=0.
\ee
So, BPS states satisfy the condition
\be
(\e_\a Q^\a + \e_{\dot\a} Q^{\dot\a})|\bps\>=0,
\ee
with
\be
\e^a = p^i\g_i^{{\dot\a} a}\e_{{\dot\a}}.
\ee
Precisely the same argument holds for the heterotic string. So we see
that $8$ of the $16$ left-moving supercharges annihilate these states.
Consequently, for the type II and the heterotic strings these are $1/4$
and $1/2$ BPS states respectively. 

The degeneracies of these BPS states is entirely given by the
right-moving states, that can be arbitrary as long as level matching is
satisfied. So we find that the number $D(p)$ of perturbative BPS states with
given charge $p$ are given simply by 
\be
D(p)=\option {d(\half p^2)} {\quad\hbox{\it type II}}
{c(\half p^2)} {\quad \hbox{\it heterotic}}
\ee

\newsubsection{D-brane states}

A second class of BPS states that are now reasonable well-understood are the
D-brane states, that can appear in the type II and type I string. We will
concentrate on the type II states here. 

In general these D-brane BPS states are believed to appear as follows
(see \eg \cite{bsv})
In a compactification on a Calabi-Yau space $X$ there
exist special embedded subspaces $C\subset X$ called
supersymmetric cycles. To such a cycle $C$ we can associate a charge $q$
in the homology lattice 
\be 
q=[C]\in H_*(X,\Z).
\ee
We can further associate to the cycle $C$ a moduli space $\cM_C$ that parametrizes
the inequivalent supersymmetric embeddings.
The BPS states should appear by ``quantizing'' this classical moduli
space $\cM_C$. In a naive way quantization means computing
the ground states of the supersymmetric particle moving on the moduli
space. These ground states are represented as
harmonic differential forms. So roughly we have 
\be
 \cH_q \approx H^*(\cM_C)
\ee
However, there are various subtleties in this reasoning that we will
partly explain and partly sweep under the rug (see \eg
\cite{morrison,harvey-moore-II} for a critical analysis).

What kind of cycles are appropriate? Here we have to distinguish between
the type IIA and IIB theories. Recall that we had two sets of bosonic
fields: NS fields and RR fields. Consequently we have a
decomposition of the charge lattice as 
\be
\G = \G_{NS} \oplus \G_{RR}
\ee
In the NS sector we find the graviton $G_{\m\n}$, 2-form
$B_{\m\n}$ and the dilaton $\F$. In the RR sector we have a series of
generalized $k$-forms gauge fields $A^{(k)}$, where $k$ can take
all odd or even values in the IIA or IIB theory respectively. Moreover
the total RR curvature 
\be
F=\sum dA^{(k)}
\ee
satisfied the self-duality condition $*F=F$.

If we compactify on $X$ we can have abelian gauge fields out of both
sectors. The NS sector only gives a contribution if $X$ is the product
with a torus $T^n$. Then we have the following charges: $n$ momenta
(from the metric $G$) and $n$ winding numbers (from the 2-form field
$B$) giving\footnote{There are two exceptions to this rule: When we
compactify down to five dimensions we can dualize the $B$ field to give
one more magnetic charge. In compactifications to four dimensions all
gauge fields can be dualized, doubling the lattice.}
\be
\G_{NS} = \G^{n,n} .
\ee
In the RR sector we produce after compactification an abelian gauge
field out of a $(p+1)$ form $A^{p+1}$ for every element $C_p\in
H_p(X,\Z)$. The ``push-down''
\be
A = \int_{C_p} A^{p+1}
\ee
gives a one-form in the uncompactified space-time. So
the RR charge lattice is given by 
\be
\G_{RR} = \option {H_{even}(X)}{\ \ \ \ type IIA}
{H_{odd}(X)}{\ \ \ \ type IIB}
\ee
In the type IIA theory the most general D-brane state with charge
\be
p \in H_{even}(X)
\ee
is realized as a sum of irreducible even-dimensional
cycles $C_i$ with multiplicity
$n_i$
\be
p = \sum_i n_i [C_i].
\ee
In the Type IIA theory we demand that the cycles $C\subset X$ are
holomorphically embedded for a complex structure on $X$ compatible with
the given Ricci flat metric \cite{beckers-strominger}. 
We further need the additional date of a
choice of holomorphic vector bundle $E_i \ra C_i$ of rank $n_i$ over each
irreducible component. There is a concrete proposal for what the
relevant moduli space $\cM_C$ of such cycles should be: the moduli space
of simple, semi-stable coherent sheaves $\cS$ with Mukai vector
\cite{morrison,harvey-moore-II}
\be
Ch (\cS) \,\sqrt{{\rm td}\, X} = p .
\ee

In the type IIB theory we are dealing with odd dimensional cycles. Let
us assume for a moment that we are dealing with a proper Calabi-Yau
three-fold that is simply connected, so that $H_{odd}(X)=H_3(X)$. The
correct notion of a supersymmetric three-cycle is now special Lagrangian
\cite{beckers-strominger,syz,ooy}. 

A special Lagrangian $L$ satisfies two
conditions \cite{calibrations}: (1) it is a Lagrangian, \ie the K\"ahler
symplectic form vanishes when restricted to $L$, and (2) it minimalizes
the volume in its homology class. More precisely, it has the property
that (with an appropriate phase factor $e^{i\th}$) the period of the
holomorphic three-form $\W$ satisfies
\be
\int_L e^{i\th}\W = \vol(L).
\ee
For an irreducible BPS cycle we also have to pick a line bundle over
$L$, as we did in the IIA case. The special Lagrangian calibration was introduced because its
properties closely match those of holomorphic cycles that we met on
the IIA side. In fact, Strominger, Yau and Zaslow \cite{syz}
have been able to
proof that the moduli space $\cM_L$ has some very nice properties. It is
always a K\"ahler manifold, even a Calabi-Yau space, and its complex
dimension is given
by
\be
\dim_\C \cM_L = b_1(L).
\ee

We will now consider a few interesting examples of compactifications
where we know how to compute the BPS spectra.

\newsubsection{Example --- Type IIA on $K3$ = Heterotic on $T^4$}

Our first example concerns the compactification of the Type IIA theory
on $K3$ \cite{vafa-gas,sen}. 
By string duality this is equivalent to a compactification of
the heterotic string on the four-torus. This compactification produces a
supersymmetric theory in six dimensions with $N=2$ supersymmetry. The
charge lattice is of signature $(4,20)$ and even, self-dual
\be
\G = \G^{4,20}.
\ee
This lattice can be understood as the Narain lattice from the point of
view of the heterotic string. 

Viewed from the type IIA theory it is
the (even) homology lattice of a $K3$ surface $X$
\be
\G = H_*(X,\Z),
\ee
which labels the RR charges. It indeed carries an intersection 
form that is even and self-dual. (There are no perturbative charged states
in this compactification.) The duality group is 
\be
G = O(4,20,\Z) = \Aut(\G) .
\ee
In the heterotic description these transformations are all perturbative
$T$-duality symmetries. In the $K3$ compactification they correspond to
quantum automorphisms of the $K3$ manifold \cite{k3}.

For $K3$ surfaces the quantum moduli space is completely known \cite{k3}. 
The duality group is
\be
G = O(4,20,\Z),
\ee
which is the automorphism group of the unique even, self-dual
lattice of signature $(4,20)$
\be
\G^{4,20} = H \oplus H \oplus H \oplus (-E_8) \oplus (-E_8).
\ee
The quantum moduli space is given by the Narain space
\be
\cM_{K3} =  G \backslash O(4,20,\R)/O(4,\R) \times O(20,\R).
\ee
One of the implication of string duality is that both compactifications
should also have the same moduli space, which is indeed the case, since
the above Narain space is well-known to classify the heterotic
compactifications (see the discussion in \S7.4).

The BPS spectrum has a straightforward perturbative description in terms
of the heterotic string, it is simply a Fock space built on the lattice
$\G$. So the
degeneracy $D(p)$ for a state with charge $p\in\G$ is given by
\be
D(p)=c(\half p^2)
\ee
with generating function $\eta^{-24}$, see (\ref{eta}). According to the
string duality hypothesis, these states should have an alternative
description in terms of quantizing D-branes on $K3$.

This idea has been in fact (partly) confirmed. There are various cases
where the moduli space of D-branes can be quantized and does give the
required degeneracies. We have seen that in IIA theory a general D-brane
configuration is a set of holomorphic\footnote{Holomorphic for a complex
structure on $X$ that is compatible with the hyperk\"ahler structure
that is parametrized by the moduli space $\cM_X$.} subvarieties
$C_i\subset X$ with multiplicities $n_i$ 
\be
p = \sum_i n_i [C_i],
\ee
and the moduli space $\cM_C$ of such cycles
is some set of appropriate sheaves. The strong version of string 
duality says that
any two elements $C,D$ of $H_*(X)$ with the same
self-intersection number give rise to isomorphic moduli spaces
\be
\cM_C \cong \cM_{D}
\ee
related by the duality group $O(4,20,\Z)$.
Of course for the group $O(3,19,\Z)$, which is the image of $\Diff(X)$
in the second cohomology group, this is a trivial statement. As it
stands, the strong duality statement is definitely not true with the
above definition of $\cM_C$ in terms of coherent sheaves, it most likely
needs quantum corrections.

One can show that under certain conditions the
moduli space $\cM_C$ is isomorphic to the Hilbert scheme $X^{[N]}$ of
subschemes of length $N$ on an algebraic $K3$ surface $X$,
\be
\cM_C \cong X^{[N]},\qquad N=\half p^2.
\ee
Crucial in the correspondence with the heterotic description is the
result of G\"ottsche \cite{goettsche}, who computed the Hodge numbers of
these Hilbert schemes. The Hilbert scheme can be understood as a
resolution of the orbifold symmetric product spaces
\be
X^{[N]} \opra^\pi S^n X = X^N/S_N.
\ee
For algebraic surfaces, the Hilbert scheme is always smooth and its
cohomology can be computed unambiguously. For the symmetric products we
can do a similar computation, but we must be careful that we use the
appropriate orbifold Euler character \cite{hirzebruch,vafa-witten}. 
Anyhow, the result is the generating function
\be
\sum_{N\geq 0} q^N \chi(S^NX) = \prod_{n>0} {1\over (1-q^n)^{24}} =
{q\over \h(q)^{24}}.
\label{gen}
\ee

 There is a clear intuitive picture
why we get a Fock space. Let $\a_{-1}^\m$ ($\m=1,\ldots,24$)
be a basis of $H_*(X)$. The suffix ${}_{-1}$ will be explained shortly.
Now there are obvious cohomology classes that we can construct
in the $N$-th symmetric product of $X$, one simply takes symmetric
products of the classes on $X$ 
\be
\a_{-1}^{\m_1} \cdots \a_{-1}^{\m_N}.
\ee
If this was everything the generating function (\ref{gen}) would
read $1/(1-q)^{24}$. However, there are an infinity of mirror images of
the classes $\a_{-1}^\m$. In the diagonal $X$ in the symmetric product $S^2X$
we find a second copy of this class, denoted as $\a_{-2}^\m$. Similarly,
if a point on $X^N$ is left invariant by a cycle of length $n$ we get a class
$\a_{-n}^\m$. The (orbifold) cohomology is the symmetric algebra on
all these generators (creation operators) $\a_{-n}^\m$, \ie a Fock space.

One case where the correspondence with Hilbert schemes
is made rather easy, is the situation of $N$ zero-branes
and one four-brane. In that case we have $p^2=2N$, and the 
moduli space is indeed given by
\be
\cM_C = X^{[N-1]},
\ee
so that we compute
\be
D(p)=\chi(X^{[N-1]})=c(N),
\ee
which checks with the heterotic prediction $c(p^2/2)$.

A particular other interesting case has been studied in \cite{bsv}
and by Yau and Zaslow in \cite{yz}, namely the case
that $C$ is an irreducible curve of genus $g$ with class $p=[C]\in
H_2(X)$. In that case the adjunction formula expresses the genus as
\be
p^2 = 2g -2.
\ee
The moduli space $\cM_C$, that parametrizes such holomorphic
curves together with a choice of a
holomorphic line bundle, is fibered over the local system $\P^g$ that
parametrizes just the family of curves
\be
\cM_C \opra^\pi \P^g.
\ee
The fiber of this map is the Jacobian ${\rm Jac}(C)$ of line bundles on
$C$. This Jacobian is generically a torus, ${\rm Jac}(C) \cong T^{2g}$.
Therefore the Euler character $\chi(\cM_C)$ seems naively to vanish,
since the Euler number of the fiber is zero. There is a catch in this
argument, however, if the curve is singular and
degenerates to a curve of lower genus with nodes.
Over these exceptional subspaces the dimension of the fiber drops. Now in
the case of a total degeneracy to a genus zero curve, the fiber is a
point. Only these points,  corresponding to nodal rational curves, can
contribute to the Euler number. Therefore there is a beautiful prediction
from string duality for the number of rational curves with $g$ nodes on
a $K3$ surface: it is given by the coefficient $c(g-1)$ of the
generating function $1/\eta^{24}$. This checks with the explicit
computations that have been done for low number of nodes \cite{yz}.

\newsubsection{Example --- Type II on $T^4$}

Our second example concerns the compactification of the Type IIA or IIB
string on a four-torus. (T-duality will relate the two types.)
This model has $N=4$ supersymmetry in six dimensions. It has both NS
and RR charged states. In fact, the duality group is
\cite{hull-townsend}
\be
G = O(5,5,\Z),
\ee
that acts on the rank 16 charge lattice as a chiral ${\bf 16}$
{\it spinor} representation.
Of course, there is a perturbative T-duality subgroup
\be
T=O(4,4,\Z) \subset G,
\ee
under which the charge lattice decomposes as
\be
\G = \G_{NS} \oplus \G_{RR}.
\ee
Here both lattices are isomorphic to the even, self-dual
Narain lattice $\G^{4,4}$. For
the NS lattice this is true by construction. Note that
 it transforms as a ${\bf 8}_v$ under the
T-duality group. The RR lattice we obtain (say in the Type IIA) theory)
as
\be
\G_{RR} = H_{even}(T^4,\Z).
\ee
This has indeed an even, self-dual
 intersection product of signature $(4,4)$, as we have seen in \S9.1. 
So we also
obtain $\G_{RR} \cong \G^{4,4}$. Note however that this lattice transforms
as a spinor ${\bf 8}_s$ under the T-duality group $O(4,4,\Z)$. (The Type
IIB lattice $H_{odd}(T^4,\Z)$ transforms as the conjugate spinor ${\bf
8}_c$.) 

Now we can make two computations. The perturbative states, with
$p\in\G_{NS}$, have a direct computable degeneracy in terms of a
left-moving superstring partition function
\be
D(p) = d(p^2/2),
\ee
with generating function (\ref{super}). The full duality group $G$
allows us to transform these perturbative states in pure D-brane states
with $p\in\G_R$. Here we have to make a similar computation as we did
for the case of $K3$. 

There is a conjecture with ample support that the relevant moduli space
of torsion free coherent sheaves is indeed always a hyperk\"ahler
deformation of the Hilbert scheme or symmetric product $S^N T^4$
\cite{ama}. The relevant formula is now that the cohomology of these
symmetric products produces the chiral superstring partition function.
This is indeed the case, since
\be
\sum_{N\geq 0} q^N \chi(S^NT^4) = 
\prod_{n>0} \left( {1+q^n\over 1-q^n}\right)^8.
\ee
This is again a powerful check on string duality. 

For a general
state with charge $p$ in the full charge lattice $\G=\G_{NS}\oplus\G_R$
the degeneracy is given by the following formula
\cite{5brane}. There is natural
bilinear map ${\bf 16}\otimes {\bf 16} \ra {\bf 10}$ in $O(5,5,\Z)$ given by the
Dirac matrices. So $\half p^2$ is naturally a ten-dimensional vector in the
Narain lattice 
\be
\half p^2 \in \G^{5,5},
\ee
with components $\half p^\a\g^\m_{\a\b}p^\b$, $\m=1,\ldots, 10$. In
general such a vector will be a multiple $N$ of a primitive
vector. The complete duality invariant formula for the degeneracy is now
\be
D(p)= d(N).
\ee

\newsubsection{Example --- Type II on $K3\times S^1$ = Heterotic on $T^5$}

Our following example is a compactification down to five dimensions,
best understood first from the heterotic perspective. In a compactification
on $T^5$ we have clearly a
 Narain lattice $\G^{5,21}$ of perturbative charges. However, the
full charge lattice is one dimension higher,
\be
\G = \G^{5,21} \oplus \Z,
\ee
because of the following reason. The $B$ field, which has a three-form
field strength, can be dualized in five dimensions to produce an extra
gauge field $A$ with $dA=*dB$. We will denote this extra charge as $m$.
States with non-zero $m$ charge are magnetic monopoles for the $B$
field. Elementary string states can only carry electric $B$ field
charge. So, a generically charged BPS state with non-zero magnetic
charge $m$ cannot be described in terms of perturbative heterotic string
states and we are stuck. (States with $m=0$ can of course be described
in CFT terms, but they are special. For example, their degeneracies do
not grow as fast as the generic states with $m\not=0$.)

On the other hand, there is a nice description of these states from the
type IIA formulation in terms of strings and D-brane states on the
manifold $S^1\times K3$. Here we have a different decomposition of the
charge lattice \cite{strominger-vafa}. From the NS charges we first have
the $\G^{1,1}$ Narain lattice associated to the $S^1$ factor. The
type IIA $B$-field gives an additional magnetic charge
\be
\G_{NS}= \G^{1,1}\oplus \Z
\ee

In the type IIA  formulation the RR charges come from the homology of
the $K3$ factor,
\be
\G_{RR} = H_*(K3,\Z) \cong \G^{4,20}
\ee

We first consider the moduli space associated to a D-brane state with R
charge $p_R \in \G_{RR}$. As we have seen, this moduli space is
conjectured to be isomorphic to (a hyperk\"ahler resolution) of the
symmetric product $S^NK3$, with $N=\half p_R^2-1$. In section \S13.4 we
computed the cohomology of this moduli space, or in physical terms we
considered supersymmetric quantum mechanics on it. Now we are dealing
with a fiber bundle $K3\times S^1$. In the limit that the $K3$ fiber is
much smaller then the circle\footnote{Recall that we are computing an
integer topological invariant, so we are allowed to pick a convenient
metric.}, we might assume by an adiabatic argument that the D-branes are
supersymmetric for each point of the circle. In this way we obtain a map
\be
S^1 \ra S^N K3,
\ee
\ie an element of the loop space $\cL(S^NK3)$.
We now want to quantize this loop space. Note that it has a
natural circle action, corresponding to rotations of the $S^1$, and the
appropriate object to compute will be the $S^1$-equivariant Euler
character of the loop space. In physics terms this means we have to
consider a two-dimensional $N=2$ supersymmetric
sigma model with as target space this
symmetric product \cite{strominger-vafa}. 

BPS states will correspond to states that are in right-moving
ground states. For any target space $X$, the spectrum of such states is
computed by the elliptic genus \cite{ell-genus}
\be
\chi(X;q,y) = \Tr_\cH\left((-1)^F y^{F_L}q^{L_0-{c\over 24}}\right),
\ee
where $\cH$ is the Hilbert space of the sigma model with RR boundary
conditions on the fermions. If $X$ is a Calabi-Yau space of complex
dimension, $\chi(X;q,y)$ has beautiful modular properties. It is a weak
Jacobi form of weight zero and index $d/2$ (with, as usual, $q=e^{2\pi i
\t},$ $y=e^{2\pi i z}$). This implies that the elliptic genus has an
expansion in non-negative powers of $q$
\be
\chi(X;q,y) = \sum_{n\geq 0,\ k} c(n,k) q^n y^k.
\ee
The coefficients $c(n,k)$ have a topological definition in terms of
characteristic classes on $X$. 
One result we will need is the elliptic genus of $K3$.
It has many representations. One that is easy to derive in the orbifold
limit $K3=T^4/\Z_2$ is the following expression in classical Jacobi
theta functions
\be
\chi(K3;q,y) = 8 \sum_{\a=2,3,4} \left({\vartheta_\a(\t,z) \over
\vartheta_\a(\t,0)}\right)^2
\ee

Since $K3$ is a CY space, all its symmetric products will also be CYs.
We can thus consider their elliptc genera. In fact, there is a nice
formula for the generating function of the elliptic genera of the
symmetric products, which generalizes G\"ottsche's result \cite{dmvv}
\be
\sum_{N\geq 0} p^N \chi(S^NK3;q,y) =\prod_{n>0,m,k}
(1-p^nq^my^k)^{-c(nm,k)}.
\ee
Here $c(m,k)$ are the expansion coefficients of the elliptic genus of
$K3$. Let us expand the RHS in Fourier coefficients
\be
\sum_{n,m,k} d(n,m,k) p^n q^m y^k.
\ee

According to the D-brane picture of the BPS states
the final formula for the degeneracy of a BPS state with $p_R\in\G_R$ and 
magnetic charge $m$ and spin $J$ is now given in terms of the elliptic
genera of the symmetric products $S^NK3$ as
\be
d(\half p_R^2,m,J).
\ee
Using the above explicit formula, one can evaluate the asymptotic of
these degeneracies and finds the expected entropy \cite{dyon}
\be
S(p_R,m,J) \sim 2\pi \sqrt{\half m p_R^2 - J^2}
\ee

\newsubsection{Example --- Type IIA on $X$ = Type IIB on $Y$}

In our last example we compactify the IIA or IIB theory on a Calabi-Yau
three-fold to four dimensions with $N=2$ supersymmetry.

One of the most interesting applications of D-brane BPS states has been
to the issue of mirror symmetry. The usual statement of mirror symmetry
tells us that the Type IIA theory on a Calabi-Yau three-fold $X$ is
equivalent to the Type IIB theory compactified on a dual Calabi-Yau
three-fold $Y$, with $X$ and $Y$ related by the mirror map. Indeed, we
have seen how Type IIA and Type IIB were simply related by flipping the
sign of the right-moving $U(1)$ charge. The two string theories are
believed to be equivalent in the perturbative sector, which is the only
thing CFT arguments will teach us anything about. It is no more than
natural to conjecture that this equivalence extends to the
nonperturbative sectors, and in particular to the D-brane BPS states. 

Now we have also seen that the description of D-brane states was very
different in the IIA and IIB theories. In the IIA setup we had to find a
holomorphic subvariety $C\subset X$ with charge
\be
[C] \in H_{even}(X).
\ee
The actual BPS states were obtained by quantizing the moduli space
$\cM_C$ associated to $C$. This moduli space is most likely the space of
some appropriate sheaves, possibly with ``quantum'' corrections.

In the Type IIB theory the same BPS states should arise by quantizing
the moduli space $\cM_L$ that parametrizes the inequivalent special
Lagrangians $L\subset Y$ with charge
\be
[L] \in H_3(Y).
\ee

Now the ``strong mirror conjecture'' claims that not only do the BPS state
spaces obtained from $\cM_C$ and $\cM_L$ agree, the actual moduli spaces
should be isomorphic
\be
\cM_C \cong \cM_L.
\ee

This conjecture has an immediate and powerful implication \cite{syz}: it
allows us to directly construct the manifold $X$ from the geometry of
the mirror manifold $Y$. The argument is extremely simple and clear:
There is one very special supersymmetric cycle on the IIA manifold $X$
--- a point $P\in H_0(X)$. The moduli space all possible ``zero-branes''
$P$ is of course the original space $X$ itself,
\be
\cM_P = X.
\ee
So if mirror symmetry holds this
 means that on the IIB manifold $Y$ there must be one very
special Lagrangian three-cycle $L$ with the property that
\be
\cM_L \cong X.
\ee
The moduli space of $L$ must therefore be a Calabi-Yau three-fold. As we mentioned
the complex dimension of $\cM_L$ is given by $b_1(L)$. So this condition
forces the three-cycle $L$ to be a three-torus
\be
L \cong T^3.
\ee
So we find that, in order for a mirror pair $(X,Y)$ to exist, the Calabi-Yau space
$Y$ must allow a fibration by three-tori over some
 (real) three-dimensional base manifold $B$,
\be
Y \opra^\pi B,\qquad \pi\inv(x) \cong T^3.
\ee
$B$ parametrizes the different special Lagrangian embeddings of $L$.
The full moduli space $\cM_L$ also takes into account the choice of a
line bundle on $L$. But the space of line bundles on a torus $T^3L$ is
nothing but the Jacobian or dual torus $Jac(T^3)=\hat T^3$. So this gives
$\cM_L$ and thus also the IIA manifold $X$ as a dual fibration
\be
X = \cM_L \opra^{\hat\pi} B,\qquad \hat\pi\inv(x) \cong \hat T^3.
\ee
Now going from $T^3$ to the dual torus $\hat T^3$ is nothing but a
T-duality, which explains the title of \cite{syz} ``mirror symmetry is
T-duality.'' This picture has been confirmed in examples
\cite{mirror-examples}.

\end{document}